\documentclass[sigconf,balance=false]{acmart}
\usepackage{popets}

\setcopyright{popets}
\copyrightyear{YYYY}

\acmYear{YYYY}
\acmVolume{YYYY}
\acmNumber{X}
\acmDOI{XXXXXXX.XXXXXXX}
\acmISBN{}
\acmConference{Proceedings on Privacy Enhancing Technologies}
\usepackage{comment}
\usepackage{cancel}
\usepackage{subcaption}
\usepackage{xspace}
\usepackage{xcolor}
\usepackage{hhline}
\usepackage[utf8]{inputenc}

\usepackage{tikz}
\usepackage{paralist}
\usepackage{tabularx}
\usepackage{booktabs}
\usepackage{multirow}
\usepackage{makecell}
\usepackage{url}
\usepackage{xurl}  %

\newcommand{\parheading}[1]{\vspace{2pt}\noindent{}\textbf{{#1}}}
\newcommand{\point}[1]{\vspace{2pt}\noindent{}\textit{{#1}}}
\newcommand{\eg}{\textit{e.g.,}}
\newcommand{\ie}{\textit{i.e.,}}
\newcommand{\etc}{etc.}
\newcommand{\etal}{et al.}
\newcommand{\wrt}{w.r.t.}
\newcommand{\system}{\textsc{BehaVR}}
\newcommand{\steam}{SteamVR}
\newcommand{\users}{20}
\newcommand{\apps}{20}
\newcommand{\vrdevice}{Quest Pro}
\newcommand{\alvr}{ALVR}
\newcommand{\bodydata}{body motion}
\newcommand{\eyedata}{eye gaze}
\newcommand{\handdata}{hand joints}
\newcommand{\facedata}{facial expression}
\newcommand{\Bodydata}{Body Motion}
\newcommand{\Eyedata}{Eye Gaze}
\newcommand{\Handdata}{Hand Joints}
\newcommand{\Facedata}{Facial Expression}

\newcommand{\appadv}{app adversary}
\newcommand{\devadv}{device adversary}
\newcommand{\Appadv}{App Adversary}
\newcommand{\Devadv}{Device Adversary}

\newcommand*\circled[1]{\tikz[baseline=(char.base)]{
            \node[shape=circle,draw,inner sep=1pt] (char) {\footnotesize#1};}}
\usepackage[normalem]{ulem}

\begin{document}

\title{\system: User Identification Based on VR Sensor Data }

\author{Ismat Jarin*†}
\orcid{0009-0002-6406-0603}
\email{ijarin@uci.edu}
\affiliation{%
  \institution{University of California, Irvine}
  \city{} %
  \state{} %
  \country{} %
}

\author{Yu Duan*}
\orcid{0009-0005-9376-7175}
\email{duany12@uci.edu}
\affiliation{%
  \institution{University of California, Irvine}
  \city{} %
  \state{} %
  \country{} %
}

\author{Rahmadi Trimananda}
\orcid{0000-0002-9900-7506}
\email{rtrimana@uci.edu}
\affiliation{%
  \institution{University of California, Irvine}
  \city{} %
  \state{} %
  \country{} %
}

\author{Hao Cui}
\orcid{0000-0002-7574-2004}
\email{cuih7@uci.edu}
\affiliation{%
  \institution{University of California, Irvine}
  \city{} %
  \state{} %
  \country{} %
}

\author{Salma Elmalaki}
\email{salma.elmalaki@uci.edu}
\affiliation{%
  \institution{University of California, Irvine}
  \city{} %
  \state{} %
  \country{} %
}

\author{Athina Markopoulou}
\orcid{0000-0003-1803-8675}
\email{athina@uci.edu}
\affiliation{%
  \institution{University of California, Irvine} 
  \city{} %
  \state{} %
  \country{} %
}

\renewcommand{\shortauthors}{Jarin, Duan et al.}

\begin{abstract}
Virtual reality (VR) platforms enable a wide range of applications, however, pose unique privacy risks.  In particular, VR devices are equipped with a rich set of sensors that collect personal and sensitive information (\eg{} \bodydata{}, \eyedata{}, \handdata{}, and \facedata{}).
The data from these newly available sensors can be used to uniquely identify a user, even in the absence of explicit identifiers.
In this paper, we seek to understand the extent to which a user can be identified based solely on VR sensor data, {\em within and across} real-world apps from diverse genres.
We consider adversaries with capabilities that range from observing APIs available within a single app (app adversary) to
observing all or selected
sensor measurements across multiple apps on the VR device (device adversary). 
To that end, we introduce \system{}, a framework for collecting and analyzing data from {\em all} sensor groups collected by {\em multiple} apps running on a VR device. We use \system{} to collect data from real users that interact with \apps{} popular real-world apps. We use that data to build machine learning models for user identification within and across apps, with features extracted from available sensor data. We show that these models can identify users with an accuracy of up to 100\%, and we reveal the most important features and sensor groups, depending on the functionality of the app and the  adversary.  %
To the best of our knowledge, \system{} is the first to analyze user identification in VR comprehensively, \ie{} considering all sensor measurements available on consumer VR devices,
collected by multiple real-world, as opposed to custom-made, apps.

\end{abstract}

\maketitle
\def\thefootnote{*}\footnotetext{The two authors made equal contributions and share first authorship.}\def\thefootnote{\arabic{footnote}}
\def\thefootnote{†}\footnotetext{Corresponding author.}\def\thefootnote{\arabic{footnote}}

\section{Introduction}
\label{sec:introduction}

\begin{figure} [t!]
\centering
\includegraphics[width=0.4\textwidth]{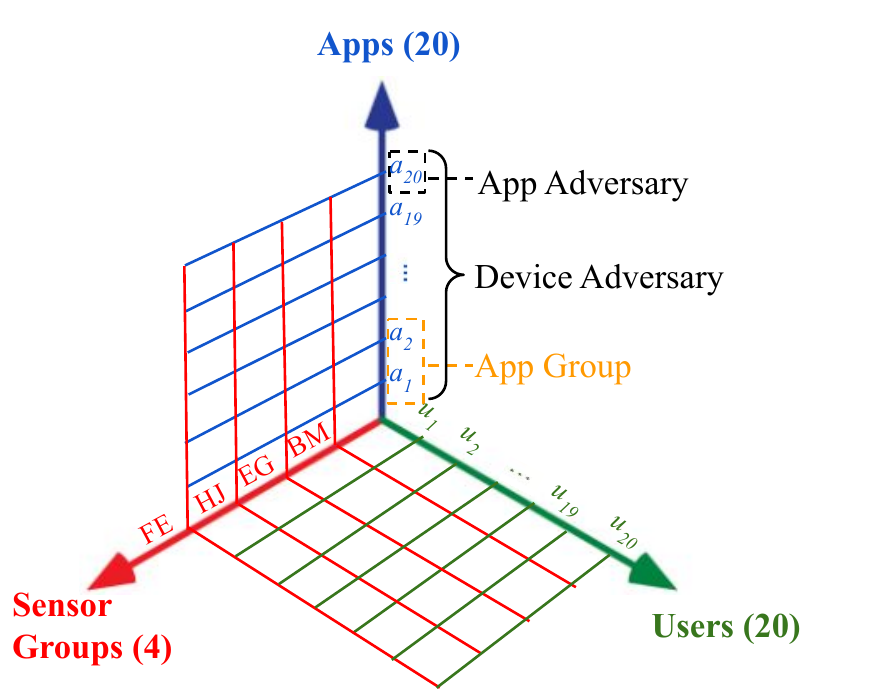}
\caption{\system{} problem space  spans several dimensions: users, apps, and sensors. We consider four sensor groups: \bodydata{} (BM), \eyedata{} (EG),  \handdata{} (HJ), \facedata{} (FE). We consider \apps{} real-world apps covering vast domains of VR apps. We have two types of adversaries: the \appadv{} has access only to one app; the \devadv{}  has access across multiple apps. We further define App Groups as having similar activities and emotional states.
}
\label{fig:problem_space_fig}
\end{figure}

\begin{figure*}[t!]
	\centering
 \includegraphics[width=.98\linewidth]{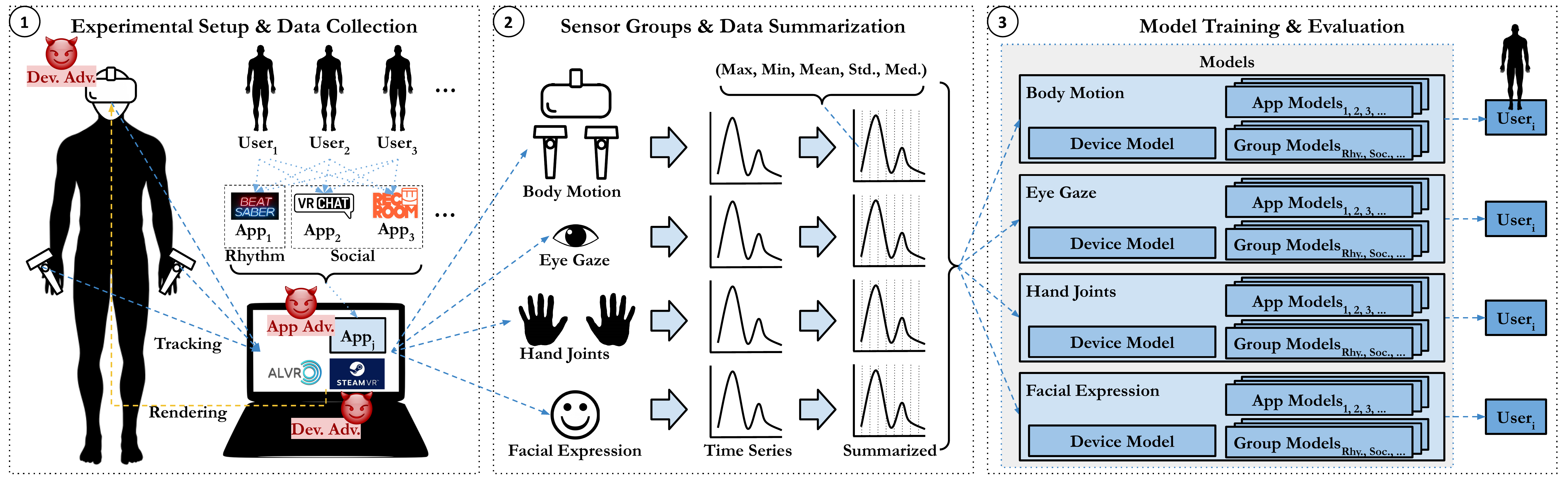}
	\caption{\textbf{Overview of \system{}.} 
        (1) \textbf{Data Collection Setup:} every user interacts with each app using \vrdevice{}; each app (\eg{} Beat Saber) runs on a PC and its VR environment is rendered on the \vrdevice{} headset; this enables the recording of sensor data sent from \vrdevice{} to the PC; apps are grouped based on similarity of activities and emotional states.
        (2) \textbf{Data Processing:} there are four groups of sensors, namely 
        \textit{\bodydata{}}, \textit{\eyedata{}}, \textit{\handdata{}}, and \textit{\facedata{}}; we divide the time series generated by every sensor group into blocks, and we compute 5 \textit{statistics} per block as features.
        (3) \textbf{Model Training \& Evaluation:} using the previous features per block, we train different models (using data per app, across apps, even per group of apps) that an adversary can use to uniquely identify users.
        }
	\label{fig:behavr-workflow}
\end{figure*}

Virtual reality (VR) is a large and growing market~\cite{vrgrowth} that enables a wide range of apps, from gaming to education~\cite{vrforeducation}, and work~\cite{vrforwork}.
Meta Quest, one of the most popular consumer VR devices, has sold nearly 20 million units as of February 2023~\cite{quest_sale_count}.
\steam{}, the largest VR gaming platform, has over 7,800 VR apps as of May 2024~\cite{steamvrnumberofapps}.
The VR ecosystem also comes with privacy concerns.
Recent work showed that Oculus VR and its apps already collect personally identifying information~\cite{trimananda2022ovrseen}, and can further infer sensitive attributes~\cite{nair2022metadata}.
Some of this tracking and profiling are similar to practices in other app ecosystems, such as mobile~\cite{taintdroid, shuba2018nomoads}, smart TV~\cite{varmarken2020tv, mohajeri2019watching}, web~\cite{degeling2018we}, \etc{} with some differences: the VR ecosystem is younger, more centralized, and not driven by ads, yet~\cite{trimananda2022ovrseen}. 

\parheading{User Identification.}VR has access to a rich set of sensors that capture sensitive, personal information. Consumer VR devices (\eg{} Meta \vrdevice{}), including their headsets and controllers, are equipped with sensors that collect measurements about head and body motion (``BM'')~\cite{bodytracking, openxrcoordinate}, eye gaze (“EG”)~\cite{eyetracking, eyegazepositionrotationkhronos}, hand joints (“HJ”)~\cite{handtracking, openxrhandtracking}, and facial expression (``FE'')~\cite{facetracking, openxrfacetracking}. 
All these measurements are available on the device itself (\eg{} \vrdevice{}), can be sent to the platform (Meta), and a subset can be made available to app developers via APIs. Recent works~\cite{miller2020personal, nair2023unique} have shown that some of these measurements can indeed be used for unique identification.
The privacy implication is that a user's behavior in VR creates {\em implicit identifiers}%
\footnote{The combination of features extracted from the VR sensor data streams can produce unique 
fingerprints, such as behavioral biometrics (or ``behaviometrics'') \cite{stephenson2022sok,pfeuffer2019behavioural}.}
that can be used to identify users in the virtual world,
even in the absence of \textit{explicit identifiers} (\eg{} device IDs or user accounts) that are often well protected by permissions.
Such implicit identifiers based on sensor measurements may remain effective in the case of shared devices (\eg{} shared among family members, coworkers, or public platforms), multiple accounts or devices per user, anonymized and released VR sensor dataset \etc{} (more details in Section \ref{sec:threat-model}). Identification using implicit identifiers has explored in mobile platforms~\cite{de2013unique} and recently in VR \cite{nair2023unique,miller2020personal}. %

In this paper, we broadly refer to the sensor measurements collected on VR devices, as well as to features extracted from them, as VR sensor data.
We are interested in understanding {\em to what extent a user can be uniquely identified based on VR sensor data across \apps{} real-world apps} from diverse genres, from social (\eg{} VRChat) to education/training (\eg{} X-Plane 11), from entertainment (\eg{} BeatSaber) to virtual offices (\eg{} Job Simulator) among others; and which are the {\em top features, across real-world apps and sensor groups} , for an adversary that wants to uniquely identify a user with minimal effort. In particular, we consider two types of adversaries, depending on their vantage point for access to sensor data: (1) the {\em \appadv{}} mimics an app developer who has access to sensor data from APIs available within the app; and (2) the {\em \devadv{}} can have access to sensor data collected across multiple apps (see Section \ref{sec:threat-model}). The full problem space we consider is depicted in Fig.~\ref{fig:problem_space_fig}.

\parheading{Comparison with Prior Work.} 
Prior work has considered only parts of our problem space.
In terms of sensor groups, user identification has been demonstrated based on \bodydata{} (\eg{}~ positional and rotational) sensor data from VR devices~\cite{miller2020personal,nair2023unique,tricomi2022you,nair2022metadata}, \ie{} the BM sensor group in our problem space.
Newer VR devices, such as the Meta \vrdevice{} and Apple Vision Pro, are equipped with more sensors that track other body parts, including eyes, hands, and face~\cite{facetracking,eyetracking,visionproblog}.
Privacy aspects of these sensors have been studied before \cite{kaleido_USENIX21,nair2022metadata}, but their use for identification has not been, and neither has been their comparison for identification purposes.
\system{}, for the first time, explore all available sensor groups for identification.
In terms of the experimental setup, prior work has focused on either one specific app and task (\eg{} Beat Saber in~\cite{nair2023unique}), or custom apps specifically designed for their studies~\cite{miller2020personal,tricomi2022you,nair2022metadata}.
In \system{} experiments, participants interact with \apps{} unmodified commercial VR apps, under limited guidance, better representing real-world scenarios of user identification.

\parheading{Approach.} 
We introduce \system{}, a framework for collecting and analyzing data from all available sensor groups (\ie{} \bodydata{}, \eyedata{}, \handdata{}, and \facedata{}), and performing user identification within (\ie~ \appadv{}) and across (\ie~ \devadv{}) apps. %
To the best of our knowledge, \system{} is the first to analyze user identification in VR comprehensively, \ie{} considering {\em all} sensor measurements available on a VR device, and across multiple commercial apps.
Fig.~\ref{fig:behavr-workflow} presents the overview of \system{}. Next, we describe the \system{} components and we highlight methodological contributions along the way.

\point{(1) Collection of sensor data from real-world apps on the VR device.}
We develop an approach to observing, for the first time, all the sensor data in real-time during gameplay.
We instrument ALVR, an open-source streaming app that is essential for Meta Quest devices (and other popular VR headsets) to play \steam{} apps~\cite{alvr}, to record all sensor data by listening to the API calls.
This gives visibility into data collected by real-world apps running unmodified on the VR device, which was not previously possible. Using the \system{} setup, we perform a user study and collect a comprehensive dataset that covers all four sensor groups, consisting of around 400 sensor data records from users interacting with \apps{} popular apps on the \steam{} store (see Section~\ref{subsec:data-collection}). %

\point{(2) Sensor Data Analysis and Feature Engineering.}
\system{} is the first to explore all available sensor groups for identification in real-world apps.
The comprehensive sensor data and diverse app genres pose unique challenges for data processing.
First, unlike prior work that focuses on specific tasks or deals with fixed time blocks~\cite{miller2020personal} %
\system{} dataset exhibits high variability in session time across users, apps, and sensors.
To process variable-length time series of sensor data into time blocks %
we propose a new %
time block-division approach that is robust to the variability (see Section~\ref{subsec:time-series-abstraction}). 
Second, in addition to the standard features extracted from the VR sensor data \cite{miller2020personal,nair2023unique}, we introduce data augmentation and selection that are adapted for new sensor groups: for \eyedata{}, we add new features that correlate left and right eye's data; for \facedata{}, we explore facial elements and their combinations that represent users' emotions \cite{facsexplained, openxrfacetracking} when interacting with an app. In addition, our feature analysis reveals how activities and valence/arousal states in different apps generate key identification features. %

\point{(3) Identification Models and Evaluation.}
We evaluate user identification in diverse real-world scenarios that cover different adversary capabilities.
We train a Machine Learning model per sensor group for user identification.
The identification model predicts on each time block and maximum voting~\cite{miller2020personal} is used to produce the final label per user. %
Depending on the adversary's capability, the model is trained on sensor data from one app (app adversary), one app group, or all apps (device adversary) and evaluated on the same or different setting (\textit{open-world-setting}) of the same app; or a completely different app (\textit{zero-day scenario}) from same or different app groups (Section \ref{sec:evaluation}). We discuss the generality of \system{} (Section~\ref{subsec:discussion_behavrPerspective}) and provide recommendations for privacy practitioners %
(Section \ref{subsec:recommendation}).

\parheading{Identification Insights.} 
Section~\ref{sec:evaluation} presents a comprehensive evaluation across sensor groups, apps or groups of apps, and adversary models, guided by the following research questions (RQs):

\point{RQ1: How well a user can be identified using VR sensor data?} %
We find that the adversaries can achieve up to $100$\% accuracy for many apps, especially using data from \facedata{} and \bodydata{} that perform better than \eyedata{} and \handdata{}. 

\point{RQ2: How long does it take to identify a user?}
The app adversary generally requires around $18-20$ seconds of data across \bodydata{} and \facedata{} for up to 90\% , and $\sim$50 seconds for \eyedata{} for up to $85$\% accuracy. 
The device adversary requires less data ($\sim$9 seconds on average), since it combines data across apps. 

\point{RQ3: What are the top features for identification \wrt{} various apps and adversaries?}
We observe that for the unique identification, the top features describe the unique interaction between users and VR environment as well as user's physical characteristics (\eg{} height).

\point{RQ4: Can we identify a user across different settings of same app or across different apps?} 
We find that \appadv{} can provide 60-100\% accuracy with new app settings (\ie{} open-world-settings). \Devadv{} can identify users in a new app (\ie{} zero-day) from the same (70-100\%) or different (5-40\%) app group.

\point{RQ5: What are the most important sensor groups for identification?}
Apps from different groups show sensor groups importance based on app activity and emotional states. %
Knowing the relative importance of different sensor groups allows the adversary to effectively train models or help users to decide which sensor groups to share.

\parheading{Outline.} 
The rest of the paper is structured as follows:
Section~\ref{sec:background} provides background and the problem setup. %
Section~\ref{sec:experimental-setup} presents the experimental setup and data collection.
Section~\ref{sec:methodology} presents the data analysis and model training.
Section~\ref{sec:evaluation} presents the evaluation for app and device adversaries, for all sensor and app groups.
Section~\ref{sec:related-work} reviews related work. Section~\ref{sec:Discussion} and \ref{sec:conclusions} provides discussion and conclusion respectively.
The appendices provide additional results. %

\section{Background and Problem Setup}
\label{sec:background}

\subsection{VR Hardware and Platform} \label{subsec:vr_Hardware_Platform}

There are many different VR platforms and setups that require varied software and hardware combinations.
In this paper, we focus on \steam{}, the most popular VR gaming platform with over 7,800 VR apps~\cite{steamvrnumberofapps} and millions of users~\cite{steamvr_user_count}.
In the \steam{} setup, the VR apps run on a personal computer (with either Windows or Linux system), and a compatible VR device is connected to stream the graphics and track user actions via its sensors.
A streaming software needs to be installed on the PC and the VR device to transmit the graphics and sensor data~\cite{steam_link,alvr,relive_for_vr}.

In our experimental setup, we use ALVR~\cite{alvr} as the streaming app. ALVR is open-source and thus eases the instrumentation (see Section~\ref{sec:experimental-setup}).
As for the hardware, we choose Meta Quest Pro for testing
~\footnote{Note that \steam{} differs from Oculus VR, Meta's VR platform that runs VR apps natively on its Android-based system~\cite{meta_store}.},
because the headset is equipped with the most comprehensive VR sensors, including body motion and eye gaze (which are supported by older VR devices like Quest 2), as well as hand joints and facial expression data (which are increasingly supported by newer generations of VR devices). \system{} leverages \steam{} to run apps and ALVR~\cite{alvr} to record all sensor data by listening to the API calls (see Section \ref{sec:experimental-setup}).
SteamVR (and ALVR) supports for many other consumer VR devices, notably HTC Vive Focus, ByteDance Pico and Apple Vision Pro~\cite{alvr}.
We expect that our study is generalizable to any devices supported by SteamVR. 

\subsection{Sensor Groups} \label{subsec:sensor-data-types}
We explore all VR sensors available on today's consumer VR devices, \ie{} the following four groups: \bodydata{} (BM)~\cite{bodytracking, openxrcoordinate}, \eyedata{} (EG)~\cite{eyetracking, eyegazepositionrotationkhronos}, \handdata{} (HJ)~\cite{handtracking, openxrhandtracking}, and \facedata{} (FE)~\cite{facetracking, openxrfacetracking}.
These sensors are available to developers through device-independent OpenXR APIs~\cite{openxrstandard},
as well as captured by \alvr{}~\cite{alvr} in the \system{} setup.
Depending on the device and platform, additional permissions may be required to access specific sensors, \eg{} \vrdevice{} requires permissions for EG, and FE.
However, in the SteamVR setup~\cite{steamvr} that we use, apps run on PC and there is no permission check for collecting sensor data.
On the \vrdevice{}, the ALVR app~\cite{alvr} requests initial permissions for FE and EG during installation. Thereafter, it operates without additional runtime permission requests for all \apps{} apps in our experiment.

We follow the data structure definitions from the OpenXR standards~\cite{openxrstandard}. The main elements of the data structures are \textit{position}, \textit{rotation}, \textit{linear} and \textit{angular velocities}. Position, and linear and angular velocities are expressed in $x$, $y$, and $z$ values of the Cartesian right-handed coordinate system, and rotation is expressed in $x$, $y$, $z$, and $w$ values of the Quaternion coordinate system. Additional information regarding sensor groups is provided in the Appendix~\ref{app:sensor_groups}.

\subsection{VR Apps}\label{subsec:app-selection-grouping}

\begin{table*}[t!]
  \scriptsize
  \centering %
  \caption{Grouping apps ($a_1, .., a_{20}$ listed in Table~\ref{tab:list-of-apps}, Appendix~\ref{app:list-of-steamvr-apps}) based on their similarity of activities and emotional states (arousal/valence).  {\em Sensor Groups:}  BM, EG, HJ, FE. {\em Emotional States:} LA = low arousal, HA = high arousal, PV = positive valence, NV = negative valence. 
  }
  \begin{tabular}{l p{22mm} p{76mm} l l}
    \toprule
    \textbf{App Groups} & \textbf{App No.} & \textbf{App-Specific Activities}  & \textbf{Arousal/Valence} & \textbf{Important Sensors}
    \\
    \midrule
      Social & $a_{12}$, $a_{15}$, $a_{18}$ & Walking, waving, grabbing and sightseeing/exploring virtual environment & LA/PV, HA/PV  &  BM, EG, FE, HJ \\
      Flight Simulation & $a_3$,  $a_{19}$, $a_{20}$ & Holding onto the airplane control stick, interacting with control panel/buttons in an airplane cockpit & LA/NV, HA/NV, LA/PV & BM, HJ, FE\\ %
      Golfing & $a_6$ & Slow walk, holding a golf stick, and put the ball towards hole   & LA/PV, HA/PV  & BM \\ %
      Interactive Navigation & $a_{2}$, $a_{9}$, $a_{10}$, $a_{11}$, $a_{16}$, $a_{17}$  & Grabbing, moving objects, opening doors, \ie{} frequent interaction with virtual objects & Neutral, LA/PV, LA/NV & BM, EG, HJ \\
      Knuckle-walking & $a_7$   & Walking using an open fist like a gorilla, sightseeing/exploring virtual environment & LA/PV, HA/PV, LA/NV & BM, HJ, FE\\ %
      Rhythm & $a_1$  & Dancing-like moves and cutting objects in quick pace & All & BM,  HJ, FE, EG\\ %
      Shooting \& Archery & $a_{13}$, $a_{14}, a_{5}$  & Grabbing and holding a gun/arrow, aiming and shooting at objects  & LA/NV, HA/NV & BM, EG, FE, HJ \\ %
      Teleportation & $a_4$, $a_8$  & sightseeing by teleportation (instead of walking) \ie{} without extensive body movement& All & FE \\ %
    \bottomrule
  \end{tabular}
  \label{tab:Apps-Grouping}
\end{table*}

\parheading{App Selection.}
Starting from the top 100 apps from the ``Most played VR games'' list on Steam ~\cite{apprankings}, we  select \apps{} VR apps based on several criteria. 
First, we exclude apps that may cause inconvenience to most users, \eg{} horror or violence genre. This first criterion is mandated in 45 CFR § 46.111(a)(1) to minimize the experimental risk (\eg{} physical or psychological harm) on our study participants~\cite{irbminimalrisk}. Second, we exclude apps without complete VR support, like those for VR devices other than \vrdevice{} or needing both VR controllers and a PC keyboard for input.
Finally, we attempt to compile a rich set of apps from various genres, \eg{} social, entertainment, flight, gaming \etc{}
The list of \apps{} \steam{} apps is shown in Table~\ref{tab:list-of-apps} in Appendix~\ref{app:list-of-steamvr-apps}, referred to as $a_{1}$, .., $a_{20}$, throughout the paper.

\parheading{Apps Grouping.} \label{subsec:app-group} 
We group apps based on the similarity of their activities and emotional states, considering an adversarial point of view. Our motivation is to leverage app similarities for cross-app identification and zero-day scenarios, \ie{}, using data from multiple similar apps to better identify users within the same group. %
Although heuristic, our app grouping performs well (see Section \ref{subsec:OpenWorld_AppGroup} and Table \ref{tab:IdenAcc_bigTable}) and serves as proof of concept, however, an adversary can further optimize it.
We propose app grouping in Table~\ref{tab:Apps-Grouping}. 
For BM and HJ, we group apps according to similar app-specific activities; \eg{} social apps require walking, waving, and exploring, contrarily shooting apps require targeting and shooting objects -- leading to different motion patterns. Note that, even within same group, differences exist; \eg{} for shooting group, $a_{13}$ requires teleporting, while $a_{14}$ requires walking.

For \facedata{} only, app grouping further considers emotional \textit{arousal} (\eg{} how calm or active an emotion is) and \textit{valence} (\ie{} how positive or negative an emotion is) states induced by the VR environment of an app; we use the approach proposed in~\cite{kuppens2013relation,suhaimi2018modeling,bouchard2011emotions}. %
There are four types of arousal-valence states we have considered in our study, namely high arousal positive valence (HA/PV), \eg{} happiness; low arousal positive valence (LA/PV), \eg{} surprise; high arousal negative valence (HA/NV), \eg{} fear/stress; and finally low arousal negative valence (LA/NV), \eg{} sadness. Different app environment may induce any of these states. For example, we observe that social apps induce mostly joy or surprise (\ie{} HA/PV), and flying/shooting apps induce mostly fear/stress (\ie{} HA/NV). 
Furthermore, one app can induce multiple emotions (\eg{} when completing a level in a game, users feel happy if they succeed and sad if they fail). See Table \ref{tab:Apps-Grouping} for detailed lists of app groups and their associated valence/arousal states.

The last column in Table~\ref{tab:Apps-Grouping} lists the important sensor groups for each app group.  %
Sensor group importance arises from app-specific activities and emotional states, which increase with the active use of specific sensor groups or are influenced by strong valence/arousal states of the app. Consequently, these data are adequately available to the adversaries.
For example, in flight apps, users use controller/hands and induce emotions like surprise, fear %
thus, \bodydata{}, \handdata{} and \facedata{} are listed as the important sensor groups for them. The importance of these sensor groups for user identification is confirmed in the evaluation (see Section \ref{subsec:comparative-evaluation_AppVsSensor}).

\subsection{Current Practices Regarding VR Sensor Data}

Different VR platforms and apps have different practices regarding sensor data collection, use, and sharing. We looked into privacy policies and permissions to better understand those practices.

\subsubsection{VR Sensor Data in App's Privacy Policies}\label{sec:privacy-policy}
Privacy laws, such as GDPR~\cite{gdpr} and CCPA~\cite{ccpa}, require disclosure of data collection, use and sharing practices. 
Both CCPA (in §1798.140(c)) and GDPR (Article 4(14)) define \textit{behavioral characteristics} as part of ``biometric information'' or ``biometric data'' that can uniquely identify a person. This motivates us to %
look into real-world VR apps and platforms and their disclosure of VR sensor data. %

We look at the top 100 apps from the ``Most played VR games'' list on Steam~\cite{apprankings} and download their privacy policies. 
As of May 2023, only 60 apps provided a privacy policy. 
Our authors first read and check all privacy policies to understand how they disclose the collection of VR sensor data. We looked for statements on ``biometric data'' or ``sensory data'', as well as more specific types (\eg{} ``head movement'') in any of the sections. Then, we used string matching and ChatGPT to scan the whole text again to ensure that we do not miss any content. We found that only a few (10 out of 60) privacy policies discuss the collection of VR sensor data, and some make conflicting statements. Additional details can be found in Appendix~\ref{app:privacy-policy}.
These observations are aligned with the findings in~\cite{trimananda2022ovrseen}, that many VR apps did not provide a privacy policy or did not disclose VR sensory data collection adequately. Meta, the maker of \vrdevice{}, indicates in its privacy policy that they collect data and use it for personalization~\cite{metapolicy}.
Unity, the top game engine that many VR apps build on, claims the collection of biometric information for the purpose of identifying an individual. Unity explicitly mentions ``hand and face geometry'' (HJ and FE sensor groups) as examples of biometrics that may be collected~\cite{unity-policy}.
This further motivates us to study how well real-world VR apps and VR platforms can identify users based on their VR activities.

\subsubsection{VR Sensor Data Permissions}\label{app:permission} 

What sensor groups an app has permission to access depends on the platform.

The \system{} study in this paper, is based on \apps{} apps from the SteamVR Store \cite{steamvrstore}. Our review of each app's website on SteamVR revealed no information about which sensor data are collected or what their collection purposes are. Furthermore, as detailed in Section ~\ref{subsec:vr_Hardware_Platform}, SteamVR apps do not have runtime permission constraints that prevent apps from reading any sensor data, whether the app needs them for functionality or not.

We also looked at how these same \apps{} apps are used beyond the SteamVR Store, on other popular platforms, particularly  MetaQuest Store ~\cite{metastore}. We found 10 of our \apps{} apps available there. Of these, 6 apps disclose the types of sensor data they collect. All 6 reports are collecting BM data by default. One app, VRChat ($a_{18}$) discloses collecting EG, HJ, and FE data, and another app, RecRoom ($a_{15}$), collects FE data only. We also identified that the Job Simulator app ($a_9$) discloses the collecting of HJ data. Meta apps have runtime permission checks to protect FE, HJ, and EG ~\cite{metapolicy}, while BM is more widely available without permission checks.

\subsection{Threat Model} \label{sec:threat-model}

\subsubsection{VR Threat Scenarios.}\label{subsec:ThreatModel_Threat_Scenarios} 
The adversary's goal is to identify users based on VR sensor data. During training, the adversary observes users in VR apps, records and analyzes their sensor data streams, and creates models for user identification. During evaluation, the adversary observes new sensor data streams and uses the trained models to identify the user who generated them.

It is worth noting that if explicit identifiers (such as device IDs, user accounts, or software IDs) are available, they are straightforward to use for user identification. Instead, our focus is on using VR sensor data alone as implicit identifiers for user identification.
Unique identification without explicit identifiers has been studied using different data in the past, ranging from mobile location data ~\cite{de2013unique}, to body motion data in VR recently \cite{nair2023unique,miller2020personal}. %
It has also privacy implications: an adversary can identify and track a user based on the VR sensor data {\em alone}, w/o necessarily having access to the explicit identifiers; \ie{}
the adversary can obtain access to VR sensor data in various ways: directly (an honest-but-curious developer or 3rd party library without access to device IDs as they are often well protected by permissions), or by compromising any of the above, %
or through an anonymized and released dataset. 
The question has also implications for anonymity (or lack thereof) in the virtual world\footnote{Several scenarios where implicit identifiers can be useful are:(1) unlike mobile devices that are personal, VR devices and user accounts can be shared by a group of people (\eg{} among family members, friends, public VR game-stores or education platforms~\cite{IEEEVRMeusum19}, among coworkers \cite{VRforTrainingSH21});  
(2) one user may use multiple  accounts for one or multiple apps, multiple devices or avatars for privacy or other reasons. }: even if a user changes their VR device, account, app, or avatar, they can still be (re-)identified based on their sensor data. On a positive note, the uniqueness of VR sensor data can potentially be used for authentication ~\cite{luooculock}.

\subsubsection{\system{} Adversaries.}
\label{subsec:ThreatModel_Adversaries}Next, we define two types of adversaries, depending on their vantage points and sensor data access, described next and depicted in ``\circled{\textbf{1}}'' in Fig.~\ref{fig:behavr-workflow}. %
Both adversaries build models using individual sensor groups for identification. This emphasizes the importance of considering scenarios where an adversary may have access to only one sensor group rather than all sensor groups. Other scenarios include the adversary aiming to minimize its effort by utilizing fewer data or users may not use certain sensor groups; \eg{} for HJ models, we assume the adversary only utilizes HJ sensor groups for identification, not other groups such as BM, EG, or FE. {\footnote{The adversary cannot utilize BM and HJ simultaneously since users use the controller and hand alternatively in VR. As a result, the adversary may receive zero or corrupted values from controllers (part of the BM sensor group) while using a hand. In addition, we project the scenario where FE/EG can be disabled by users as well. }} \system{} adversaries are as follows: %

\parheading{\Appadv{}.} The app adversary ($adv_{app}$) has access to sensor data collected from APIs available within a single app. This mimics an app developer and any third party that has the same permissions and receives the data from the app, \eg{} Unity\cite{Unity_Analytics}. When ALVR app \cite{alvr} grants full sensor permissions, the SteamVR apps do not request any runtime permission. Therefore, we assume that SteamVR apps may access or collect all sensor groups (see more details in Section \ref{subsec:vr_Hardware_Platform} and \ref{subsec:sensor-data-types}). The \appadv{} corresponds to the client adversary in the taxonomy in~\cite{garrido2023sok} and has been previously studied for unique identification~\cite{nair2023unique, miller2020personal,tricomi2022you,nair2022metadata}.

\parheading{\Devadv{}.} The \devadv{} ($adv_{dev}$) has access to all four sensors collected from multiple VR apps. Realistic examples of the \devadv{} include the device manufacturer, a game company that releases multiple apps, third parties with access to multiple apps' data (\eg{} Unity \cite{Unity_Analytics}, SteamVR\cite{steamvr}), or malware that records sensor data.
Additionally, its capabilities are available to the VR device (\eg{} \vrdevice{}) and its operating system, as well as to the PC in our \steam{} setup or a compromised library with functionality similar to ALVR.
The \devadv{} corresponds to the hardware or client adversaries %
in the taxonomy in~\cite{garrido2023sok}.
To the best of our knowledge, it has {\em not} been previously studied for identification in VR, since collecting sensor data across multiple real-world apps was not previously possible.

Both adversaries collect users' sensor data trace $D$ from the four sensor groups discussed in Section~\ref{subsec:sensor-data-types}. Both
train models to identify a user $u_i$ among $n$ users ($u_1,.., u_n$). Their main difference is, that the \appadv{} has access to data from one app, while the \devadv{} has access to {\em multiple} apps' data. %
Both adversaries train one model per sensor group, which predict the label for each sensor data block and can combine all labels of blocks through max voting per user~\cite{MaxVote}. 
Furthermore, the \devadv{} may %
train on data from all apps, or groups of similar apps as defined in Section~\ref{subsec:app-selection-grouping}.
See Section~\ref{subsec:model-building} for details on the models \footnote{We use the terms ``adversary'' and ``model'' interchangeably.} %
and~\ref{sec:evaluation} for their evaluation.

\section{Experimental Setup}%
\label{sec:experimental-setup}

\subsection{Using \alvr{} as a Vantage Point}\label{subsec:steamvr-alvr}
In our study, we used \vrdevice{}, the latest state-of-the-art VR device from Meta, released in November, 2022. While other VR devices also work with \steam{}~\cite{steamvr},  \vrdevice{} collects \eyedata{} and \facedata{} data~\cite{eyetracking,facetracking}, in addition to \bodydata{} and \handdata{} data~\cite{handtracking}----collected by older VR devices; \eg{} Meta Quest 2. 

The \system{} data collection system is depicted in ``\circled{\textbf{1}}'' in Fig.~\ref{fig:behavr-workflow}. Although Oculus OS on \vrdevice{} can natively run VR apps, we use the \steam{} setup \cite{steamvr} that allows us to intercept and record sensor data. In \system{}, \vrdevice{} sends sensor data to a PC that runs a VR app and the \vrdevice{} and PC are connected via WiFi. \system{} integrates the \alvr{} \cite{alvr}, an open-source software that can run VR apps on a PC. With the help of \steam{}, \alvr{} can run Steam apps that provide VR support on \vrdevice{}: the sensor data sent from \vrdevice{} are received by \alvr{} and become input to the app to process and render the app's VR environment in real-time. Finally, the app sends the rendering results back to the \vrdevice{}, so the headset displays the VR environment to the user. 
While, the \steam{} was intended to enhance VR performance by performing heavy tasks on PC, we use it for passive data monitoring, for the first time.
We instrument parts of \alvr{}'s source code that receive sensor data from the \vrdevice{} (\ie{} by creating hooks on the four sensor groups data streams) and save the data 
as time series.

\subsection{User Study and Data Collection}\label{subsec:data-collection}

We conducted an IRB-approved user-study from our institution's IRB review committee. 
We recruited participants aged 20-40, with an equal gender split to better represent the diverse demographic of VR users~\cite{userstatistics}. Please see the participant distribution summary in Appendix \ref{app:statistics}. 
The data collection was performed by three authors and required $\sim5-6$ hours per participant (including briefing and training) to collect $\sim400$ sensor data records (\apps{} real-world apps per user, $\sim$$3$ months in total for \users{} users). Each participant was compensated \$10/h, declared in IRB and participant consent form.

Each user was asked to wear a \vrdevice{} headset, and interact with all \apps{} apps in our corpus  
(see Table~\ref{tab:list-of-apps} in Appendix~\ref{app:list-of-steamvr-apps}). 
A research team member provided rough prompts to the VR user during user-app interaction.\footnote{\label{footnote:prompts}For example, in Golf It! ($a_{6}$), we give users the prompt: ``Please putt the golf ball into the hole, in the beginning with controller, then with bare hands''. It is up to the user in what way and how many times they putt the ball.}
These prompts guide users in interacting with each app, according to the purpose of the app, but users have freedom to interact with the app in their own  way and pace. Meanwhile, \system{} (\ie{} the instrumented version of \alvr{}) was running and recording sensor data from \vrdevice{}. For each app, a user completes the app-specific activity twice (\ie{} two sessions) 
and we collect two data traces, whose duration was typically $\sim3-4$ minutes: the first trace is for model training \& validation. For evaluation, we utilize few/all (seconds) data from second session of the same app, or from new/different settings of the same app, or from different apps based on our adversarial set-up %
(see Section \ref{subsec:model-building}).

\subsubsection{Dataset Summary and Size.}\label{subsec:UserStudy_study_size}
The number of participants in our user study (\users{}) is on-par with most prior user studies that collected data from participants, \eg{} \cite{miller2022combining,liebers2021understanding,pfeuffer2019behavioural,tricomi2022you,nair2022metadata} considered 16-50 participants. This number is smaller than 500 participants in \cite{miller2020personal}, who however, performed simple tasks compared to our work that considers multiple real-world apps. %
It is also smaller than 50K users in~\cite{nair2023unique}, in which the authors considered one popular real-world app (Beat Saber) and body motion data, in a dataset provided by BeatLeader~\cite{BeatLeader}; this number of users is obviously impossible to involve in in-person user studies. In terms of duration, we recorded around $3-4$ minutes per user for two sessions (see Fig.~\ref{fig:session-durations}), which compares to~\cite{nair2023unique} (median of $\sim$3 minutes per session) and~\cite{miller2020personal} (five 20-second videos for a total of 1 minute 40 seconds). %

\section{Data Analysis and Model Training} 
\label{sec:methodology}
This section presents the \system{} pipeline for processing the sensor data (Section~\ref{subsec:time-series-abstraction}, also see ``\circled{\textbf{2}}'' in Fig.~\ref{fig:behavr-workflow}) and for model training (Section~\ref{subsec:model-building}, also see ``\circled{\textbf{3}}'' in Fig.~\ref{fig:behavr-workflow}) .
\subsection{Sensor Data Summarization}\label{subsec:time-series-abstraction}
Here we convert the sensor data streams into feature vectors, which are suitable for a non-sequential model (\eg{} Random Forest).

\begin{figure}[t!]
    \centering
    \begin{subfigure}{0.43\textwidth}
        \flushright
        \includegraphics[width=.94\textwidth]{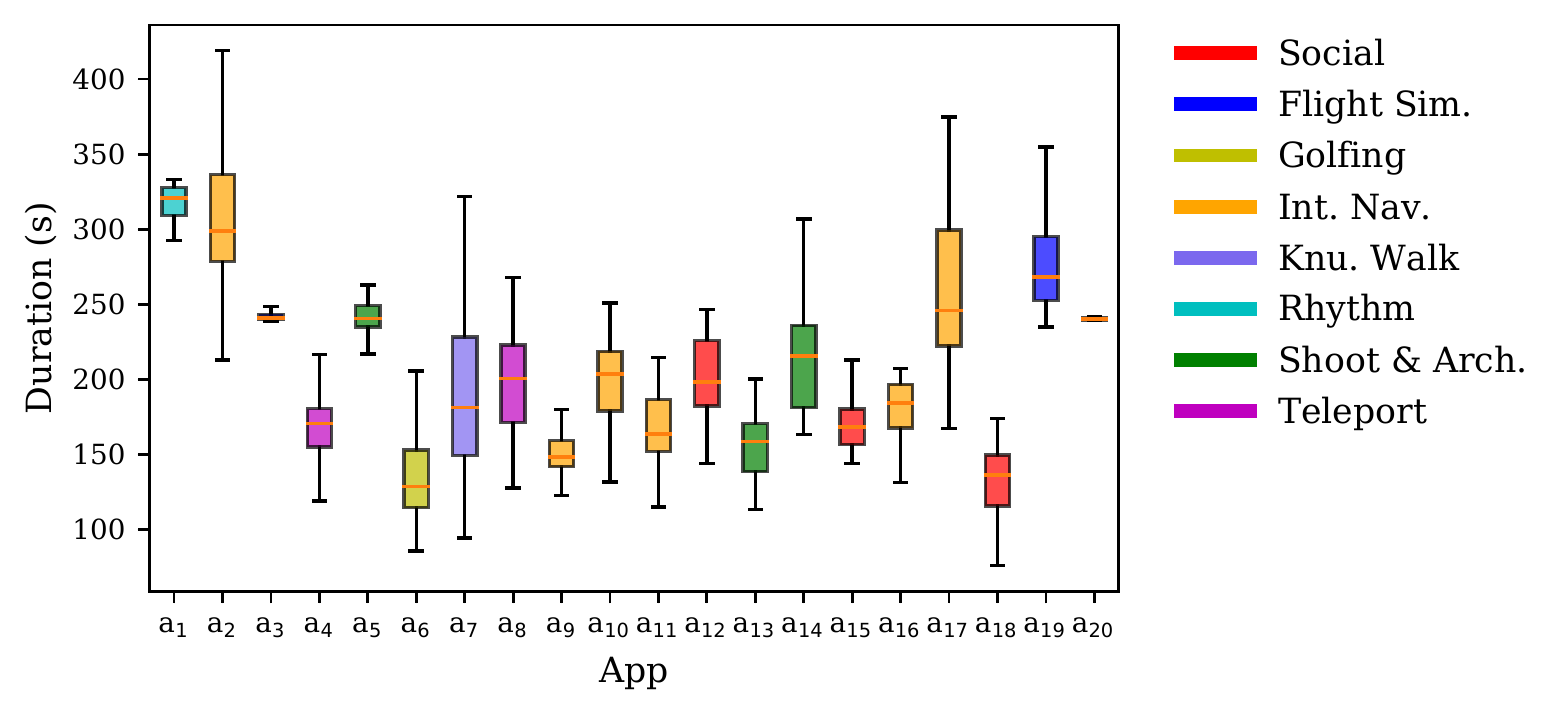}
        \caption{Total durations of sessions grouped by app}
        \label{fig:game-dynamics}
    \end{subfigure}
    \begin{subfigure}{0.38\textwidth}
        \flushleft
        \includegraphics[width=.86\textwidth]{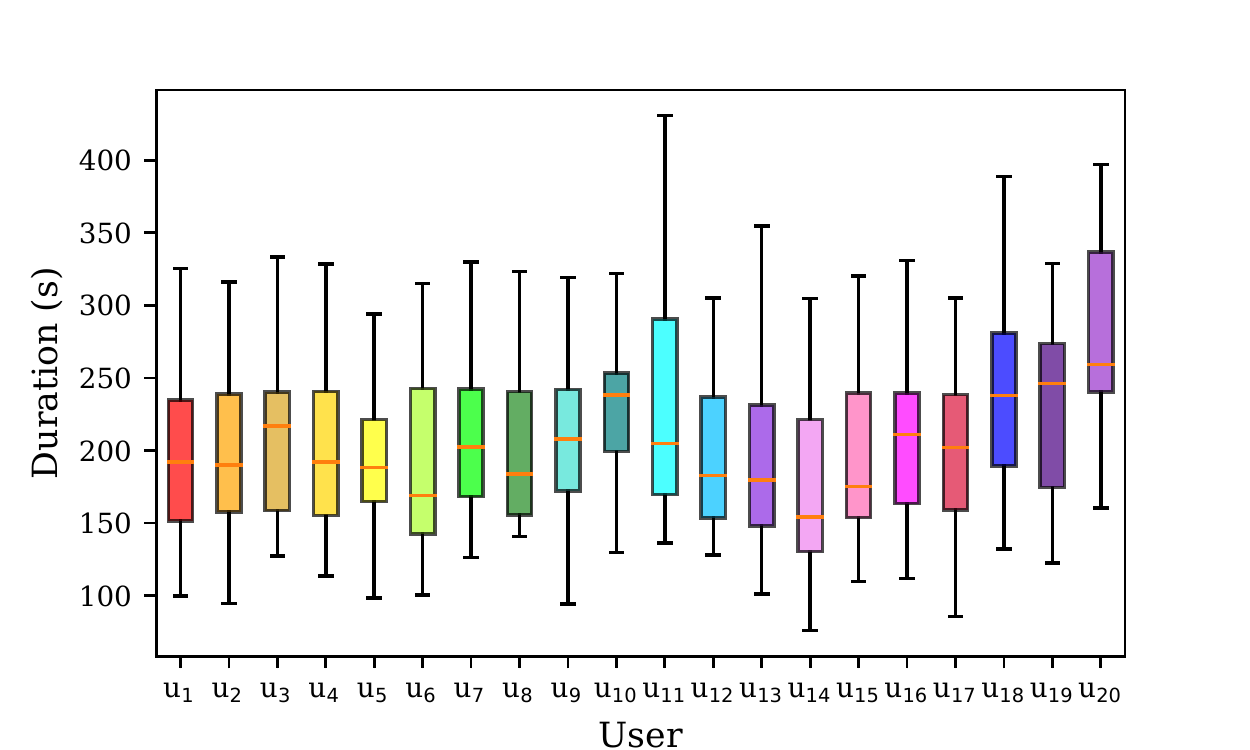}
        \caption{Total durations of sessions grouped by user}
        \label{fig:user-dynamics}
    \end{subfigure}
    \caption{Durations of sessions. There are \users{} users, each interacts with \apps{} apps. Colors represent app groups.
    }
    \label{fig:session-durations}
\end{figure}

\subsubsection{Insight: Variability}\label{subsec:dynamic-analysis}
The \system{} dataset exhibits variability across users, apps, and sensors, even when they perform the same activity \wrt{} the same app. Designing for variability was a decision we made on purpose to capture users' natural behavior.
From a {\em user} perspective, the user has the freedom on how, and at what pace, to perform the activity in each app. 
From the {\em app} perspective, variability is caused by different apps having different activities.
From the {\em sensor} perspective, variability occurs as the four sensor groups operate with different sampling rates and time spans.

To illustrate the variability in {\em session duration} across users and apps, we plot the distribution of total durations of sessions per app (Fig.~\ref{fig:game-dynamics}) and per user (Fig.~\ref{fig:user-dynamics}).
In Fig.~\ref{fig:game-dynamics}, We observe that users interact with the same app for varying durations: while average durations differ across apps, the variance for each app is relatively small.
In Fig.~\ref{fig:user-dynamics}, we observe that the average durations are closer to each other, but have larger variance for each user. 
Thus, we summarize the sensor data on per app and per sensor group basis.

\subsubsection{Data Processing}\label{subsec:data-processing}
Next, we describe data processing. The details regarding data processing are available in Appendix \ref{app:data_processing}. The goal of this step is to segment sensor data into time blocks and summarizing each as a feature vector for model input. One challenge is to choose the length of the block.
For block division, we first experimented with fixed-block length (FBL)---also been used in~\cite{miller2020personal}. FBL divides time series data into blocks: each block has a fixed length (\eg{} 1 or 2 seconds), and it ignores the variability of users, apps, and sensors. Since there is much variability in \system{} dataset, 
we develop an intuitive but robust method, refer as \textit{fixed-block amount} (or FBA), guided by previous observations on variability across apps and users. FBA divides time series into a fixed amount (number) of blocks for each sensor group per app ( \eg{} unlike FBL, FBA takes variability into account). 
FBA works comparatively better in our case (See Fig. \ref{fig:cmb_FBA_FBL_all} in Appendix \ref{app:fba-evaluation-optimization}). 

Fig.~\ref{fig:FBA-illustration} shows FBA applied to BM, processing a headset rotation value ($x$) for app $a_1$. With session durations of 3.6, 4.1, and 4.6 seconds for 3 users, the average duration is 4.1 seconds, rounded down to 4 blocks.
Each user's time series is divided into 4 blocks of $\sim$1 second.
A parameter $r \in (0, 2]$ to adjust block division: $N_{FBA_j}=r\cdot N_j$, increasing $r$ increases block count and decreases block length.
Finally, we summarize 
the time series of each reading in each block with a vector of five statistics, \ie{} maximum, minimum, mean, standard deviation, and median.
This summarization previously was used in~\cite{miller2020personal} and ~\cite{nair2023unique} for BM.  
    
\begin{figure} [t!]
\centering
\includegraphics[width=0.45\textwidth]{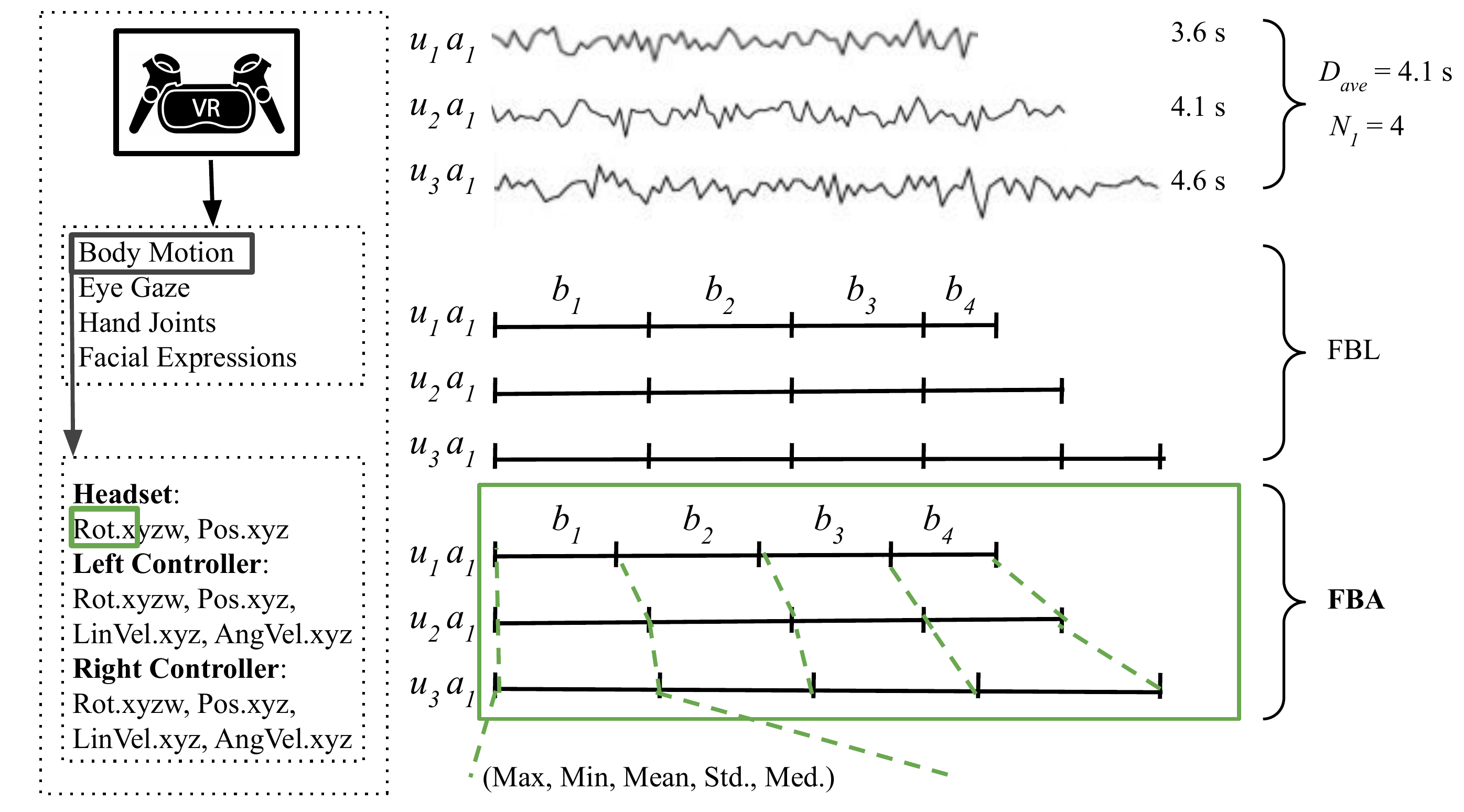}
\caption{FBA illustration for the $x$ value of headset rotation from the BM sensor group.
}
\label{fig:FBA-illustration}
\end{figure}

\subsubsection{Feature Selection and Engineering}\label{subsec:featurization}

Next, we select and augment the features, as follows:

\parheading{\Bodydata{}.}
We use all 33 BM sensor readings, including 3 position and 4 rotation readings from each controller and the headset, and 3 linear velocity and 3 angular velocity%
\footnote{Unlike prior works~\cite{miller2020personal,miller2022combining,nair2023unique} that focused \textit{only} on position and rotation, we additionally consider angular and linear velocity.} readings from each controller.
After computing five statistics (max, min, mean, stdev, median) for each reading, we obtain 165 BM features per block. %

\parheading{\Eyedata{}.}
The EG features are derived from one position and 3 rotation readings per eye%
\footnote{The EG position $y, z$ and  rotation $z$ are always zero, presumably because eyes cannot move in these directions relative to the headset.}.
From position readings, we derive the {\em interpupillary distance} (IPD, \ie{} the $x$ position difference between two eyes).
Prior work has shown that IPD is a top feature for gender identification~\cite{nair2022metadata}.
Inspired by this, we also augment rotation readings by calculating the differences in the same reading between two eyes.
By calculating the five statistics for 3 rotation readings per eye (6 total) and the 3 differential values, and including IPD as a separate feature%
\footnote{We do not compute statistics for position, since they do not change much over time.},
we obtain 46 EG features per block.

\parheading{\Handdata{}.}
There are 182 readings per hand that describe 3 position and 4 rotation readings from each of 26 joints. After calculating five statistics, 
we have 1,820 HJ features per block for two hands.
In order to limit model complexity and reduce run time, we reduce the number of features, using information gain~\cite{RF_2014} -- a popular technique for feature selection~\cite{guyon2003feature,singh2021feature} employed in prior work~\cite{miller2020personal, nair2023unique}.
We compute the information gain based on the model's performance on the training data. We exclude features with negligible importance and choose the top 400 HJ features per block.

\parheading{\Facedata{}.}
\system{} captures 64 sensor readings (see Appendix~\ref{app:sensor_groups}). After calculating five statistics per reading, we obtain 320 FE features per block for model training and evaluation (see Section \ref{subsec:facial-expression-evaluation}). Next, we select FE readings that describe emotion-related actions (\ie{} Action Units, AU) for each emotion (see Table \ref{tab:facial-emotion-actionunits})~\cite{facsexplained}. For example, there are 4 sensor readings corresponding to happiness, described by AU6 (facial elements 5 and 6) and AU12 (facial elements 33 and 34). After computing five statistics per reading for each emotion, we obtain the final features for that emotion. For example, for happiness, the final feature set consists of 20 FE features after computing the statistics. Final features for each emotion are utilized to train and evaluate individual models for each specific emotion in Section~\ref{subsec:facial_emotion_evaluation}. Additionally, we selected 25 FE readings for all emotions, resulting in 125 FE features, to evaluate a model on all facial emotions combined.

\parheading{Final Features.} After 
data processing and feature engineering, we obtain the final set of features per group, summarized in Table~\ref{tab:abstraction-dim} Appendix~\ref{app:fba-evaluation-optimization}. These are used to train and apply  the adversary models, described next.

\subsection{User Identification Models} \label{subsec:model-building} 
\subsubsection{Classification Task.} \label{subsubsec:classification}
We perform a multi-class classification task to uniquely identify user $u_i$ among the set of $n$ (\ie{} 20) users. 

\parheading{Train-Test Split.} 
We gathered sensor data from users in two sessions per app, with each session involving the completion of app-specific activities (see details in Section ~\ref{subsec:data-collection}). Data from the first session were split, with 90\% for training ($D_{train}$) and 10\% for validation ($D_{val}$). For evaluation, data from the same app (under similar or different settings) or a completely different app can be used.

\parheading{Model Architecture and Hyperparameters Tuning.} 
We explored models including Random Forest (RF)~\cite{RF}, Gradient Boosting (XGB)~\cite{XGB}, Support Vector Machine (SVM)~\cite{SVM}, and Long Short-Term Memory Networks (LSTM)~\cite{LSTM}. We found that  RF and XGB performed best on the \system{} dataset; this aligns with the closest prior works~\cite{miller2020personal,nair2023unique}. More details on algorithm selection and hyperparameter tuning can be found in Appendix ~\ref{app:fba-evaluation-optimization} and ~\ref{app:algorithm-selection}.

\parheading{Model Training.} 
FBA is applied to divide each session into $N_{FBA_j}$ blocks per user in an app $a_j$, \ie{} $[b_1, b_2, ..., b_{N_{FBA_j}}]$. The duration per session per user is $T$, and each block's duration is $t$. All blocks from the first session is used for training ($D_{train}$). For evaluation, we pick $s$ number of blocks per user from second session; $s$ represents a \textit{sub-session} that has $[b_1, b_2, ...,b_s]$ blocks, where $s \leq N_{FBA_j}$. $S_t$ for the sub-session is $s \times t$. $S_t$ allows us to investigate the minimum time we need per sensor group per user for identification. $S_t$ equals to the whole evaluation session per user when $s=N_{FBA_j}$ and $S_t=T$. Finally, we perform a classification task for each block (\ie{} predict the label for each $b_1, b_2,...,b_s$) and use maximum voting~\cite{MaxVote} across all blocks to determine the final label for each user.

\subsubsection{Different Adversaries and Their Models.} \label{subsubsec:open_world_models} 
In this section, we describe the experimental setup (\eg{} model) for different adversary.
\parheading{App Adversary Models.} The app adversary trains an {\em app model} on each app's data. Initially, we assume that the app models are trained and tested in similar app settings (\eg{} difficulty level, virtual rooms, songs) from the two different logins of the same user.

Next, we relax some constraints of app adversary and show that user identification works even in more open-world scenarios, where the app models are trained and tested in separate settings of the same app. We refer to this as {\em open world settings} for \appadv{}. We chose five representative apps from five distinct app groups. In the case of Beatsaber($a_1$), a popular rhythm app, training, and testing data were collected from different songs and difficulty levels based on users' preferences. For RecRoom ($a_{15}$), a social app, we gather training and testing data from separate virtual locations such as MacDonald's virtual restaurant, virtual campus or party venue. Similarly, in the case of Gorilla Tag ($a_7$), a knuckle-walking app, training, and testing data were collected in separate virtual spaces. For Elven Assasin ($a_5$), belonging to shoot.\& archery app group, training, and testing occur across different difficulty levels of the gameplay. Finally, for Chess ($a_{17}$), an interactive navigation app, training, and testing data were gathered across different gaming rounds, while the users freely moved chess pieces.

\parheading{\Devadv{} Models.} The device adversary has access to multiple apps' data. As \devadv{} aims to identify users across different apps, this setting undergoes \textit{open world settings} for across app evaluation.
We start from the \textit{universal model} that uses data from all apps ($a_1,..., a_{20}$); adversary can choose to train on a subset--a \textit{group model} for an app group (see details in Section~\ref{subsec:app-group}). 
Suppose an app group  (\eg{} social) has $n_g$ number of apps. The \devadv{} trains an app group model on all $n_g$ apps' training data. First, we consider the scenario where the adversary identifies a user of an app in a similar app group: the adversary can apply the model on each app's test data to identify users. We identify users across different app groups ($n_g$) by evaluating the app group model with an average data representation ($a_{avg}$).\footnote{For example, with 3 apps in a group, we use 33.33\% $S_t$ from each app for evaluation. This is to make a fair comparison using the same amount of data for evaluation.} 
 
Next, device adversary can initiate attacks under {\em zero-day scenario}, where the adversary attempts to identify a user from an app that it has not previously trained on. To that end, we train an app group model with $n_g-1$ apps' training data and test on $n_g^{th}$ app ($n_g^{th}$ app's data is not in $D_{train}$). We refer this type of attack as {\em zero-day attack}. We perform leave one out method and report the average accuracy to report effectiveness of {\em zero-day attack}.  

\parheading{Top Features.} For each model, we analyze the feature importance for RF and XGB using information gain ~\cite{RF_2014}. This helps both an adversary or privacy designer who wants to minimize its work.

\begin{table*}[ht!]
  \scriptsize
  \centering
  \caption{Identification accuracy (\%)  for \appadv{} ($adv_{app}$) and \devadv{} ($adv_{dev}$) \wrt{} sensor groups. The {\bf app adversary ($adv_{app}$)} trains and evaluates an app model on sensors data from a single app  (listed in \textbf{App No} column). The {\bf device adversary ($adv_{dev}$)} has two rows. The  first row (\eg{} $a_{12},a_{15},a_{18}$) reports results from $adv_{dev}$ training a group model on all apps in that group
  and evaluating on each individual app. The second row for $adv_{dev}$ reports results from training a group model and evaluating on average data of that group (\eg{} $a_{avg}$ indicates each app contributes 50\% of the data if $n_g=2$). %
  Each group, Golfing, Rhythm, and Knuckle-walking has exactly one app; thus, $adv_{app}$ and $adv_{dev}$ are the same (filled with ``both'' in \textbf{Adver.} column).
  }
  \label{tab:Identification_accuracy_app_dev}
  \begin{tabular}{ p{15mm}| c | c | c |c |c |c }
    \toprule
    \multirow{2}{*}{\textbf{App Group}} & \multirow{2}{*}{\textbf{ Adver.}} & \multirow{2}{*}{\textbf{App No.}} & \multicolumn{4}{c}{\textbf{Sensor Group}} \\
    \cline{4-7}
    &  &  & \textbf{BM} & \textbf{EG} & \textbf{HJ} & \textbf{FE} \\
    \midrule
    \multirow{3}{*}{Social} & \multirow{1}{*}{$adv_{app}$} & $a_{12}, a_{15}, a_{18}$ & 85, 95, 95 & 80, 90, 90 & 60, 65, 75 & 95, 100, 100 \\
    \cline{2-7}
    \multirow{3}{*}{} & \multirow{3}{*}{$adv_{dev}$} & $a_{12},a_{15},a_{18}$ & 95, 95, 95 & 75, 90, 85  & 70, 85, 75 & 100, 100, 100 \\
     & & $a_{avg}$ & 100 &  95 & 90 & 100 \\
    \midrule %
    \multirow{3}{*}{Flight Sim.} & \multirow{1}{*}{$adv_{app}$} & $a_3,a_{19},a_{20}$ & 95, 100, 95 & 85, 90, 75 & 80, 75, 75 & 100, 95, 95 \\
    \cline{2-7}
     & \multirow{2}{*}{$adv_{dev}$} & $a_3, a_{19},a_{20}$ & 95, 100, 100 & 90, 90, 90  & 80, 80, 75 & 100, 100, 95 \\
    & & $a_{avg}$ & 100 & 95  & 95& 100 \\
    \midrule %
    \multirow{2}{*}{Interactive} & \multirow{1}{*}{$adv_{app}$} & $a_2,a_9,a_{10},a_{11},a_{16},a_{17}$ & 95, 80, 95, 95, 95, 100 & 80, 80, 85, 95, 75, 80 & 60, 40, 60, 60, 70, 90 & 100, 100, 90, 95, 100, 100 \\
    \cline{2-7}
    \multirow{2}{*}{Navigation} & \multirow{2}{*}{$adv_{dev}$} & $a_2, a_9, a_{10},a_{11}, a_{16}, a_{17}$ & 95, 85, 90, 95, 95, 100 & 75, 60, 80, 80, 60, 75  & 65, 40, 60, 75, 75, 90 &100, 100, 95,100,100,100 \\
     & & $a_{avg}$ & 100 & 95 & 85 & 100 \\
    \midrule %
    \multirow{2}{*}{Shooting \&} & \multirow{1}{*}{$adv_{app}$} & $a_{5}, a_{13}, a_{14}$ & 95, 90, 100 & 85, 75, 90 & 70, 60, 80 & 90, 100, 100 \\
    \cline{2-7}
    \multirow{2}{*}{Archery} & \multirow{2}{*}{$adv_{dev}$} & $a_{5}, a_{13}, a_{14}$ & 95, 100, 100 & 85, 80, 90  & 70, 65, 80 & 90, 100, 100 \\
     & & $a_{avg}$ & 100 &  90 & 85 & 100 \\
    \midrule %
    \multirow{4}{*}{Teleport.} & \multirow{1}{*}{$adv_{app}$} & $a_4, a_8$ & 70, 80 & 90, 70 & 35, 45 & 95, 95 \\
    \cline{2-7}
    \multirow{2}{*}{} & \multirow{2}{*}{$adv_{dev}$} & $a_4, a_8$ & 75, 85 &  90, 75 & 35, 50 & 100, 95 \\
     & & $a_{avg}$ & 85 &  90 & 50 & 100 \\
    \midrule %
    Golfing & both  & $a_6$ & 80 & 70 & 50 & 90 \\
    \midrule
    Rhythm & both & $a_1$ & 95 & 90 & 75 & 100 \\
    \midrule
    Knu.-walk. & both & $a_7$ & 95 & 80 & 65 & 100 \\
    \midrule %
    \multirow{2}{*}{All} & \multirow{2}{*}{$adv_{dev}$}  & $a_1, a_2, ...,a_{20}$ & $90-100$ & $50-80$ & $45-95$ & 100 \\
    & & $a_{avg}$ & 100 & 90 & 100 & 100 \\
    \bottomrule
  \end{tabular}
  \label{tab:IdenAcc_bigTable}
\end{table*}

\section{Evaluation Results}
\label{sec:evaluation}
In this section, we evaluate the performance of \system{}'s adversaries. %
Table~\ref{tab:Identification_accuracy_app_dev} summarizes the results.
For each sensor group (in Sections~\ref{subsec:body-motion-evaluation},~\ref{subsec:eye-gaze-evaluation},~\ref{subsec:hand-joints-evaluation}, and~\ref{subsec:facial-expression-evaluation}), we evaluate \apps{} models (\ie{} one model per app) %
guided by the following research questions (RQs):
\begin{compactitem}[$\bullet$]
    \item \textit{RQ1 (Accuracy):} How well can a user be identified using different VR sensor group?
    How do these groups compare to each other?
    \item \textit{RQ2 (Sub-session Time $S_t$):} %
    How long does identification take?
    \item \textit{RQ3 (Top Features):} What are the top features for identification \wrt{} various apps and adversaries?
\end{compactitem}

In Section~\ref{subsec:open-world-exp}, we evaluate our open-world %
experiment, discussed in Sections~\ref{subsec:app-selection-grouping} and \ref{subsec:model-building}, answering the following:
\begin{compactitem}[$\bullet$]
    \item \textit{RQ4 (Open-World Setting):} Can we identify a user across different settings within same app or a user across similar or different apps (app groups)? What if the app is {\em not included} in the training of the 
    in adversary's model (\textit{zero-day} scenario)?
\end{compactitem}
Next, Section \ref{subsec:comparative-evaluation} discusses relative sensor group importance (among sensor groups and \wrt{} app groups) by answering the following: %
\begin{compactitem}[$\bullet$] 
    \item \textit{RQ5 (Sensor Group Importance):} What are the most important sensor groups in general, and as they relate to particular app groups? 
    Moreover, can combining weak sensor models help to generate a comparatively stronger attack?
\end{compactitem}

\subsection{\Bodydata{} Models Evaluation}\label{subsec:body-motion-evaluation}
\point{RQ1 (Accuracy).}
The identification accuracy for \bodydata{} app models is $100$\% in $3$ apps and $\geq$$95$\% (\ie{} at most $1$ out of $20$ users is falsely identified) in $14$ apps (see $adv_{app}$ results from BM column of Table \ref{tab:IdenAcc_bigTable}). These results are consistent with previous studies (see Section~\ref{sec:related-work}). Most apps, such as Beat Saber ($a_1$; extensively studied in~\cite{nair2023unique}), archery ($a_5$), and shooting ($a_{13}$), demand significant \bodydata{} (headset and controllers movement).
Other apps that require less \bodydata{} (\eg{} in $a_4$, users move through teleportation) provide $\sim$70-80\% accuracy. 
As the \devadv{} considers a larger amount of data and tasks for training, the accuracy is up to $100\% $ (see $adv_{dev}$, Table \ref{tab:IdenAcc_bigTable}) compared to a single app.

\point{RQ2 (Sub-session Time $S_t$).}
For the app models, accuracy is $\sim$80\% with an average user sub-session time ($S_t$) of 4s. The app models require at least $16$s of $S_t$ to reach 90\% accuracy (see Fig.~\ref{fig:AppAdv_AccBlockPerUser_motion} in Appendix~\ref{app:subsession-time-characterization}).
The device model achieves higher accuracy with similar $S_t$ by training on all \apps{} apps (see Fig.~\ref{fig:min-time-devadv} in Appendix~\ref{app:subsession-time-characterization}), accumulating comprehensive user behavior knowledge.

\point{RQ3 (Top Features).}
In Appendix~\ref{app:top-features}, Table~\ref{tab:AppAdv_FeatureImp_All} presents the top-3 features for identification for each app. 
 The top features are influenced by app-specific activities and user measurements; \eg{} %
 flight apps require users to sit and make left-right head movements to control flight making the $x$-position of the headset (left-right movements) as top feature.
Shoot.\& archery apps (\eg{}$a_5$) require both headset and controller movement/velocity that influence as top features.
For the {\em device model}, the $y$, $x$, and $z$ positions of the headset are top features (see Fig.~\ref{fig:DeviceAdv_FeatureImp_Motion} in Appendix~\ref{app:top-features}), indicating height, left-right and forward-backward extent of the head movements influence identification.
Fig.~\ref{fig:AppAdv_Acc_motion_cmp} in Appendix~\ref{app:top-features} shows the importance of headset features: 5\% to 35\% higher accuracy compared to controller features alone ($\sim$21\% on average) across all apps.

\parheading{Key Takeaways.} BM identification relies on both app-activity specific and users unique measurement (\eg{} height) features.

\subsection{\Eyedata{} Models  Evaluation}\label{subsec:eye-gaze-evaluation}
\point{RQ1 (Accuracy).} App models provide $\geq$$90$\% accuracy for 8 apps, $\geq$$85$\% for 12 apps, $\geq$$75$\% for the remaining. We observe that identification accuracy is influenced by frequent object-eye interactions, \eg{} Beat Saber ($a_1$) model gives $95$\% as users frequently look at and follow the movement of virtual objects in this app. Similarly, archery, shooting, and flight simulation app models show high accuracy due to frequent eye-object interaction. The device model's accuracy can be up to $100$\% (see EG column in Table~\ref{tab:IdenAcc_bigTable}).

\point{RQ2 (Sub-session Time $S_t$).}
The app models accuracy is $\sim$50\% with $5$s of $S_t$ per user on average. It increases to $\sim$70\%  with $19$s of $S_t$ (accuracy may vary depending on apps, see Fig.~\ref{fig:AppAdv_AccBlockPerUser_Eye} in Appendix~\ref{app:subsession-time-characterization}).  
Device model (Fig.~\ref{fig:min-time-devadv} in Appendix~\ref{app:subsession-time-characterization}) shows $80$\% with 17s of $S_t$.

\point{RQ3 (Top Features).}
For both app and device models (see Table~\ref{tab:AppAdv_FeatureImp_All}, Appendix~\ref{app:top-features} and Fig.~\ref{fig:DeviceAdv_FeatureImp_Eye} , Appendix~\ref{app:top-features}) show that augmented features contribute the most to user identification for EG. The top features are the $y$-rotation that correlates left and right eyes (\ie{} ``Quat.y Left Right''), that matches our intuition: for EG, augmented features (\ie{} $f^a_{LR}$) are important for unique identification.
Fig.~\ref{fig:AppAdv_Acc_Eye_cmp} in Appendix~\ref{app:top-features} shows that augmented features ($f^a_{LR}$) improve model accuracy significantly ($5-35$\% or $\sim$20\% on average) across all apps.

\parheading{Key Takeaways.}
Augmenting the standard features with the distance between the eyes improves identification accuracy.

\subsection{\Handdata{} Models Evaluation}\label{subsec:hand-joints-evaluation}
\point{RQ1 (Accuracy).}
The app models provide $\geq$$70$\% accuracy for 9 apps, which involve diverse hand movements and gestures 
(see Table \ref{tab:list-of-apps}); \eg{} in $a_1$ (Beat Saber), users swing light sabers using hands, involve claw position and frequent hand movements; 
for $a_{17}$ (chess), users grab and move chess pieces. Conversely, several app groups, such as teleportation, lack hand-specific activities that cause low identification accuracy;
\eg{} teleportation provides the lowest accuracy ($\sim$35\%).
For the device models, the accuracy is $\geq$85\% in most cases. %

\point{RQ2 (Sub-session Time $S_t$).}
For the app models, the accuracy is $\geq 60$\% with $S_t$ of at least $20$s (see Fig.~\ref{fig:AppAdv_AccBlockPerUser_hand} in Appendix~\ref{app:subsession-time-characterization}).
For the device model, accuracy is $90\%$ with $120$s of $S_t$ (see Fig.~\ref{fig:min-time-devadv_hand} in Appendix~\ref{app:subsession-time-characterization}).

\point{RQ3 (Top Features).}
For app models, (see Table~\ref{tab:AppAdv_FeatureImp_All},  Appendix~\ref{app:top-features}) shows that the top features are influenced by app-activities. See Table \ref{tab:HJ_features} for description.%
\eg{} for $a_1$ (Beat Saber), top features are the positions of joints 1 and 3 (thumb metacarpal and palm) of the right and joint 24 (little intermediate) of the left hand.
These joints are exercised when making a fist for holding sabers. 
For $a_{17}$ (chess), joints 22 and 25 (little metacarpal and distal) of right hand (use for moving chess pieces) are top features. For device model (see Fig.~\ref{fig:DeviceAdv_FeatureImp_hand},  Appendix~\ref{app:top-features}), positions of left-hand joints 1 (palm), 2 (wrist), 7 (index metacarpal), and right-hand joints 3 (palm), 2 (wrist) and 5 (thumb distal) are top features. They represent users natural hand positions, emphasizing joint positions (\eg{} making an open fist) and activities--emphasizing joint rotations (grabbing, waving, \etc{}). 

\parheading{Key Takeaways.} Apps with more hand-related activities show higher attack accuracy using HJ sensor group.

\subsection{\Facedata{} Models Evaluation}\label{subsec:facial-expression-evaluation}
\point{RQ1 (Accuracy).}
\Facedata{} is highly effective for user identification; the app models provide $\geq95$\% accuracy for $17$ and $90$\% for the remaining 3 apps. The device models achieve up to $100$\%, consistent with other sensor groups (see FE column of Table \ref{tab:IdenAcc_bigTable}).

\point{RQ2 (Sub-session Time $S_t$).} 
For \appadv{}, most apps provide $\geq$$85$\% accuracy with $S_t$ of only $5$s, and $\geq$$90$\% accuracy with $18$s (see Fig.~\ref{fig:AppAdv_AccBlockPerUser_face} in Appendix~\ref{app:subsession-time-characterization}).
For the device model, accuracy is $95$\% with just $3$s and 100\% within  $17$s (see Fig.~\ref{fig:min-time-devadv} in Appendix~\ref{app:subsession-time-characterization}), demonstrating the high effectiveness of FE.

\point{RQ3 (Top Features).} 
Facial features are correlated to the app-specific activities (\eg{} in $a_9$:job simulation, users eat a doughnut, relates to element 27---jaw movement as a top feature) and valence/arousal states (\eg{} social, rhythm: elements 5 and 6, which correspond to AU6---action unit for happiness) (see Table~\ref{tab:AppAdv_FeatureImp_All} in Appendix~\ref{app:top-features}). See Table \ref{tab:FE_features} for description. For the device model (see Fig.~\ref{fig:DeviceAdv_FeatureImp_face} in Appendix~\ref{app:top-features}), both emotions and natural expressions are key features, \eg{} elements 5 (cheek raiser), 6 (jaw drop), and 25 (jaw thrust) are part of user expression of joy and surprise respectively; $28$ (jaw thrust) and $51$ (lips toward) contribute to natural expressions representing the outward (lower lip) and opening (both lips) movements.

\subsubsection{Facial Emotion Models Evaluation}\label{subsec:facial_emotion_evaluation}
In this section, we focus on facial elements/AU combinations that represent an emotion (see descriptions in \cite{facsexplained,openxrfacetracking} and results in Table \ref{tab:results_emotion} in Appendix \ref{app:emotion-based-identification}), rather than all/other \facedata{}. %
We argue that arousal/valence states in VR may induce certain emotions, similar to what happens in the real world. For example, socializing, whether in-person or virtually, can make a person happy (HA/PV), or seeing a positive/new environment can induce joy/surprise (PV).
From Table~\ref{tab:Apps-Grouping}, we pick one or two apps from seven groups, representing the rest of the apps and groups to evaluate our hypothesis. %
\par Our results confirm that facial elements/AUs indicating emotions can be used for identification, correlating strongly with the app's arousal/valence states. Social apps' models use AU combinations that represent happiness and surprise, provide $\geq$$95$\%. Flight simulation or shooting apps induce mostly negative valence, thus identification based on happiness facial elements induce low accuracy, \ie{} apps $a_{14}$  (shooting) and $a_{20}$ (flight simulation), give $80$\% and $75$\% respectively, %
however, both apps provide $\geq$$90$\% based on facial elements/AUs representing fear. 
In some apps, app-specific activities and arousal/valence states may induce mixed emotions. For instance, in Beat Saber ($a_1$), users may experience happiness due to the music/beat, fear from the tension of cutting blocks or avoiding obstacles, and even sadness or anger when missing some blocks. The models for these apps achieve high accuracy by considering almost all emotions. 
Apps with mostly neutral arousal/valence states (\eg{} interactive navigation apps) may achieve low accuracy ($\sim$80\%) for high arousal emotions, such as happiness/sadness as the VR environment may not strongly induce these emotions.

\par Finally, combining all AUs representing all emotions provides high accuracy across all apps (\ie{} $\geq$$90$\% for most apps). Fig.~\ref{fig:AppAdv_Acc_face_cmp} in Appendix~\ref{app:top-features} shows that AU combinations representing emotions provide better accuracy compared to other facial expression AUs (that do not represent emotions) by $5-25$\% in most apps (with some exceptions) and $5$\% on average; \eg{} considering arousal/valence, accuracy improves by $25$\% for $a_{20}$--flight and $10$\% for $a_{14}$--shooting.

\parheading{Key Takeaways.} %
Our findings suggest that an adversary can select AU combinations that represent emotions \wrt{} the app's arousal/ valence state, to identify a user with low effort.\footnote{For example, adversary can use a few facial features (\eg{} 20 for happiness in social apps) instead of all (\ie{} 325) facial features for adequate unique identification.}

\subsection{Open-World Setting Evaluation}\label{subsec:open-world-exp}

Next, we relax some constraints of Section \ref{subsec:body-motion-evaluation}- \ref{subsec:facial-expression-evaluation} and %
train and test on data from different settings and activities within and across apps, referred as {\em Open-World setting} (see Section \ref{subsubsec:open_world_models}). We show that users can be identified, not only across similar sessions of the same app (Sections~\ref{subsec:body-motion-evaluation} - \ref{subsec:facial-expression-evaluation}), also across different settings of the same app (Section~\ref{subsec:OpenWorld_within_sameapps}) and across different apps (Section~\ref{subsec:OpenWorld_AppGroup}). 

\subsubsection{User Identification across Different Settings in the Same App}\label{subsec:OpenWorld_within_sameapps}
In this section, we have considered different settings (\eg{} difficulty levels, virtual spaces, songs) in the same app and as a proof of concept, we performed additional experiments for five representative apps from five app groups (see experiment details on \ref{subsubsec:open_world_models}). %

\parheading{\Bodydata{}} For BM, accuracy of open-world ranges from 80-100\% (close to \ref{subsec:body-motion-evaluation} results), showing that users can be reliably identified across various settings within same app. Since motion patterns are unique to users {\em and} their activities within apps can be similar across different settings (not identical which adds variability that results in negligible accuracy drops) making identification feasible.

\parheading{\Eyedata{}} For EG, the identification accuracy is between 60-80\%.

\parheading{\Handdata{}} For HJ, accuracy ranges from 60-80\%. Accuracy is slightly lower in the open-world-setting compared to \ref{subsec:hand-joints-evaluation} due to added variability from different settings of the same app as BM.

\parheading{\Facedata{}} For FE, accuracy is higher (90-100\%) compared to other sensor groups as it relies heavily on facial emotion (see Fig. \ref{fig:AppAdv_Acc_face_cmp}), influenced by app's arousal/valence (see Section \ref{subsec:app-group}). If different app settings maintain a consistent arousal-valence state, users remain in similar emotional conditions, minimizing variability for FE and thus accuracy varies less across settings.

\subsubsection{User Identification across App Groups} \label{subsec:OpenWorld_AppGroup}
\begin{figure*}[t!]
    \centering
    \begin{subfigure}{0.49\textwidth}
        \includegraphics[width=.86\linewidth]{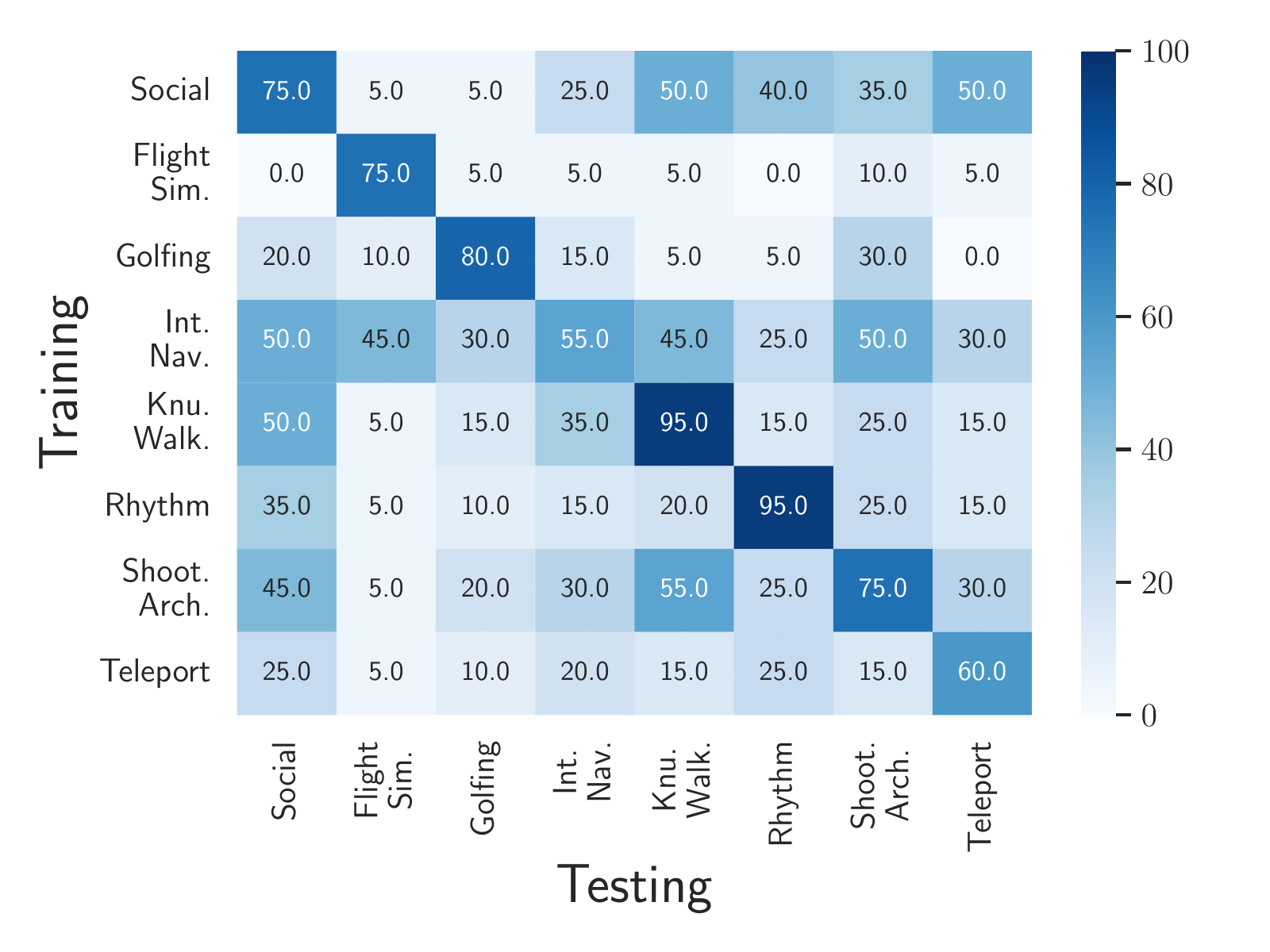}
    \caption{\Bodydata{} Sensor Group} 
        \label{fig:heatMap_Group_body}
    \end{subfigure}
    \hfill
    \begin{subfigure}{0.49\textwidth}
        \includegraphics[width=.86\linewidth]{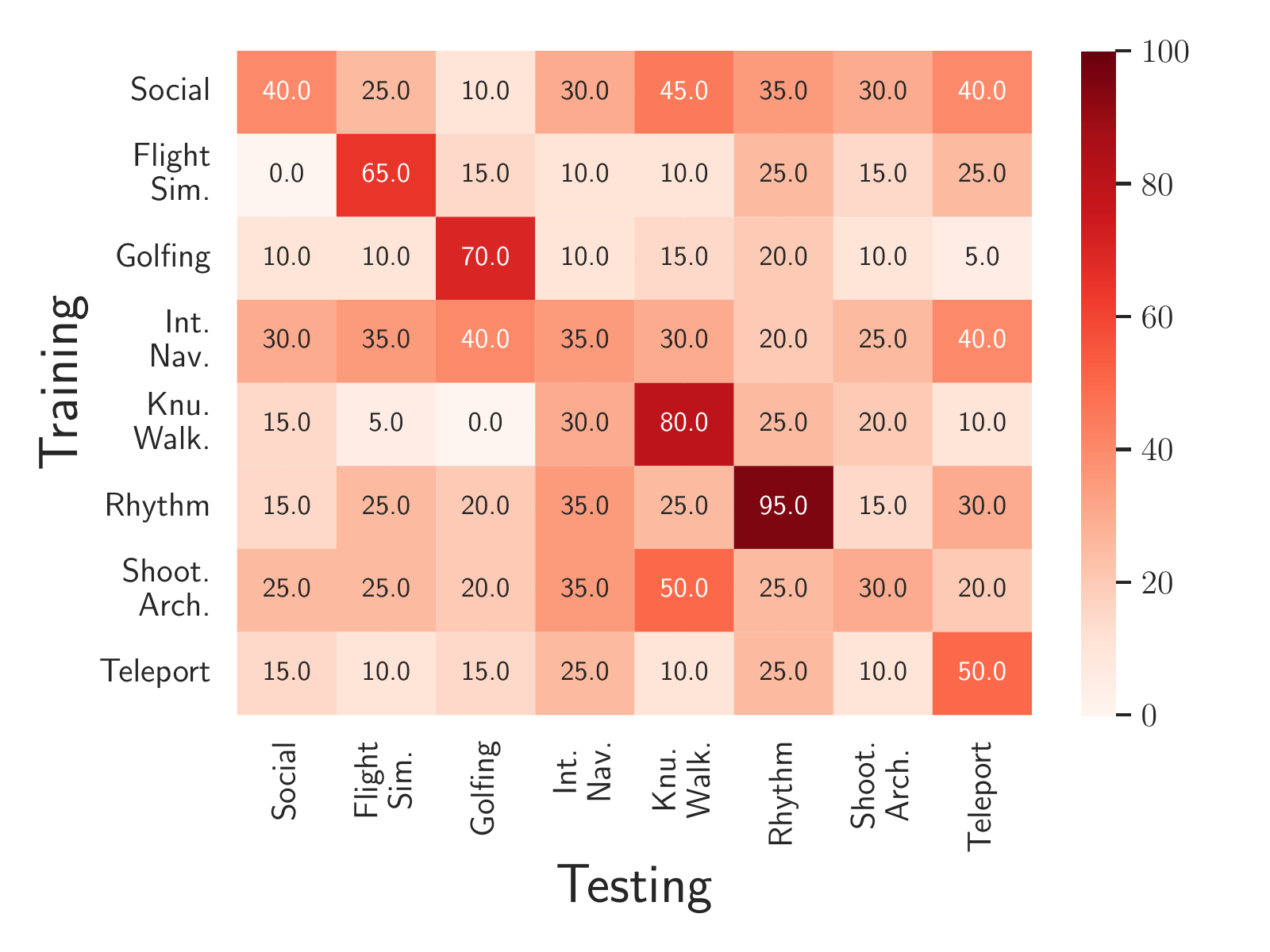}
        \caption{\Eyedata{} Sensor Group} 
        \label{fig:heatMap_Group_eye}
    \end{subfigure}
    \hfill
    \begin{subfigure}{0.49\textwidth}
        \includegraphics[width=.86\linewidth]{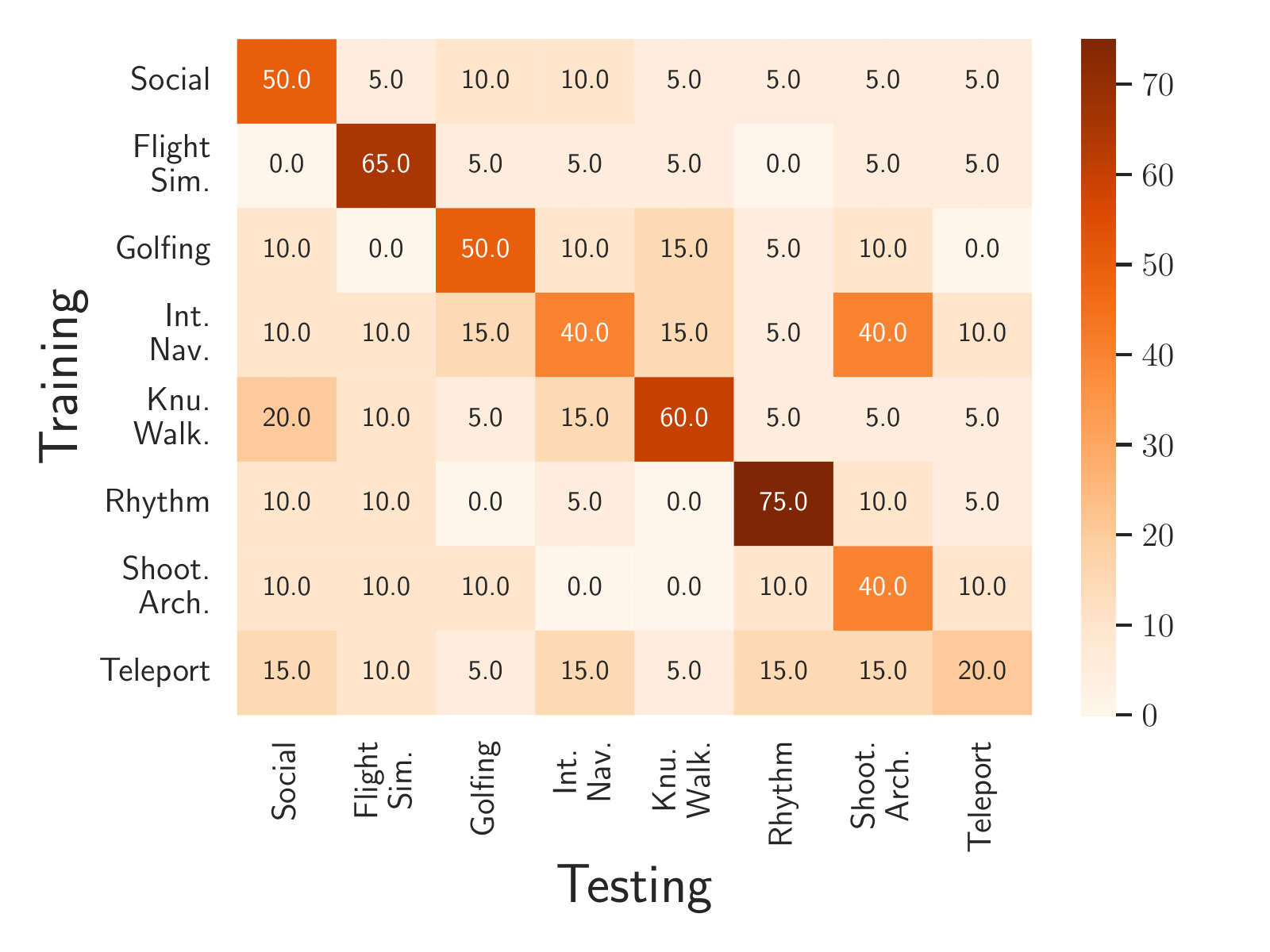}
        \caption{\Handdata{} Sensor Group}
        \label{fig:heatMap_Group_hand}
    \end{subfigure}
    \hfill
    \begin{subfigure}{0.49\textwidth}
        \includegraphics[width=.86\linewidth]{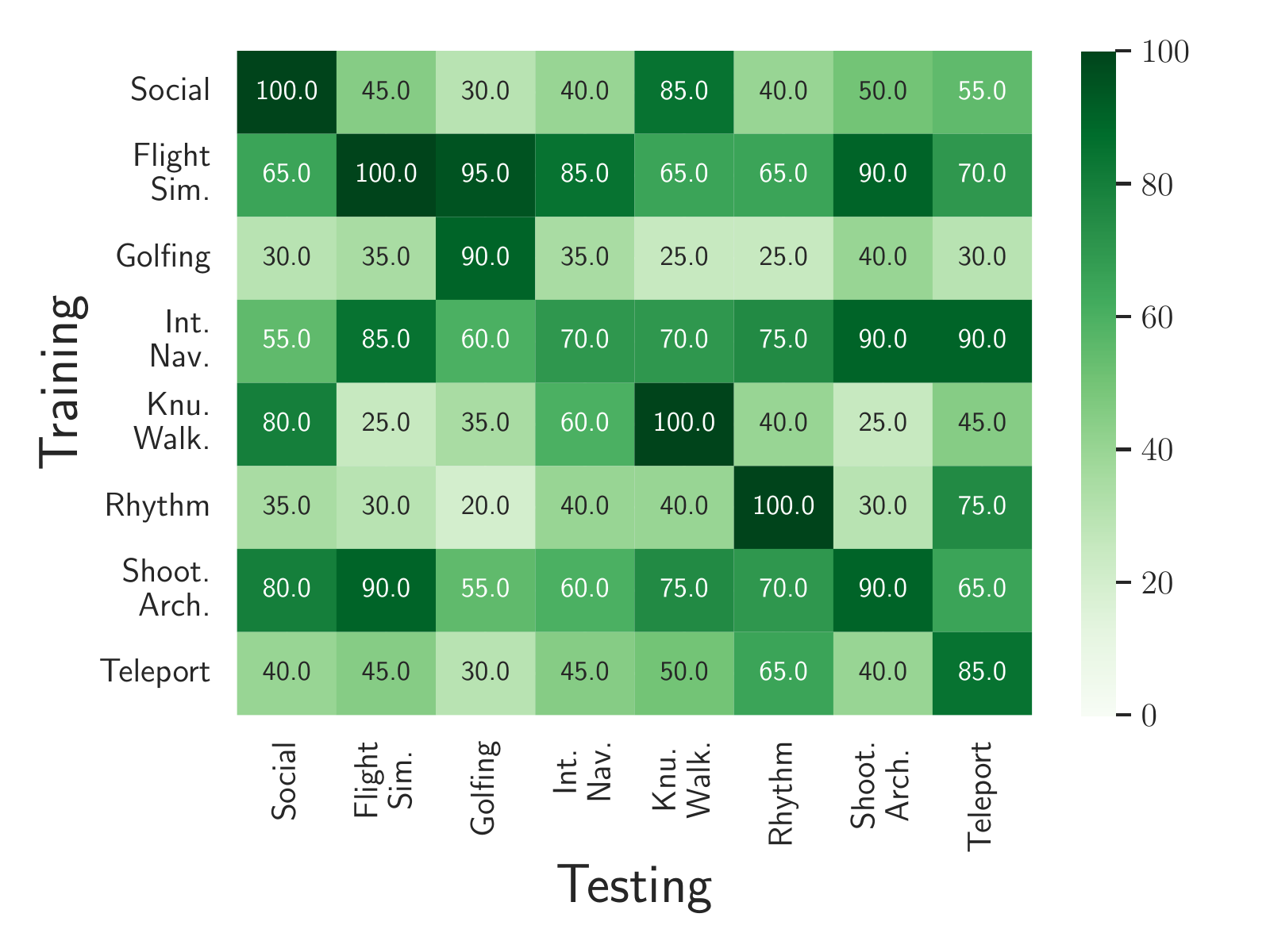}
        \caption{\Facedata{} Sensor Group} 
        \label{fig:heatMap_Group_face} 
    \end{subfigure}
    \caption{Identification accuracy (in percent) in the \textit{zero-day} scenario. The adversary trains on the data from other apps in a group, and tests in a new app (for which it did not have training data) in the same group. The \textit{diagonal} shows the accuracy for apps within the same group, whereas the \textit{other values} show the accuracy for apps from other app groups.}
    \label{fig:heatMap_Groups}
\end{figure*}

To address \textit{RQ4}, we conduct experimental evaluation (see $adv_{dev}$ row in Table~\ref{tab:IdenAcc_bigTable}) for each app group described in Table~\ref{tab:Apps-Grouping} %
and for the {\em zero-day} scenario  (see Fig.~\ref{fig:heatMap_Groups})  defined in Section~\ref{subsec:model-building}.

\parheading{\Bodydata{}.}
The performance of the {\em app group models} (trained on all apps data within a group) and the {\em universal device model} (trained on all apps) using BM is comparable, \eg{} both achieve up to $100$\% accuracy using $a_{avg}$. The group models outperform the app models in general, which encourage the \devadv{} to choose a group model over app models; \eg{} for $a_{13}$, shoot.\&arch. group model provides $10$\% higher accuracy  than app model. In the \textit{zero-day} scenario, Fig.~\ref{fig:heatMap_Group_body} shows that models evaluated on new apps in the same group perform well ($75-95$\% accuracy), %
but performs poorly on different groups (\ie{} $0-50$\% accuracy). This evaluation confirms that group models are effective in \textit{zero-day} scenarios. 

\parheading{\Eyedata{}.} No app grouping based on \eyedata{} is observed due to the lack of app-specific activity. Table~\ref{tab:IdenAcc_bigTable} indicates that group models perform similarly to app models. However, in the \textit{zero-day} scenario, grouping can be valuable as within similar groups, there is a higher occurrence of eye-object interactions that potentially help to identify users within a new app. Fig.~\ref{fig:heatMap_Group_eye} shows a diagonal in the heatmap with adequate accuracy, supporting our claim.

\parheading{\Handdata{}.} 
For \handdata{}, app group models and the universal device model have comparable performance, \eg{} using $a_{avg}$, the accuracy difference is within the range of $10-15$\% mostly. 
Akin to BM, HJ group models outperform the app models, \eg{} the social group model provides 20\% higher accuracy for $a_{15}$ than the app model. %
For \textit{zero-day} scenario, a group model evaluated on its native apps provides higher accuracy ($40-75$\%) compared to an app from different groups ($\sim$$20$\%) (see Fig.~\ref{fig:heatMap_Group_hand}).

\parheading{\Facedata{}.} 
Both the app group and the universal models achieve up to 100\% accuracy using FE data, with group models performing similarly or better than app models. 
For the \textit{zero-day} scenario, a model tested on apps from the same group provides higher accuracy ($70-100$\%) than apps from different groups ($20-65$\%) for most of the cases (see Fig.~\ref{fig:heatMap_Group_face}). However, several apps provide high accuracy within different groups; \eg{} social group model accurately predicts $a_7$ with 85\%, the shooting\&archery group model provides 90\% when tested on flight apps (\eg{} $a_{20}$) as they share similar arousal/valence states (LA/NV, HA/NV) (see Table \ref{tab:Apps-Grouping}).

\subsection{Sensor Group Importance Evaluation}\label{subsec:comparative-evaluation}

We compare the importance of different sensor groups, in general (Section~\ref{subsec:comparative-evaluation_SensorGroup}) and for specific app groups (Section~\ref{subsec:comparative-evaluation_AppVsSensor}). Finally, we evaluate whether combining multiple weaker sensor group models (ensemble) can enhance attack performance (see Section \ref{subsec:comparative-evaluation_ModelFusion}). 
\subsubsection{Model Accuracy across Sensor Groups} \label{subsec:comparative-evaluation_SensorGroup} This section partly addresses {\em RQ5}.
FE and BM sensor groups outperform the EG and HJ. For BM, $14$ out of $20$, and for FE, $17$ out of $20$ apps achieve an accuracy of $\geq95\%$.
Conversely, only $5$ out of $20$ apps offer $\geq$$90$\% for EG, which is intuitive since BM and FE cover more diverse activities compared to EG. HJ shows low attack performance in specific context: $9$ among $20$ apps provide $70-90$\% due to a lack of HJ-based activities in several apps.
Both app and device models for EG and HJ require longer $S_t$ (see \textit{RQ2} of Sections \ref{subsec:body-motion-evaluation}-\ref{subsec:facial-expression-evaluation}) than BM, FE. Therefore, from an attacker's perspective, if the goal is to minimize effort and given that the attacker has access to any of the sensor groups, FE would be the optimal choice. 
\subsubsection{Important Sensors per App Group} \label{subsec:comparative-evaluation_AppVsSensor}
 
This section partly addresses {\em RQ5} by discussing the importance of individual sensor groups relative to app groups.  
This comparison is useful: (1) for attackers to optimize which sensor groups' data to train and test models on, to efficiently utilize resources and maintain accuracy; and (2) for defense strategies regarding users' decisions to share sensor data, \eg{} users can revoke permissions for some sensor groups (including BM \cite{turnoffbodytracking}) that are not essential for that app group. %

\parheading{\Bodydata{}.} Most app groups require \bodydata{} for app-specific activities. For example, social app groups involve walking and exploring, rhythm apps require dance-like continuous movements. Consequently, BM and its associated motion features are available to adversaries in those apps and thus  provide $\geq$$85$\% accuracy in most app groups. Contrarily, BM is less crucial for certain groups \eg{} teleportation since minimal body movement occurs for teleporting to different virtual locations, providing $\leq$$80$\% accuracy. 

\parheading{\Eyedata{}.} For EG, there are no defined app-specific activities for our selected apps. 
Meta indicates EG is employed for realistic avatars %
and to estimate directions of where users are looking\cite{metaquest_policy}. Intuitively, certain app groups require frequent eye-object interactions  while performing app activities; \eg{} in shoot.\&archery or rhythm, users aim to shoot/cut, resulting in frequent eye-object interactions. Consequently, they provide high identification accuracy ($85-95$\%) and important from adversarial perspective.
Conversely, certain groups require minimal eye-object interactions as users mostly sightseeing in those apps (\eg{} knuckle-walking, provides $70-80$\% accuracy). Thus, we can argue that, EG is optional from a user's perspective for our app groups except social. Thus given the available permission system, users might disable it.

\parheading{\Handdata{}.} HJ for certain app groups that require active hand movements/gestures (\eg{} archery, flight, interactive navigation, \& rhythm) provide higher accuracy (see Table~\ref{tab:IdenAcc_bigTable}), \eg{} flight apps models provide $75-80$\% as require lots of hand activities for controlling flight. Conversely, HJ models for the teleportation group lack hand-related activities and achieve $35-45$\%. However, users may prefer to turn off HJ in certain groups where hand movements are optional (teleportation) and presumably prefer HJ in groups, where hand movements are crucial (\eg{} interactive navigation ). 

\parheading{\Facedata{}.} \Facedata{} or FE is crucial for identification in all app groups, as apps' arousal/valence states can trigger specific \facedata{}. Thus, FE achieves $\geq90$\% accuracy across all groups. Sharing FE data is particularly relevant in app-groups where user interaction is significant and realistic avatars enhance the experience (\eg{} social, job simulator). However, in groups where the realistic avatars are unnecessary, specifically in single user mode, where multi user interaction is not necessary, (\eg{} Rhythm, Flight), users may disable FE.

\subsubsection{Sensor Group Model Ensemble} \label{subsec:comparative-evaluation_ModelFusion}
So far, our study assumes the adversaries use individual sensor groups to identify users (from Section \ref{subsec:body-motion-evaluation} to \ref{subsec:comparative-evaluation_AppVsSensor}). Next, we consider settings where the adversary combines multiple sensor group models to improve attack accuracy from individual sensor group models (\ie{} $<90\%$). In this experiment, we excluded FE as an attacker may exploit FE alone for a successful attack (with $\geq 90\%$ accuracy) or users may disable FE (see Section \ref{subsec:comparative-evaluation_AppVsSensor}). Moreover, we did not consider any combination of BM and HJ since the adversary can not collect data from both simultaneously. For combining multiple models, we used the model ensemble technique~\cite{Ensemble_GanaieHMTS22}, where the final identification result is obtained through multiple sensor models using maximum voting mechanisms \cite{MaxVote} of blocks per user.

\parheading{\Bodydata{} and \Eyedata{} Model Ensemble.}
The adversary may consider combining BM and EG, given that, it has access to both BM and EG but not FE and HJ, and individual model accuracy of BM and EG are relatively low (\eg{} 80\% and 70\% for $a_6$, see Table \ref{tab:IdenAcc_bigTable}). Our results show that ensembling EG and BM together can improve attack accuracy up to 10\% (see Appendix \ref{app:model_ensemble}, Table \ref{tab:ensemble}).

\parheading{\Eyedata{} and \Handdata{} Model Ensemble.}
Adversary considers combining HJ and EG, assuming they have access to both sensor groups; and BM/FE is unavailable (\ie{} not used by users) or corrupted. Under this assumption, if individual sensor models for EG and HJ yield low identification accuracy (70\% and 45\% for \eg{} $a_8$, see Table \ref{tab:IdenAcc_bigTable}), ensemble techniques using both sensor models can enhance attack accuracy by 5-10\% (see Appendix \ref{app:model_ensemble} and Table \ref{tab:ensemble}).

\section{Related Work }\label{sec:related-work} 
\parheading{Privacy in VR.}
Adams \etal{} studied the awareness of users and developers on data collection practices on VR devices ~\cite{adams2018ethics}. %
Trimananda \etal ~\cite{trimananda2022ovrseen} analyzed the network traffic generated popular Oculus VR apps, and reported personal information (device and user identifiers and some VR sensory data) collection and their inconsistencies with app's privacy policies~\cite{trimananda2022ovrseen}.
Recently, Nair \etal{} developed an adversarial app to demonstrate tasks that can harvest users' personal information; \eg{} physical characteristics, location, gender~\cite{nair2022metadata}.
The privacy of sensor data and APIs receives growing attention. VREED 
demonstrates emotion recognition in VR through eye tracking and physiological signals \cite{VREED21}. Kal$\epsilon$ido introduces Differential Privacy (DP) for safeguarding eye tracking, emphasizing user interests revealed in eye gaze heatmaps~\cite{kaleido_USENIX21}. MetaGuard~\cite{nair2023metaguard} safeguards user privacy using DP through feature obfuscation.

\parheading{User Identification in VR.}
Most closely related to this paper is a body of prior works that focuses on identification based on sensor data collected on VR~\cite{miller2020personal,miller2022combining,liebers2021understanding,pfeuffer2019behavioural,winkler2022questsim,du2023avatars,nair2023unique,nair2022metadata,tricomi2022you}.
Most prior works focused on identifying users using BM within pre-defined tasks/custom apps. In~\cite{stephenson2022sok}, Stephenson \etal{} compared various authentication mechanisms for AR/VR, which include head/hand/eye biometrics. %
Tricomi \etal{} identified users in VR/AR based on body motion and eye movements~\cite{tricomi2022you}. 
Pfeuffer~\etal{} focused on identification in various controlled tasks (\ie{} pointing, grabbing, walking) as correlated body and eye tracking data together \cite{pfeuffer2019behavioural}. 
Miller \etal{} used \bodydata{} for identification as users randomly select and watch 360-degree videos on VR~\cite{miller2020personal}. Next, Miller \etal{} used spatial features from head and controller identification~\cite{miller2022combining}.
Nair~\etal{}~\cite{nair2023unique} analyzed a large dataset (50K users) of one commercial app (BeatSaber~\cite{beatsaber}), provided by the BeatLeader scoreboard~\cite{BeatLeader}. They utilized body motion and contextual features for identification. %

\section{Discussion
\label{sec:Discussion}}
\subsection{\system{} in Perspective} \label{subsec:discussion_behavrPerspective}
To the best of our
knowledge, \system{} is the first to analyze user identification
in VR comprehensively, \ie{} considering 
(1) all VR sensors available (including HJ, FE, in addition to BM, EG); 
(2) data we collected from several real-world, unmodified apps %
and (3) considering identifiability within and across different apps, allow us covering a wide range of adversarial settings. 

\parheading{Generalization.}
Although our study is limited to \apps{} real-world apps, we believe that our methodology and evaluation results are generalizable.
First, as explained in Sections~\ref{subsec:vr_Hardware_Platform} and \ref{subsec:steamvr-alvr}, \system{} is capable of collecting {\em all} sensor data from {\em any} of the thousands of \steam{} apps that are compatible with ALVR setup.
Second, \system{} analyzes sensor data through device-independent standard OpenXR APIs. Thus, our user identification models and evaluations work with any apps and VR platforms that support the APIs.
Finally, {\em zero-day-attack} and sensor and feature importance analysis can be applied to other apps if they fall under our app groups (based on activities and emotional states) as described in Section~\ref{subsec:app-group}.

\subsection{Recommendations for Mitigation} \label{subsec:recommendation}

Based on our experience with \system{}, we provide some recommendations for best practices and mitigation, including setting up permissions across VR platforms, auditing sensor data collection to offer users recommendations on data sharing practices, and implementing privacy-preserving mechanisms.

\parheading{Use Permissions  on all  Platforms.}
While modern VR platforms such as Oculus VR have provided additional permission checks to protect FE, HJ and EG \cite{radway_permissionOculus23, metapolicy}, SteamVR apps lack any permission system and disclosure about sensor data collection in their websites (see Section~\ref{app:permission}), leaving users with no control over these sensor data (see Section \ref{subsec:sensor-data-types}). We recommend that all VR platforms and app developers (specifically SteamVR platform and apps) should implement permission systems for collecting sensor data, similar to Oculus VR. Additionally, we recommend that developers should disclose clearly which sensor data they  collect, and limit that collection to what is needed for the functionality of the apps.

\parheading{Provide Privacy Recommendation Systems for Users.}
Not all sensor groups are necessary for users to share for every app group (see Section \ref{subsec:comparative-evaluation}). For example, FE is crucial for generating realistic avatars, is important for social apps but not for flight or interactive navigation apps. Moreover, certain sensor groups pose high identification accuracy; \eg{} FE,
thus users may avoid sharing FE in general (for privacy reasons) or in later app groups (not necessary for app activities). 
Users can also decide to share less privacy-sensitive sensor groups; %
\eg{} for flight simulation, controller (parts of BM) can be replaced by HJ as later shows low attack accuracy (see Fig. \ref{fig:AppAdv_Acc_motion_cmp} and \ref{fig:AppAdv_Acc_hand_cmp}).
Default recommendations can be offered to users via privacy nudges \cite{privacyNudges_online17} or implementation of privacy recommendation systems based on static or policy analysis of apps (to analyze what they actually collect for which purpose)~\cite{GuoDLZXH24_userRecommendation} guided by \system{}.

\parheading{Need for Privacy Preserving Mechanisms.}
We hope that our observations, particularly our feature analysis across different apps and sensor groups (see Section \ref{sec:evaluation}), can guide the design of defense mechanisms.
One potential defense strategy could be to obfuscate sensor data. This can be implemented, for example, locally through local differential privacy (LDP) \cite{dwork2006differential} either at the (1) device firmware level, before the sensor data leaves the device, or (2) software level, before the sensor data is transmitted to the server, as outlined in the framework described in \cite{garrido2023sok}. The design depends on the adversary type: if the device is trusted, (2) is sufficient, if the adversary can intercept the device, (1) needs to be implemented, to be effective against the threat model described in Section~\ref{subsec:ThreatModel_Adversaries}.
Guided by our feature analysis (Section \ref{sec:evaluation}), LDP can be applied to top features, which significantly contribute to user identification. For example, obfuscating the y-axis positional readings from the headset in the BM sensor group, which is the top feature, can significantly reduce identification accuracy.
Future work can optimize the privacy-utility trade-off of this defense approach.

 \subsection{Limitations and Future Work}\label{subsec:limitation_futureWork}
\parheading{Study Size.}
 A limitation of the user study, described in Section~\ref{subsec:data-collection},  is the number of participants (20). This number is on par with similar studies \cite{miller2022combining,tricomi2022you,nair2022metadata}, but smaller than in \cite{miller2020personal} (500 users, one task) or \cite{nair2023unique} (50K users, crowd-sourced)\footnote{We show that accuracy does not drop significantly with varying participant numbers, align with prior studies with 500 \cite{miller2020personal} or 5000 \cite{nair2023unique} participants (see Appendix \ref{app:statistics}).}. The limitation in the number of participants comes from the time-consuming nature of our experiments, interacting with several real-world apps for a significant amount of time, in-person,  under IRB guidelines; see Section \ref{subsec:data-collection} for details. A related limitation is that our dataset may not be representative of all  VR users, such as younger users (age $\leq$ 18), older adults (age > 40), or \wrt{} other demographics. These factors may introduce bias and limit the generality of the results. %

While acknowledging the study size as a limitation, we hope this work provides new insights into VR privacy by expanding the problem space in other dimensions, \ie{}  several sensors (4 groups, 475 readings) and  diverse real-world apps (20 from 8 groups). We will make the \system{} system available to enable future research to expand the study to a larger scale, if so desired.

\parheading{Auditing Data Collection.}
On \steam{}, apps are neither restricted by permissions nor required to disclose the collection of sensor data on their store pages (see Sections \ref{subsec:vr_Hardware_Platform} and~\ref{app:permission}).
Therefore, an app adversary may technically collect any sensor data without restrictions from the platform.
However, we do not claim that individual apps do so.
Auditing the data collection would involve network or program analysis~\cite{trimananda2022ovrseen} beyond the scope of \system{}.
We leave this as future work.

\parheading{Advanced User Identification and Profiling.}
Although RF and XGB models can already achieve good performance (see Section~\ref{sec:evaluation}), %
an adversary may minimize its work by feature minimization or leveraging more powerful models.
Furthermore, given the rich behavioral information embedded in the sensor data, an adversary may go beyond identification and draw more inferences about (\ie{} {\em profile}) users, such as demographics, physical conditions, and preferences. A natural next step is to exploit our dataset for profiling.

\section{Conclusion} %
\label{sec:conclusions}

We present \system{}, a framework for collecting and analyzing VR sensor data from four sensor groups. We applied it to \vrdevice{} and conducted a user study where real users interacted with real-world VR apps. We build models that an adversary can use to identify users within similar or different settings of an app, across different apps, or within a group of similar apps. We show that these models perform well, 
and we compare their performance across different sensor groups and apps. We also investigate the minimum time and top features for identification, and the importance of sensor groups on the apps or app groups. Additionally, we provide insights on how \system{} can be generalized and effective for diverse studies in VR, and recommend strategies for privacy practitioners, including setting permissions and implementing privacy measures.

\begin{acks}
We would like to thank Diana Romero for her help with part of our VR app selection process. We would also like to thank Professor Habiba Farrukh for her valuable discussions, including on data access and permissions for VR apps.
This work was supported in part by the National Science Foundation under award numbers 1956393, 1900654 and 2339266, and a gift from the Noyce Initiative.

\end{acks}

\bibliographystyle{ACM-Reference-Format}
\bibliography{references}
\appendix

\section{Details on Sensor Groups} \label{app:sensor_groups}
In this appendix, we expand on Sections \ref{subsec:sensor-data-types} and \ref{subsec:featurization} and provide additional details and discussion of the sensor groups. 

\parheading{\Bodydata{} (BM).} %
\system{} captures the position $(x, y, z)$, rotation $(x, y, z, w)$, angular $(x, y, z)$ and linear velocity $(x, y, z)$ from the two controllers and only position and rotation from the headset~\cite{openxrcoordinate}. %
This sensor group has received much attention in prior work~\cite{miller2020personal,winkler2022questsim,miller2022combining,liebers2021understanding,nair2023unique}. However, the focus was \textit{only} on position and rotation values. 

\parheading{\Eyedata{} (EG).} \system{} captures the position and rotation of \eyedata{} for both left and right eyes (7 values per eye)~\cite{eyegazepositionrotationunity,eyegazepositionrotationkhronos}. Some of the prior work has also looked into eye data, but from different angles~\cite{pfeuffer2019behavioural,tricomi2022you}. In~\cite{pfeuffer2019behavioural}, the authors analyzed \eyedata{} data together with \bodydata{} data. Meanwhile,~\cite{tricomi2022you} looked into eye parameters (\ie{} pupil size and eye openness). In \system{}, we analyze \eyedata{} as an independent sensor group (see Section~\ref{subsec:featurization}).

\parheading{\Handdata{} (HJ).}  The OpenXR standard tracks the motion of each hand as a composition of \textit{26 individually articulated joints}.
See Table \ref{tab:HJ_features} for the full list and descriptions that follows data structure of the \texttt{XrHandJointEXT} in~\cite{openxrhandjointsconvention}. \system{} captures the position and rotation of each joint~\cite{openxrhandtracking} %
for each hand. 

\parheading{\Facedata{} (FE).} The OpenXR standard tracks 64 facial elements. See Table \ref{tab:FE_features} to find full list and descriptions derived from the data structure of the \textit{XrFaceExpressionFB}~\cite{openxrfacetracking}.
The 64 facial elements can be mapped to 31 Action Units (AUs) as per the Facial Action Coding System (FACS)~\cite{Ekman1978FACS}.
Each AU in the FACS standard represents one facial muscle movement. The combinations of the AUs may correspond to a particular emotion. For example, the combination of AU6 (Cheek Raiser) and AU12 (Lip Corner Puller) may indicate a person smiling, which can be correlated with the emotion happiness~\cite{facsexplained}. Details regarding OpenXR facial expression elements \cite{openxrfacetracking} mapping to emotion AUs are in Table~\ref{tab:facial-emotion-actionunits}.

\begin{table*}[ht!]
	\fontsize{6.8pt}{7pt}\selectfont
	\centering
	\caption{List of 26 joints in the \handdata{} data structure per OpenXR convention~\cite{openxrhandtracking}.}
	\begin{tabular}{r l l }
	    \toprule
		\textbf{No.} & \textbf{OpenXR Data Structure} & \textbf{Joint Name}\\
		\midrule
                1. & XR  HAND  JOINT  PALM  EXT                     & Palm                   \\ 
                2. & XR  HAND  JOINT  WRIST  EXT                    & Wrist                  \\ 
                3. & XR  HAND  JOINT  THUMB  METACARPAL  EXT        & Thumb Metacarpal       \\ 
                4. & XR  HAND  JOINT  THUMB  PROXIMAL  EXT          & Thumb Proximal         \\ 
                5. & XR  HAND  JOINT  THUMB  DISTAL  EXT            & Thumb Distal           \\ 
                6. & XR  HAND  JOINT  THUMB  TIP  EXT               & Thumb Tip              \\ 
                7. & XR  HAND  JOINT  INDEX  METACARPAL  EXT        & Index Metacarpal       \\ 
                8. & XR  HAND  JOINT  INDEX  PROXIMAL  EXT          & Index Proximal         \\ 
                9. & XR  HAND  JOINT  INDEX  INTERMEDIATE  EXT      & Index Intermediate     \\ 
                10. & XR  HAND  JOINT  INDEX  DISTAL  EXT           & Index Distal           \\ 
                11. & XR  HAND  JOINT  INDEX  TIP  EXT              & Index Tip              \\ 
                12. & XR  HAND  JOINT  MIDDLE  METACARPAL  EXT      & Middle Metacarpal      \\ 
                13. & XR  HAND  JOINT  MIDDLE  PROXIMAL  EXT        & Middle Proximal        \\
	    \bottomrule
	\end{tabular}
    \hspace{10pt}
	\begin{tabular}{ r l l}
	    \toprule
		\textbf{No.} & \textbf{OpenXR Data Structure} & \textbf{Joint Name}\\
		\midrule\textbf{}
                14. & XR  HAND  JOINT  MIDDLE  INTERMEDIATE  EXT    & Middle Intermediate    \\ 
                15. & XR  HAND  JOINT  MIDDLE  DISTAL  EXT          & Middle Distal          \\ 
                16. & XR  HAND  JOINT  MIDDLE  TIP  EXT             & Middle Tip             \\ 
                17. & XR  HAND  JOINT  RING  METACARPAL  EXT        & Ring Metacarpal        \\ 
                18. & XR  HAND  JOINT  RING  PROXIMAL  EXT          & Ring Proximal          \\ 
                19. & XR  HAND  JOINT  RING  INTERMEDIATE  EXT      & Ring Intermediate      \\ 
                20. & XR  HAND  JOINT  RING  DISTAL  EXT            & Ring Distal            \\
                21. & XR  HAND  JOINT  RING  TIP  EXT               & Ring Tip               \\
                22. & XR  HAND  JOINT  LITTLE  METACARPAL  EXT      & Little Metacarpal      \\
                23. & XR  HAND  JOINT  LITTLE  PROXIMAL  EXT        & Little Proximal        \\
                24. & XR  HAND  JOINT  LITTLE  INTERMEDIATE  EXT    & Little Intermediate    \\
                25. & XR  HAND  JOINT  LITTLE  DISTAL  EXT          & Little Distal          \\
                26. & XR  HAND  JOINT  LITTLE  TIP  EXT             & Little Tip             \\
	    \bottomrule
	\end{tabular}
	\label{tab:HJ_features}
\end{table*}

\begin{table*}[h!]
	\fontsize{6.8pt}{7pt}\selectfont
	\centering
	\caption{List of elements in the \facedata{} data structure as per OpenXR convention~\cite{openxrfacetracking} mapped into Action Units (AU). There are a total of 64 elements of \facedata{} that are mapped into 31 AUs.}
	\begin{tabular}{@{}r@{\hskip 4pt}l@{\hskip 4pt}l@{\hskip 4pt}l@{}}
	    \toprule
		\textbf{No.} & \textbf{Facial Elements in OpenXR Data Structure} & \textbf{Action Unit (AU)} & \textbf{AU\#} \\
		\midrule
                1.  & XR  FACE  EXPRESSION  BROW  LOWERER  L  FB            & Brow Lowerer         & AU4 \\
                2.  & XR  FACE  EXPRESSION  BROW  LOWERER  R  FB            &                      & \\
            \hline
                3.  & XR  FACE  EXPRESSION  CHEEK  PUFF  L  FB              & Cheek Puff           & AU34 \\
                4.  & XR  FACE  EXPRESSION  CHEEK  PUFF  R  FB              &                      & \\
            \hline
                5.  & XR  FACE  EXPRESSION  CHEEK  RAISER  L  FB            & Cheek Raiser         & AU6 \\
                6.  & XR  FACE  EXPRESSION  CHEEK  RAISER  R  FB            &                      & \\
            \hline
                7.  & XR  FACE  EXPRESSION  CHEEK  SUCK  L  FB              & Cheek Suck           & AU35 \\
                8.  & XR  FACE  EXPRESSION  CHEEK  SUCK  R  FB              &                      & \\
            \hline
                9.  & XR  FACE  EXPRESSION  CHIN  RAISER  B  FB             & Chin Raiser          & AU17 \\
                10. & XR  FACE  EXPRESSION  CHIN  RAISER  T  FB             &                      & \\
            \hline
                11. & XR  FACE  EXPRESSION  DIMPLER  L  FB                  & Dimpler              & AU14 \\
                12. & XR  FACE  EXPRESSION  DIMPLER  R  FB                  &                      & \\
            \hline
                13. & XR  FACE  EXPRESSION  EYES  CLOSED  L  FB             & Eyes Closed          & AU43 \\
                14. & XR  FACE  EXPRESSION  EYES  CLOSED  R  FB             &                      & \\
            \hline
                15. & XR  FACE  EXPRESSION  EYES  LOOK  DOWN  L  FB         & Eyes Look Down       & AU64 \\
                16. & XR  FACE  EXPRESSION  EYES  LOOK  DOWN  R  FB         &                      & \\
            \hline
                17. & XR  FACE  EXPRESSION  EYES  LOOK  LEFT  L  FB         & Eyes Look Left       & AU61 \\
                18. & XR  FACE  EXPRESSION  EYES  LOOK  LEFT  R  FB         &                      & \\
            \hline
                19. & XR  FACE  EXPRESSION  EYES  LOOK  RIGHT  L  FB        & Eyes Look Right      & AU62 \\
                20. & XR  FACE  EXPRESSION  EYES  LOOK  RIGHT  R  FB        &                      & \\
            \hline
                21. & XR  FACE  EXPRESSION  EYES  LOOK  UP  L  FB           & Eyes Look Up         & AU63 \\
                22. & XR  FACE  EXPRESSION  EYES  LOOK  UP  R  FB           &                      & \\
            \hline
                23. & XR  FACE  EXPRESSION  INNER  BROW  RAISER  L  FB      & Inner Brow Raiser    & AU1 \\
                24. & XR  FACE  EXPRESSION  INNER  BROW  RAISER  R  FB      &                      & \\
            \hline
                25. & XR  FACE  EXPRESSION  JAW  DROP  FB                   & Jaw Drop             & AU26 \\
            \hline
                26. & XR  FACE  EXPRESSION  JAW  SIDEWAYS  LEFT  FB         & Jaw Sideways         & AU30 \\
                27. & XR  FACE  EXPRESSION  JAW  SIDEWAYS  RIGHT  FB        &                      & \\
            \hline
                28. & XR  FACE  EXPRESSION  JAW  THRUST  FB                 & Jaw Thrust           & AU29 \\
            \hline
                29. & XR  FACE  EXPRESSION  LID  TIGHTENER  L  FB           & Lid Tightener        & AU7 \\
                30. & XR  FACE  EXPRESSION  LID  TIGHTENER  R  FB           &                      & \\
            \hline
                31. & XR  FACE  EXPRESSION  LIP  CORNER  DEPRESSOR  L  FB   & Lip Corner Depressor & AU15\\
                32. & XR  FACE  EXPRESSION  LIP  CORNER  DEPRESSOR  R  FB   &                      & \\
            \bottomrule
	\end{tabular}
    \hspace{2pt}
	\begin{tabular}{@{}r@{\hskip 4pt}l@{\hskip 4pt}l@{\hskip 4pt}l@{}}
	    \toprule
		\textbf{No.} & \textbf{Facial Elements in OpenXR Data Structure} & \textbf{Action Unit (AU)} & \textbf{AU\#} \\
		\midrule
                33. & XR  FACE  EXPRESSION  LIP  CORNER  PULLER  L  FB      & Lip Corner Puller    & AU12 \\
                34. & XR  FACE  EXPRESSION  LIP  CORNER  PULLER  R  FB      &                      & \\
            \hline
                35. & XR  FACE  EXPRESSION  LIP  FUNNELER  LB  FB           & Lip Funneler         & AU22 \\
                36. & XR  FACE  EXPRESSION  LIP  FUNNELER  LT  FB           &                      & \\
                37. & XR  FACE  EXPRESSION  LIP  FUNNELER  RB  FB           &                      & \\
                38. & XR  FACE  EXPRESSION  LIP  FUNNELER  RT  FB           &                      & \\
            \hline
                39. & XR  FACE  EXPRESSION  LIP  PRESSOR  L  FB             & Lip Pressor          & AU24 \\
                40. & XR  FACE  EXPRESSION  LIP  PRESSOR  R  FB             &                      & \\
            \hline
                41. & XR  FACE  EXPRESSION  LIP  PUCKER  L  FB              & Lip Pucker           & AU18 \\
                42. & XR  FACE  EXPRESSION  LIP  PUCKER  R  FB              &                      & \\
            \hline
                43. & XR  FACE  EXPRESSION  LIP  STRETCHER  L  FB           & Lip Stretcher        & AU20 \\
                44. & XR  FACE  EXPRESSION  LIP  STRETCHER  R  FB           &                      & \\
            \hline
                45. & XR  FACE  EXPRESSION  LIP  SUCK  LB  FB               & Lip Suck             & AU28 \\
                46. & XR  FACE  EXPRESSION  LIP  SUCK  LT  FB               &                      & \\
                47. & XR  FACE  EXPRESSION  LIP  SUCK  RB  FB               &                      & \\
                48. & XR  FACE  EXPRESSION  LIP  SUCK  RT  FB               &                      & \\
            \hline
                49. & XR  FACE  EXPRESSION  LIP  TIGHTENER  L  FB           & Lip Tightener        & AU23 \\
                50. & XR  FACE  EXPRESSION  LIP  TIGHTENER  R  FB           &                      & \\
            \hline
                51. & XR  FACE  EXPRESSION  LIPS  TOWARD  FB                & Lips Toward          & AU8 \\
            \hline
                52. & XR  FACE  EXPRESSION  LOWER  LIP  DEPRESSOR  L  FB    & Lip Depressor        & AU16 \\
                53. & XR  FACE  EXPRESSION  LOWER  LIP  DEPRESSOR  R  FB    &                      & \\
            \hline
                54. & XR  FACE  EXPRESSION  MOUTH  LEFT  FB                 & Mouth Stretch        & AU27 \\
                55. & XR  FACE  EXPRESSION  MOUTH  RIGHT  FB                &                      & \\
            \hline
                56. & XR  FACE  EXPRESSION  NOSE  WRINKLER  L  FB           & Nose Wrinkler        & AU9 \\
                57. & XR  FACE  EXPRESSION  NOSE  WRINKLER  R  FB           &                      & \\
            \hline
                58. & XR  FACE  EXPRESSION  OUTER  BROW  RAISER  L  FB      & Outer Brow Raiser    & AU2 \\
                59. & XR  FACE  EXPRESSION  OUTER  BROW  RAISER  R  FB      &                      & \\
            \hline
                60. & XR  FACE  EXPRESSION  UPPER  LID  RAISER  L  FB       & Upper Lid Raiser     & AU5 \\
                61. & XR  FACE  EXPRESSION  UPPER  LID  RAISER  R  FB       &                      & \\
            \hline
                62. & XR  FACE  EXPRESSION  UPPER  LIP  RAISER  L  FB       & Upper Lip Raiser     & AU10 \\
                63. & XR  FACE  EXPRESSION  UPPER  LIP  RAISER  R  FB       &                      & \\
                \hline
                64. & XR FACE  EXPRESSION  COUNT FB & Count  \\
	    \bottomrule
	\end{tabular}
	\label{tab:FE_features}
\end{table*}

\begin{table*}[h!]
  \scriptsize
  \centering
 \caption{Mapping between emotions, arousal/valence states (LA = low arousal, HA = high arousal, PV = positive valence, NV = negative valence), from Table \ref{tab:FE_features} derived from OpenXR facial expression elements \cite{openxrfacetracking}, and Action Units (AU) \cite{ekman2002facial}.}
  \begin{tabular}{l l p{65mm} p{55mm}}
    \toprule
    \textbf{Emotion} & \textbf{Arousal/Valence} &\textbf{Facial Element No.}& \textbf{AUs No.} \\
    \midrule
    Happiness & HA/PV & $(5,6) + (33, 34)$ & AU6 + AU12 \\
    Surprise & LA/PV & $(23,24)+(58,59)+(60,61)+(25)$ & AU1 + AU2 + AU5 + AU26 \\
    Anger & HA/NV & $(1,2)+(60,61)+(29,30)+(49,50)$ &  AU4 + AU5 + AU7 + AU23 \\
    Contempt & HA/NV & $ 33 + (11,12) $ &  AU12 + AU14 \\
    Disgust & HA/NV & $(56,57)+ (31,32) + (52,53) $ &  AU9 + AU15 + AU16 \\
    Fear & LA/NV & $(23,24)+(58,59)+(1,2)+(60,61)+(29,30)+(43,44)+(25)$ & AU1 + AU2 + AU4 + AU5 + AU7 + AU20 + AU26 \\
    Sadness & LA/NV & $(23,24)+(1,2)+(31,32) $ & AU1 + AU4 + AU15 \\
    All & All & All emotion elements & All AUs \\
    
 \bottomrule
  \end{tabular}
  \label{tab:facial-emotion-actionunits}
\end{table*}
\section{List of \steam{} Apps}\label{app:list-of-steamvr-apps}
In Section~\ref{sec:experimental-setup}, we discuss our experimental setup that includes how we choose \apps{} \steam{} apps from the list of top 100 \steam{} apps. Table~\ref{tab:list-of-apps} lists the \apps{} \steam{} apps and the activity that users perform during data collection.

\begin{table*}[ht!]
        \scriptsize
	\centering
	\caption{List of \apps{} VR apps in the \system{} app corpus. 
 }
	\begin{tabular}{r | p{34mm} | p{123mm}}
	    \toprule
		\textbf{App No.} & \textbf{App Title} & \textbf{Tasks}
		\\
		\midrule
                $a_1$ & Beat Saber                            & Cut objects with light-sabers: with the controllers and then with bare hands. \\ 
                $a_2$ & BONEWORKS                             & Explore the welcome scene; in front of a shelf, the user is prompted to grab dumbbells with bare hands and exercise. \\ 
                $a_3$ & DCS World Steam Edition               & Fly a military aircraft: the user first control the aircraft with controllers and then with bare hands. \\ 
                $a_4$ & Derail Valley                         & Explore the scene in a train station; in front of a table, the user is interacting with a book and a walkie-talkie using bare hands. \\ 
                $a_5$ & Elven Assasin                         & Shoot arrows to monsters: with the controllers and then with bare hands. \\ 
                $a_6$ & Golf It!                              & Putt a golf ball with the controllers; once it gets close to the hole, the user is prompted to continue with bare hands. \\ 
                $a_7$ & Gorilla Tag                           & Perform gorilla movement (walk like gorilla to explore the environment): first with the controllers, then with bare hands.\\ 
                $a_8$ & Hot Dogs, Horseshoes \& Hand Grenades & Explore a virtual park; in front of a vending machine, the user will interact with it with bare hands. \\ 
                $a_9$ & Job Simulator                         & Explore office-worker simulation; The user is to interact with a virtual office objects with controllers and then with bare hands. \\ 
                $a_{10}$ & Keep Talking and Nobody Explodes     & Defusing a bomb with the controllers; then, the user is prompted to defuse the bomb with bare hands. \\ 
                $a_{11}$ & McOsu                                & Explore the welcome scene; The user is also asked to interact with the virtual objects with bare hands. \\ 
                $a_{12}$ & Neos VR                              & Explore a futuristic building; the user interacts with books in a bookshelf first using the controllers and then with bare hands. \\ 
                $a_{13}$ & No Man's Sky                         & Explore an unknown planet by teleporting; the user interacts with a laser gun (shoot targets) with controllers and then with bare hands. \\ 
                $a_{14}$ & Pavlov VR                            & Play \& practice the basic and shootings; The user is interacting with a panel with bare hands. \\ 
                $a_{15}$ & Rec Room                             & Explore a school or a McDonald or virtual recreation center; The user will wave their hands at an avatar with bare hands. \\ 
                $a_{16}$ & Space Engine                         & Explore a virtual planetarium by teleporting to space objects (\eg{} planets, stars, \etc{}); the user is asked to interact with the planetarium first with the controllers and then with bare hands. \\ 
                $a_{17}$ & Tabletop Simulator                   & Move chess pieces: first with the controllers and then with bare hands. \\ 
                $a_{18}$ & VRChat                               & Explore the virtual scene by walking around; The user will wave or greet with bare hands. \\ 
                $a_{19}$ & VTOL VR                              & Fly a helicopter; The user interact with the control panel and stick with controller and then with bare hands. \\ 
                $a_{20}$ & X-Plane 11                           & Fly a civilian aircraft and interact with the virtual objects with the controllers and with bare hands. \\ 
	    \bottomrule
	\end{tabular}
	\label{tab:list-of-apps}
\end{table*}

\section{Privacy Policy Reading}
\label{app:privacy-policy}

In Section~\ref{sec:privacy-policy}, we discuss our findings in privacy policies of 100 most played VR games. Here, we present additional details. %

\parheading{Availability of privacy policies.} 
For the top 100 apps from the ``Most played VR games'' list on Steam, we manually visit their websites and locate the link to their privacy policies. We find that 60 of them provide privacy policies.

\parheading{Reading privacy policies.}
We read each privacy policy and look for statements on ``biometric data'' or ``sensory data'', as well as specific types (\eg{} ``head movement'') in relevant sections about data collection, use and sharing. However, human reading is prone to omission due to the lengthy text and extra work required to reveal some contents (\eg{} collapsible text). In addition, 4 privacy policies are not in English. To complement the reading, we write a script to call the ChatGPT (GPT-4) model to read the whole text and ask it to report any statements about data types of interest. We also use simple string matching to search for relevant content.

\parheading{VR sensor data in apps' privacy policies.}
Among 60 games that provide privacy policies, we find that 10 of them discuss the collection of sensor data.
Table~\ref{tab:behaviometrics-privacy-policies} shows the list of 10 apps and what sensor data they disclose to collect.
We find that only seven privacy policies \textit{clearly} mention the collection of ``biometric data'' and/or ``sensory data''. Some of them mention more specific data types, such as ``head and hand movement'', ``facial expressions'', and ``skeletal tracking''.
In addition, three privacy policies give \textit{ambiguous} statements about the collection of these data. For example, in VRChat's privacy policy~\cite{vrchatpolicy}, ``California Resident Privacy Notice'' table marks no collection and disclosure in the row ``sensory information'', but the ``Disclosure of Personal Information'' section states ``sensory information'' is shared with vendors.

\parheading{VR sensor data in Meta's privacy policy.}
We additionally read the privacy policy of Meta, the vendor of \vrdevice{}.
In contrast to the scarce discussion of sensor data in VR apps' privacy policies, Meta provides a long list of sensor and biometric data that are collected in its privacy policy and supplemental articles~\cite{metapolicy,handtracking,facetracking}. 
In the paragraph of ``Physical characteristics and movements'', it discloses the collection of the position and orientation of the headset and controllers, the speed of controller movement, hand tracking, eye tracking, facial expression and other data types.
The list clearly covers the types of sensor data explored in this work. Indeed, as the platform, Meta has the best vantage point to collect these data, which can potentially be used for user identification, \ie{} personalization~\cite{metapolicy}.

\begin{table}[t]
\footnotesize
\centering
\caption{Sensory and biometric data types discussed in the privacy policies of the top 100 VR apps on Steam.}
\label{tab:behaviometrics-privacy-policies}
\renewcommand{\arraystretch}{0.90}
\begin{tabular}{@{}l@{\hskip 4pt}l@{}}
\toprule
\textbf{App}      & \textbf{Collected Data Types}     \\ \midrule
iRacing            & sensory data, biometric data      \\ \midrule
Arizona Sunshine &
  \begin{tabular}[c]{@{}l@{}}motion sensor information, motion tracker information\end{tabular} \\ \midrule
Rec Room &
  \begin{tabular}[c]{@{}l@{}}sensory data, head movement, facial expressions\end{tabular} \\ \midrule
DeoVR Video Player & motion sensor events              \\ \midrule
Gorilla Tag        & movement data (hands and head)    \\ \midrule
\begin{tabular}[c]{@{}l@{}}Microsoft Flight Simulator\\ (2 app versions)\end{tabular} &
  skeletal tracking data, sensor data \\ \midrule
VRChat &
  \begin{tabular}[c]{@{}l@{}}sensory information, biometric information (ambiguous)\end{tabular} \\ \midrule
One-armed Cook     & biometric information (ambiguous) \\ \midrule
WGT Golf           & biometric information (ambiguous) \\ \bottomrule
\end{tabular}
\end{table}

\section{More about the User Study Participants}
\label{app:statistics}
\begin{figure}[t!]
    \begin{subfigure}{0.49\columnwidth}
        \centering
        \includegraphics[width=\linewidth,trim={2mm 0 13mm 14mm},clip]{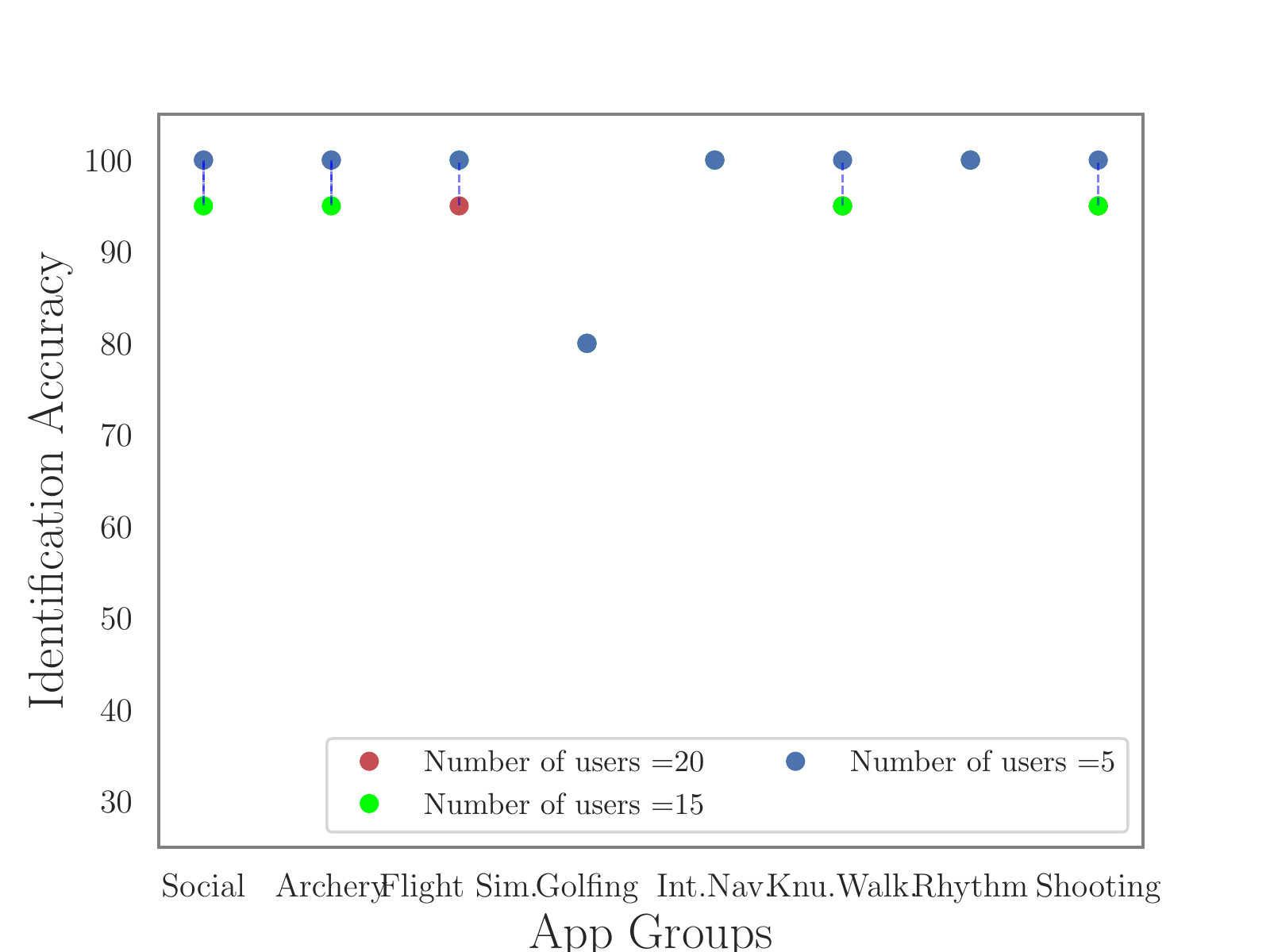}
        \caption{\Bodydata{}}
    \end{subfigure}\hfill%
    \begin{subfigure}{0.49\columnwidth}
        \centering
        \includegraphics[width=\linewidth,trim={2mm 0 13mm 14mm},clip]{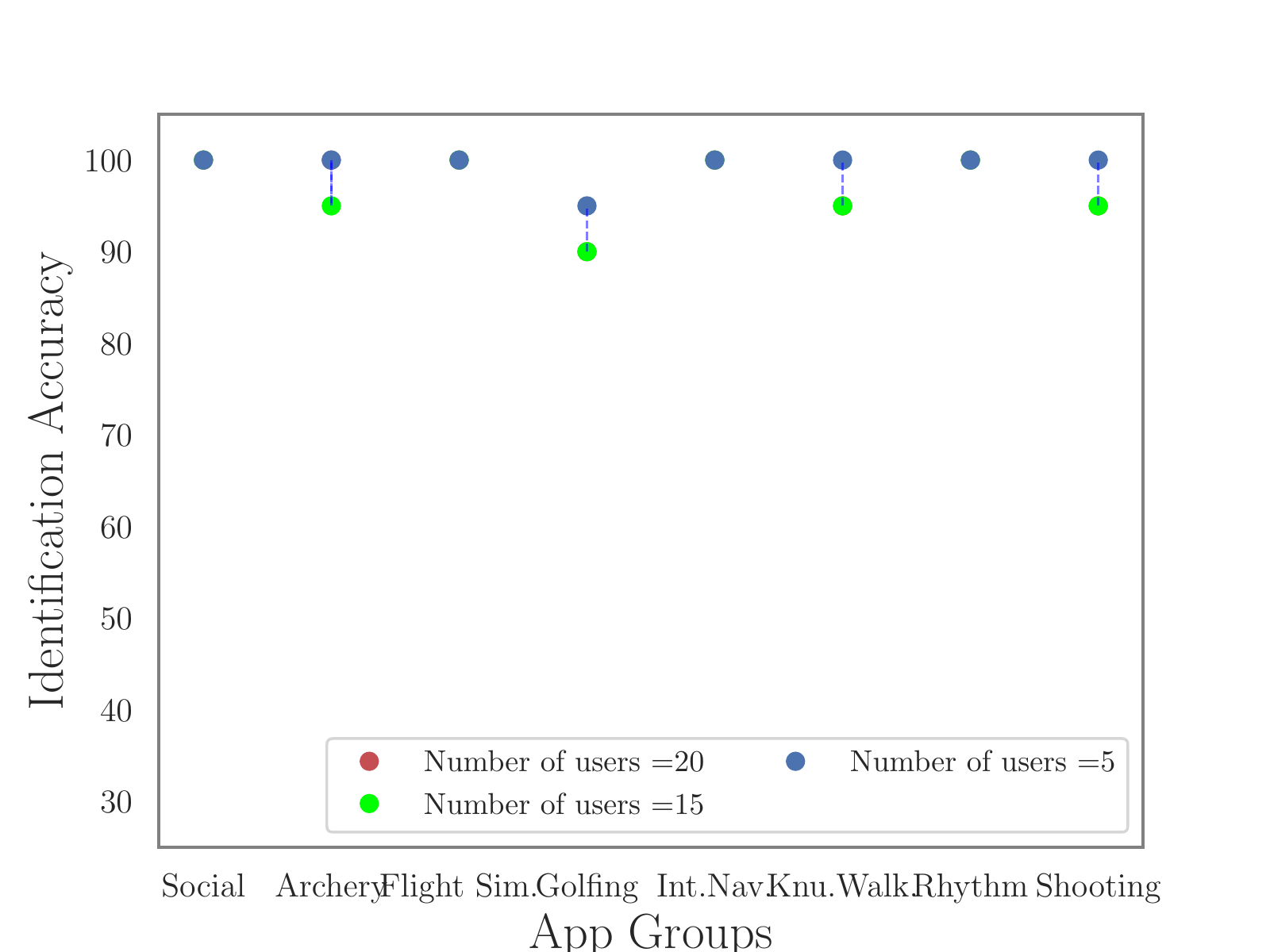}
        \caption{\Facedata{}}
    \end{subfigure}\hfill
    \caption{Visualization of identification accuracy changes by varying number of users.}
    \label{fig:acc_varying_usernumber}
\end{figure}

In this appendix, we expand Section \ref{subsec:data-collection} and provide additional details regarding \system{} user study participants. 

The demographic distributions of the participants are as follows: female is 9 (45\%), male is 11 (55\%).  The age ranges for the participants is between 20-40 with a median age of 26 and mean age of $\sim28$. The nationality of the participants are 4 (20\%) Indian , 3 (15\%) Chinese , 6 (30\%) other Asian , 3 (15\%) American, 2 (10\%) European and 2 (10\%) Undisclosed. Height distributions of the users are, 4 users (20\%) <160cm, 9 (45\%) between 160 to 175cm, 5 (25\%) >175cm and 2 (10\%) undisclosed. Dominant hand (using mostly left or right hand to interact with virtual objects)  of the users are 19 (95\%) right-handed and 1 ambidextrous (5\%). 

Among them, 10 (50\%) of users have prior VR experiences , 9 (45\%) of them was trained during our study by the authors  and 1 (5\%) did not disclose his/her experience.
Prior works show that even with 500 \cite{miller2020personal} or 5000\cite{nair2023unique} users, the identification accuracy does not drop with a significant amount using a simple RF or XGB models. As a proof of concept, we conducted experiments on both the BM and FE groups (for one app from each app group), varying the number of users. Our results show that accuracy does not fluctuate largely through varying numbers of users (See Fig. \ref{fig:acc_varying_usernumber}).

\begin{figure}[t!]
    \begin{subfigure}{0.49\columnwidth}
        \centering
        \includegraphics[width=\linewidth,trim={2mm 0 15mm 14mm},clip]{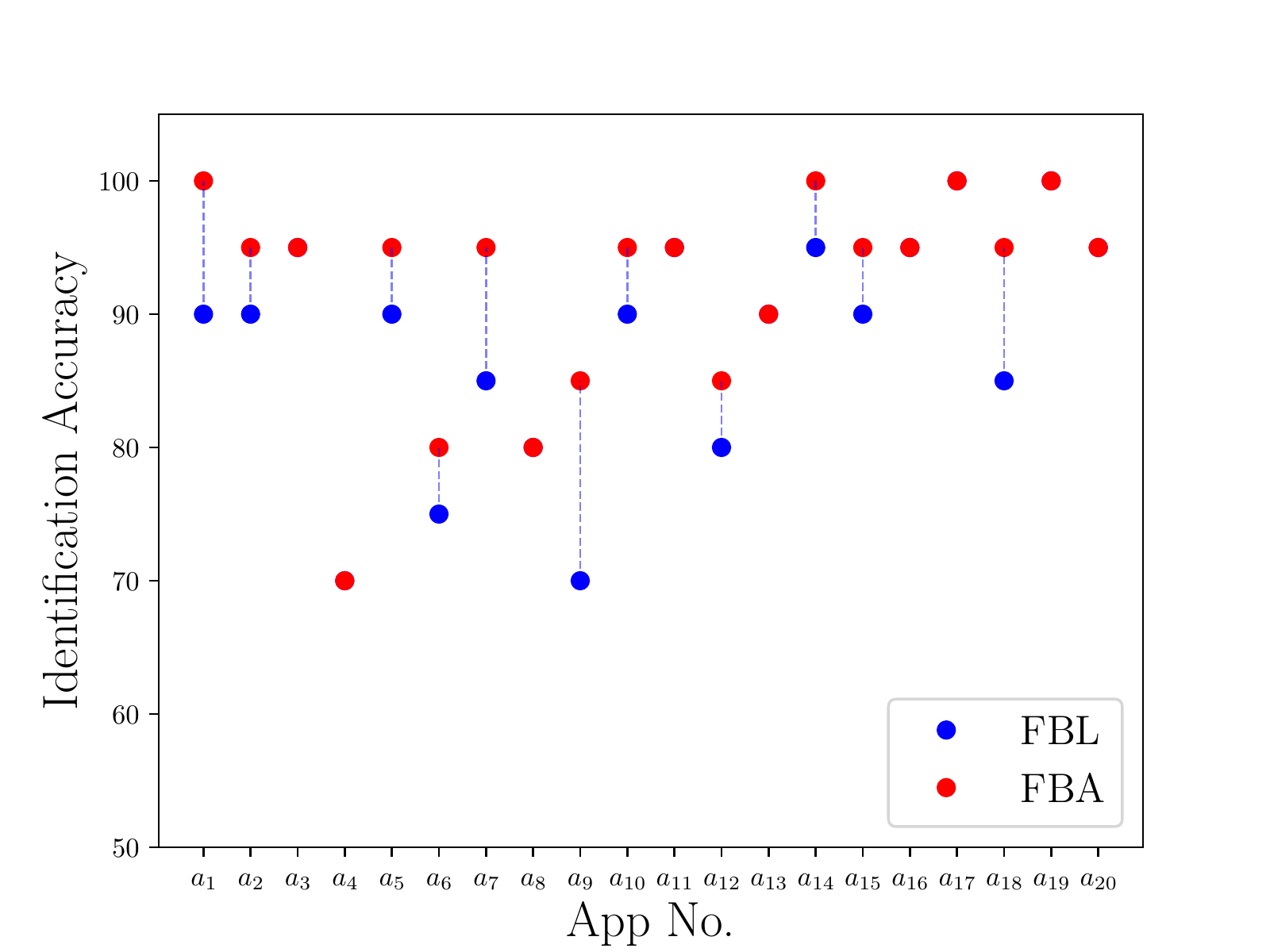}
        \caption{\Bodydata{}}
        \label{fig:cmb_FBA_FBL_motion}
    \end{subfigure}\hfill%
    \begin{subfigure}{0.49\columnwidth}
        \centering
        \includegraphics[width=\linewidth,trim={2mm 0 15mm 14mm},clip]{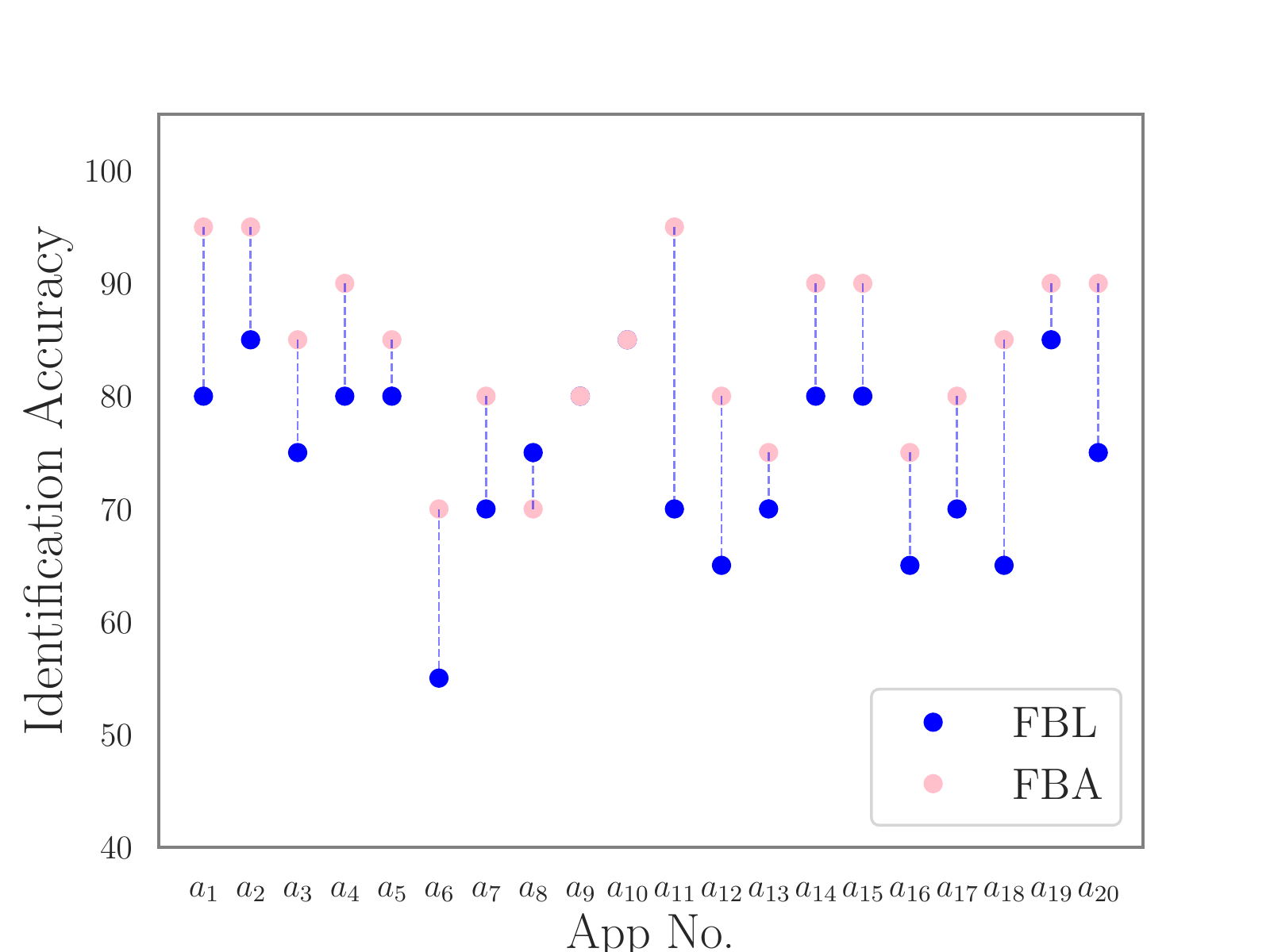}
        \caption{\Eyedata{}}
        \label{fig:cmb_FBA_FBL_eye}
    \end{subfigure}\\\bigskip
    \begin{subfigure}{0.49\columnwidth}
        \centering
        \includegraphics[width=\linewidth,trim={0 0 15mm 14mm},clip]{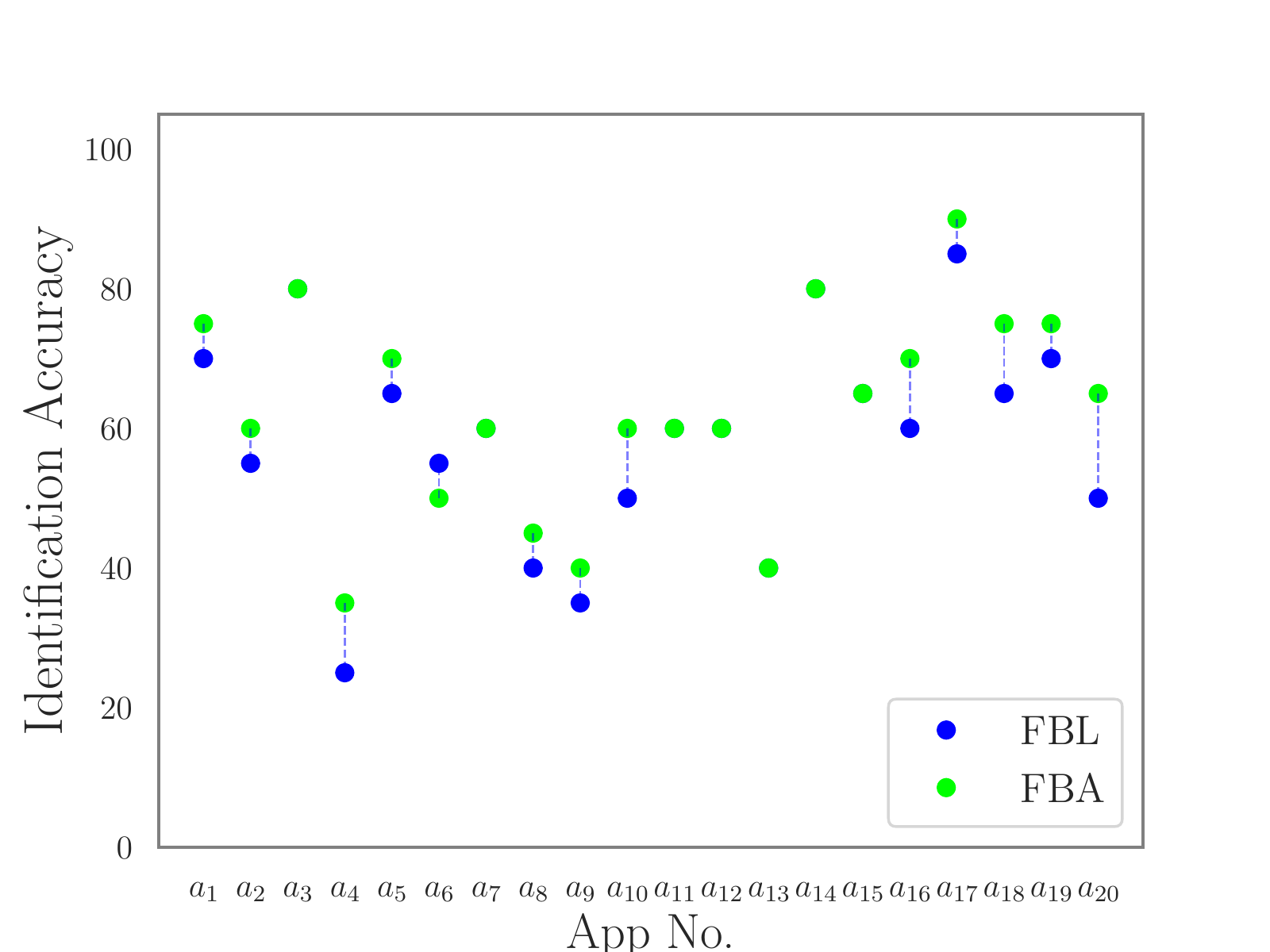}
        \caption{\Handdata{}}
        \label{fig:cmb_FBA_FBL_hand}
    \end{subfigure}\hfill%
    \begin{subfigure}{0.49\columnwidth}
        \centering
        \includegraphics[width=\linewidth,trim={0 0 15mm 14mm},clip]{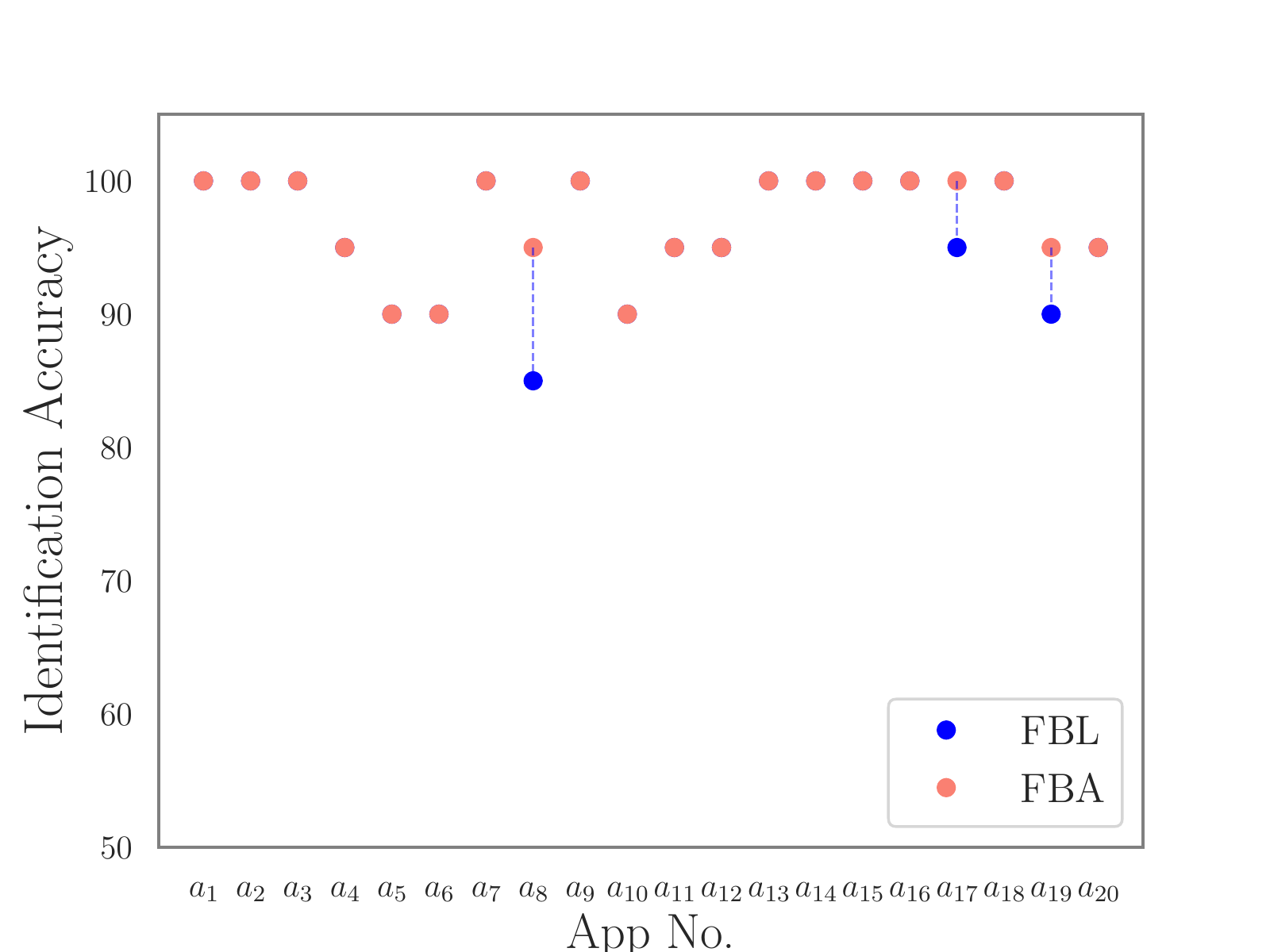}
        \caption{\Facedata{}}
        \label{fig:cmb_FBA_FBL_face}
    \end{subfigure}
    \caption{Identification accuracy comparison between FBA and FBL methods for the four sensor groups.}
    \label{fig:cmb_FBA_FBL_all}
\end{figure}

\section{More Details on Data Processing and User Identification Models}\label{app:model-building} 
This appendix expands Section ~\ref{subsec:data-processing}, where we outlined the process of converting time series data into feature blocks, and Section~\ref{subsec:model-building}, where we discussed building the \system{} models.  Additionally, we provide insights into optimizing FBA and how we select specific model architecture for user identification in \system{}.

\subsection{More about Data Processing}\label{app:data_processing}
\parheading{Pre-processing.}
This step aims to obtain valid time series data with unique timestamps. First, we de-duplicate timestamps and delete invalid columns (\eg{} columns with only zeros). Next, we check any data corruption (\eg{} rows that contain error messages) and replace the invalid values using neighboring rows.

\parheading{Block Division.}
In order to be able to divide the time series in more or less number of blocks, with much shorter or longer duration than 1 second accordingly, we introduce parameter $r \in (0,2]$, which controls the final amount of blocks (``final block amount'') for each app $a_j$: $N_{FBA_j}=r\cdot N_j$. When we increase the ratio $r$, we increase the final block amount while decreasing the block length (amount of time per block). Thus, to align $r$ values across all sensor groups, we choose $r=1$. The key insight here is, unlike FBL, FBA takes into account the variability across users to scale the number of blocks for each app, $N_{FBA_j}$, as to align similar user-app interactions in the time series.

\parheading{Summarization.} We summarize the information in the time series of each block with a vector of 5 statistics, \ie{} maximum, minimum, mean, standard deviation, and median within each block, which will serve as features next. 
This summarization was originally proposed in~\cite{miller2020personal} for \Bodydata{} and was also used in~\cite{nair2023unique}.

\parheading{Block Post-Processing.}
In this step, we verify the block's validity by checking each block (rows) and then each feature (columns). Initially, we eliminate invalid blocks and estimate missing values (e.g., filling missing values in HJ data with related ones). Finally, we refine the feature vectors for the four sensor groups by removing undesirable features (e.g., those with all zero/one values or irrelevant to the classification task).

\subsection{FBA Evaluation and Optimization}\label{app:fba-evaluation-optimization}
We evaluate and compare FBL and FBA in Figures~\ref{fig:cmb_FBA_FBL_motion},~\ref{fig:cmb_FBA_FBL_eye},~\ref{fig:cmb_FBA_FBL_hand} and~\ref{fig:cmb_FBA_FBL_face}. We can observe that FBA improves app model identification accuracy ($5-15$\% for \bodydata{}, $5-25$\% for \eyedata{}, and $5-10$\% for \handdata{}) for most apps compared to FBL, supporting the decision to use FBA over FBL across our experiments.

\parheading{Hyperparameter Tuning.} In \system{}, for RF, first we tune hyperparameters by varying \textit{n-estimators} and \textit{max-depth} from $(50, 200)$ and $(1, 20)$ respectively in five iterations. Then, we select the best model based on five-fold cross-validation. Finally, based on the accuracy obtained from the primary analysis, we choose the optimal point of FBA ratio $r$: our evaluations rely on this final model. %

\parheading{Choosing Optimal Ratio $r$.}
The main challenge when using FBA is finding the optimal FBA block division ratio $r$. If $r$ is too high, summarized data become noisy. On the contrary, if $r$ is too low, important information would be missing from the summary. Finding the right balance is crucial to preserve relevant information. We perform a preliminary experiment based on $7$ participants to find the optimal value of $r$. For \bodydata{} and \eyedata{}, the results suggest $r=1$. For \handdata{}, the results suggest $r=0.5$; this is intuitive as a meaningful hand gesture can be captured in a longer block length. For \facedata{}, the results suggest that any $r$ values will be optimal.

\begin{table}[t!]
   \scriptsize
   \centering %
   \caption{Feature dimensions and block counts for summarized sensor data using the FBA method for different $r$, the parameter that adjusts the block numbers as described in Section \ref{subsec:data-processing}. We report the (number of blocks, number of features) for each sensor group and corresponding $r$.}
  \begin{tabular}{l| r r r r r}
   \toprule
   \textbf{Sensor Group} & \textbf{$r$ = 2} & \textbf{$r$ = 1} & \textbf{$r$ = 0.5} & \textbf{$r$ = 0.2} & \textbf{$r$ = 0.1}\\
   \midrule
    Body Motion & (150658, 165) & (75342, 165) & (37834, 165) & (15133, 165) & (7468, 165)  \\
    Eye Gaze & (168400, 46) & (84200, 46) & (41920, 46) & (16520, 46) & (8080, 46) \\
    Hand Joints & (58480, 400) & (29240, 400) & (14360, 400) & (5480, 400) & (2520, 400) \\
    Facial Expression & (168400, 320) & (84200, 320) & (41920, 320) & (16520, 320) & (8080, 320)  \\
    \hline
    
  \end{tabular}
  \label{tab:abstraction-dim}
\end{table}

\begin{table}[t!]
  \centering 
  \footnotesize
  \caption{Performance analysis for algorithm selection.}
  \begin{tabular}{l |r| r r r r}
    \toprule
    \textbf{Algorithm} & \textbf{App No.} & \multicolumn{4}{c}{\textbf{Accuracy (\%)}}\\
    \cline{3-6}
    & & \textbf{BM} & \textbf{EG} & \textbf{HJ} & \textbf{FE} \\
    \midrule
    RF &  $a_1$ & 100 & 100 & 100 & 100 \\
    RF &  $a_{15}$ & 100 & 100 & 100 & 100\\\hline
    XGB &  $a_{1}$ & 100 & 100 & 100 & 100 \\
    XGB &  $a_{15}$ & 85.71 & 85.71 & 71.42 & 100 \\\hline
    SVM &  $a_{1}$ & 57.14 & 57.14 & 85.71 & 100 \\
    SVM &  $a_{15}$ & 38.23 & 38.23 & 71.42 & 38.23 \\
    \bottomrule
  \end{tabular}
  \label{tab:Alg_acc}
\end{table}

\begin{figure*}[t!]
    \centering
    \begin{subfigure}{0.33\textwidth}
        \centering
        \includegraphics[width=.9\linewidth,trim={0 0 0 14mm},clip]{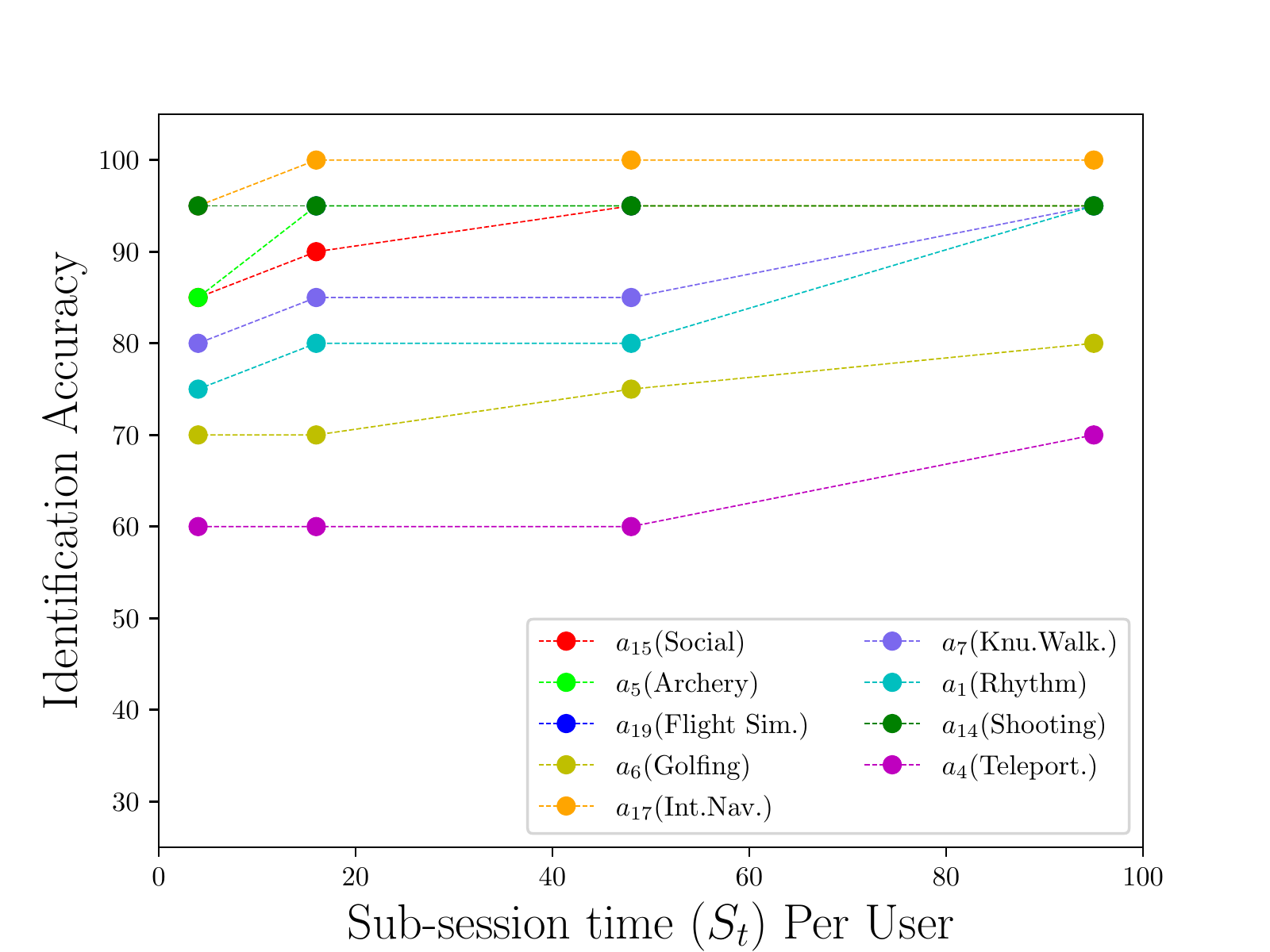}
        \caption{\Bodydata{} ($adv_{app}$)}
        \label{fig:AppAdv_AccBlockPerUser_motion}
    \end{subfigure}\hfill
    \begin{subfigure}{0.33\textwidth}
        \centering
        \includegraphics[width=.9\linewidth,trim={0 0 0 13mm},clip]{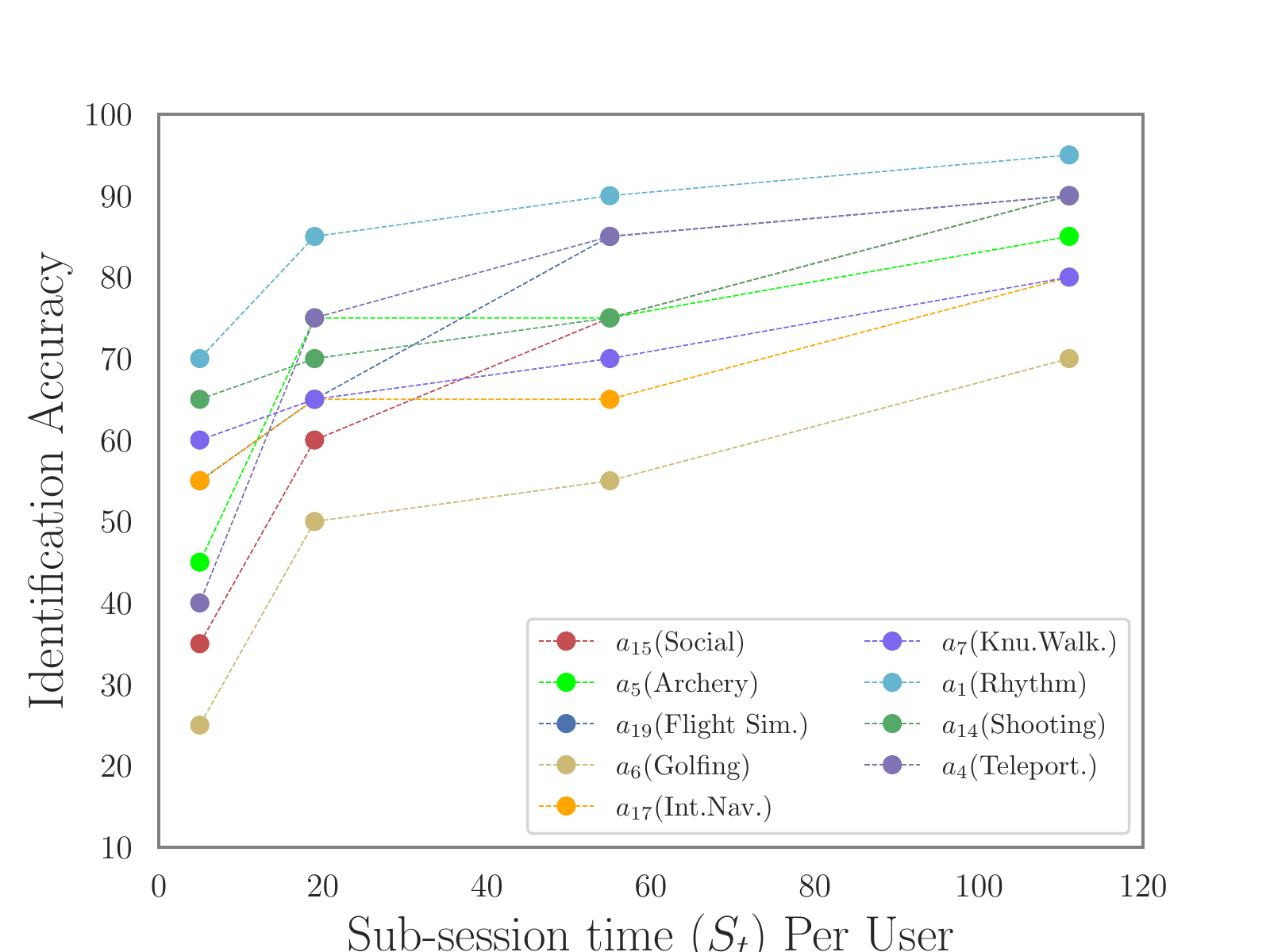}
        \caption{\Eyedata{} ($adv_{app}$)}
        \label{fig:AppAdv_AccBlockPerUser_Eye}
    \end{subfigure}
    \begin{subfigure}{0.33\textwidth}
        \centering
        \includegraphics[width=.9\linewidth,trim={0 0 0 14mm},clip]{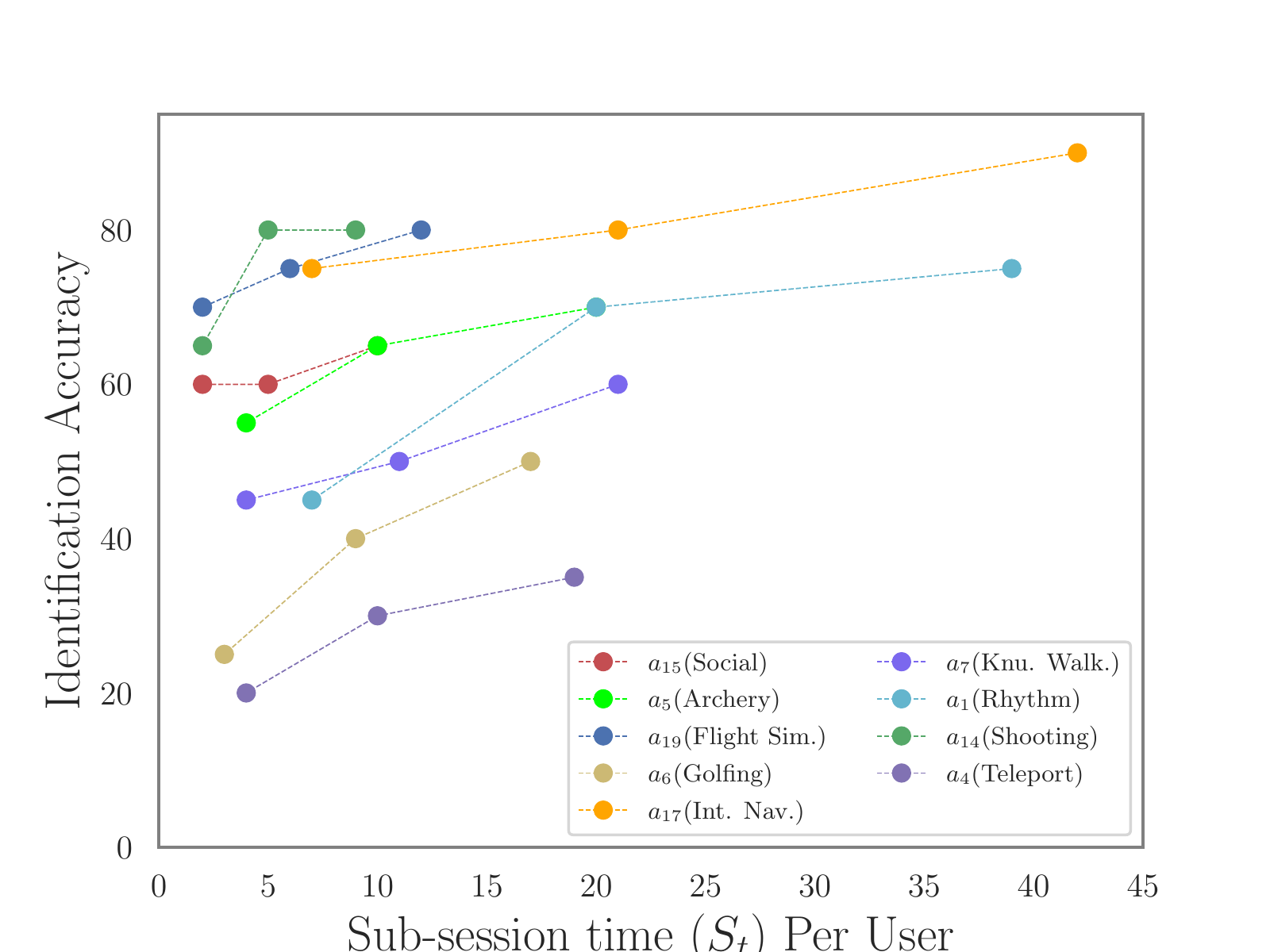}
        \caption{\Handdata{} ($adv_{app}$)}
        \label{fig:AppAdv_AccBlockPerUser_hand}
    \end{subfigure}\hfill
    \begin{subfigure}{0.33\textwidth}
        \centering
        \includegraphics[width=.9\linewidth,trim={0 0 0 14mm},clip]{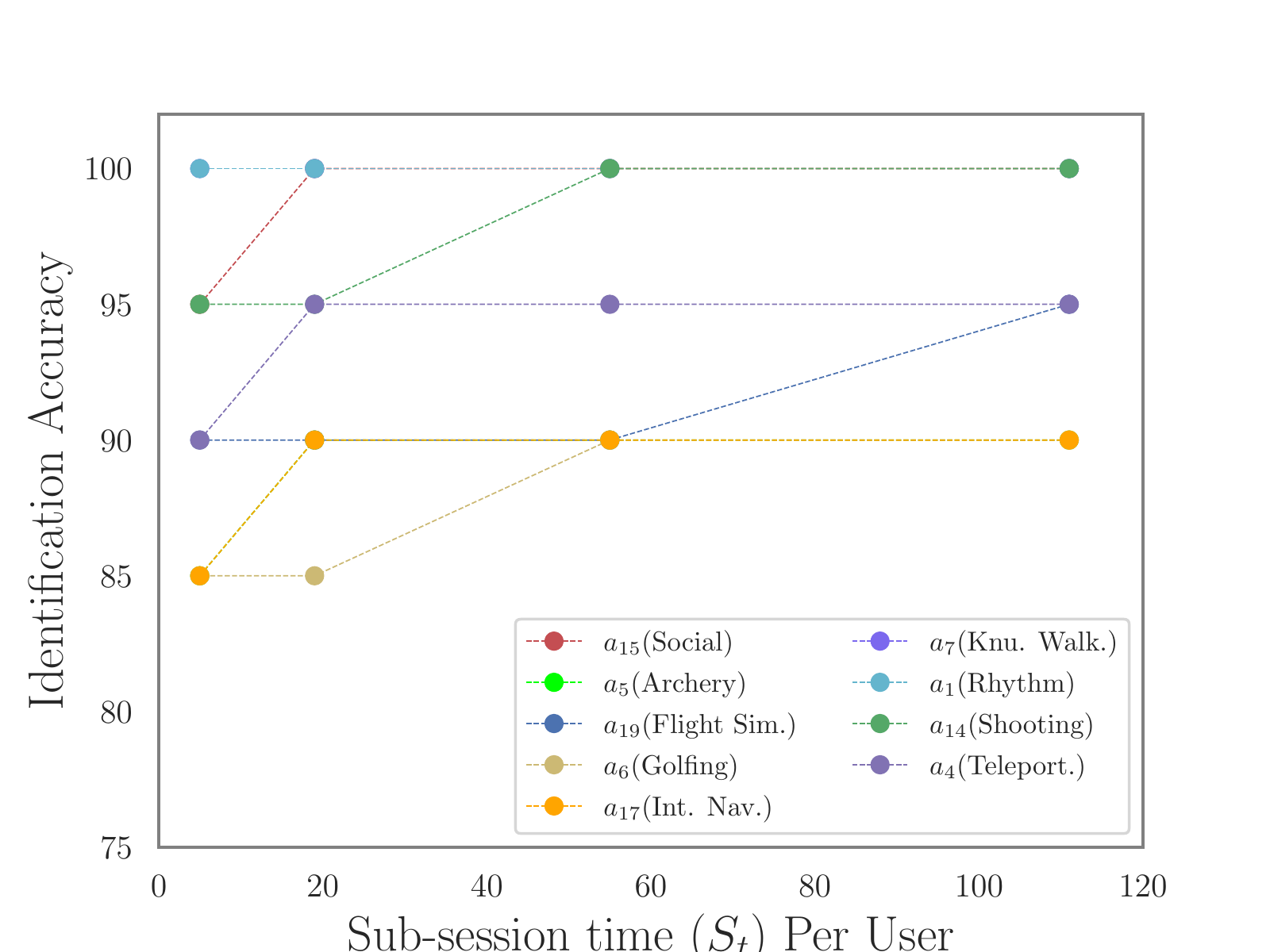}
        \caption{\Facedata{} ($adv_{app}$)}
        \label{fig:AppAdv_AccBlockPerUser_face}
    \end{subfigure}\hfill
     \begin{subfigure}{0.33\textwidth}
        \centering
        \includegraphics[width=.9\linewidth,trim={0 0 0 14mm},clip]{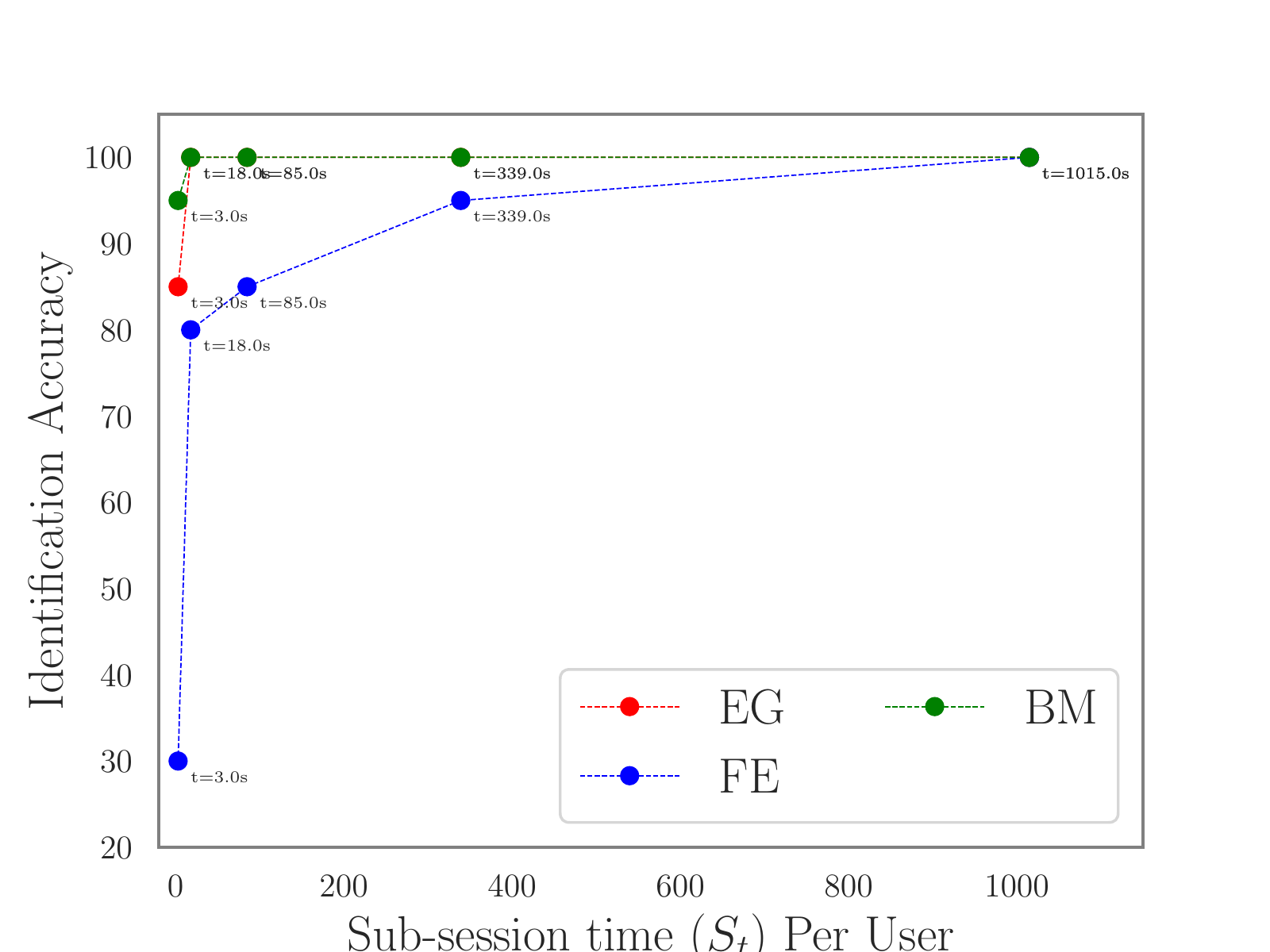}
        \caption{BM, EG and, FE ($adv_{dev}$)}
        \label{fig:min-time-devadv}
    \end{subfigure}
    \hfill
    \begin{subfigure}{0.33\textwidth}
        \centering
        \includegraphics[width=.9\linewidth,trim={0 0 0 14mm},clip]{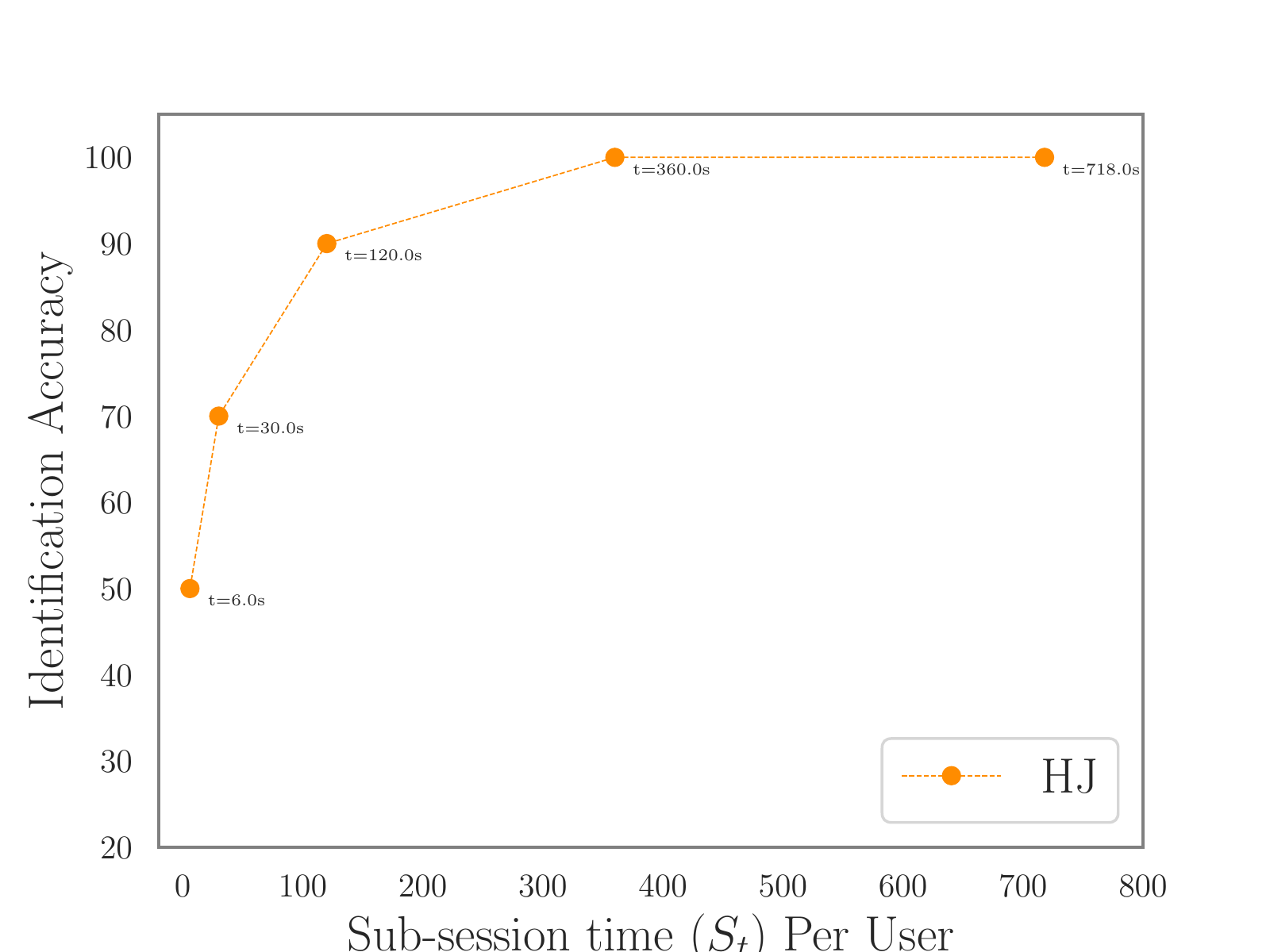}
        \caption{\Handdata{} ($adv_{dev}$)}
        \label{fig:min-time-devadv_hand}
    \end{subfigure}
    \hfill
    \caption{User identification accuracy for app and device models across four sensor groups, with respect to the average sub-session time ($S_t$ in seconds) per user.}
    \label{fig:all_FBA_opt}
\end{figure*}

\begin{figure*}[t!]
    \begin{subfigure}{0.243\textwidth}
        \centering
        \includegraphics[width=.93\linewidth]{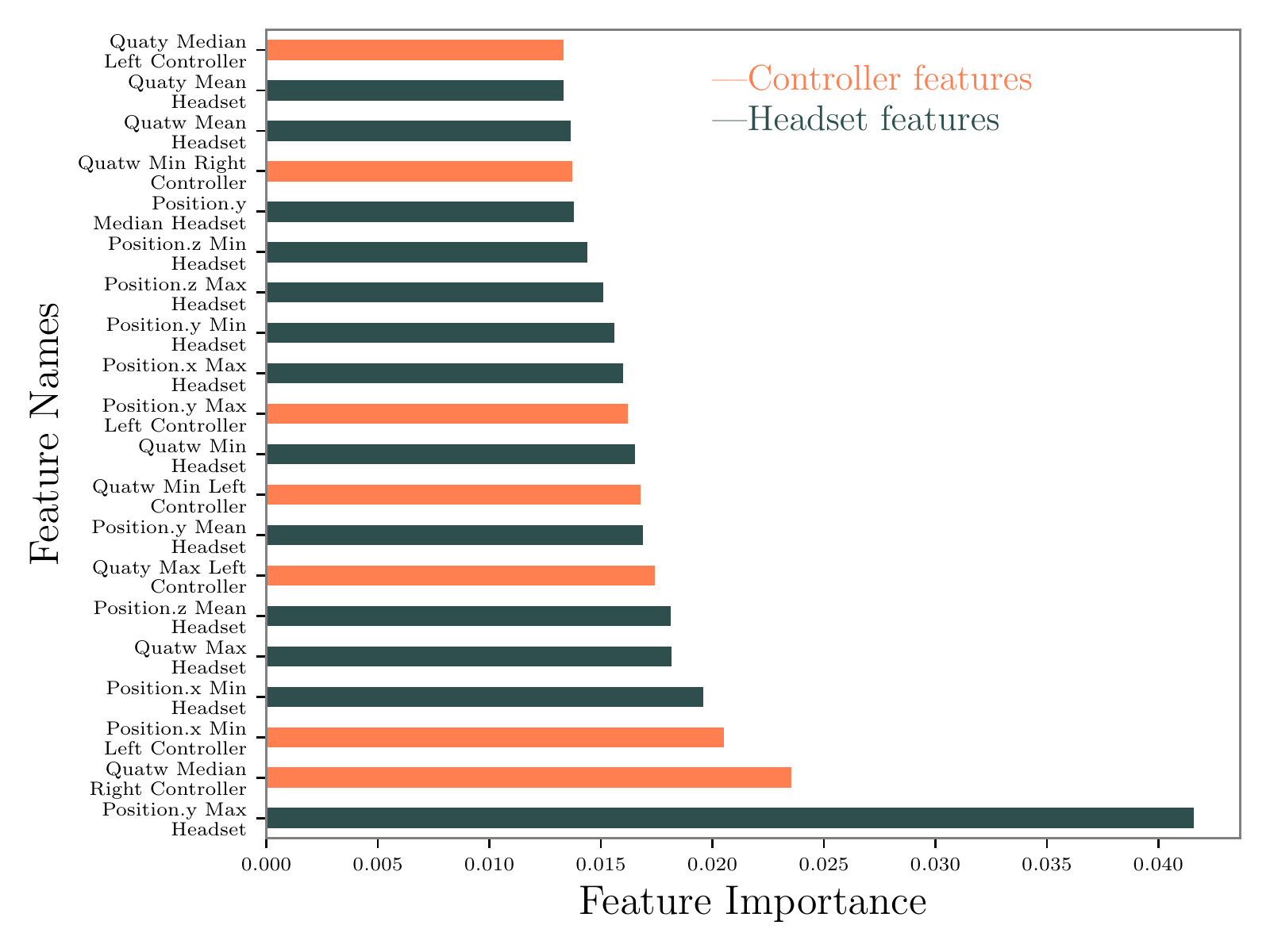}
        \caption{Top-20 features for BM}
        \label{fig:DeviceAdv_FeatureImp_Motion}
    \end{subfigure}\hfill
    \begin{subfigure}{0.243\textwidth}
        \centering
        \includegraphics[width=.95\linewidth]{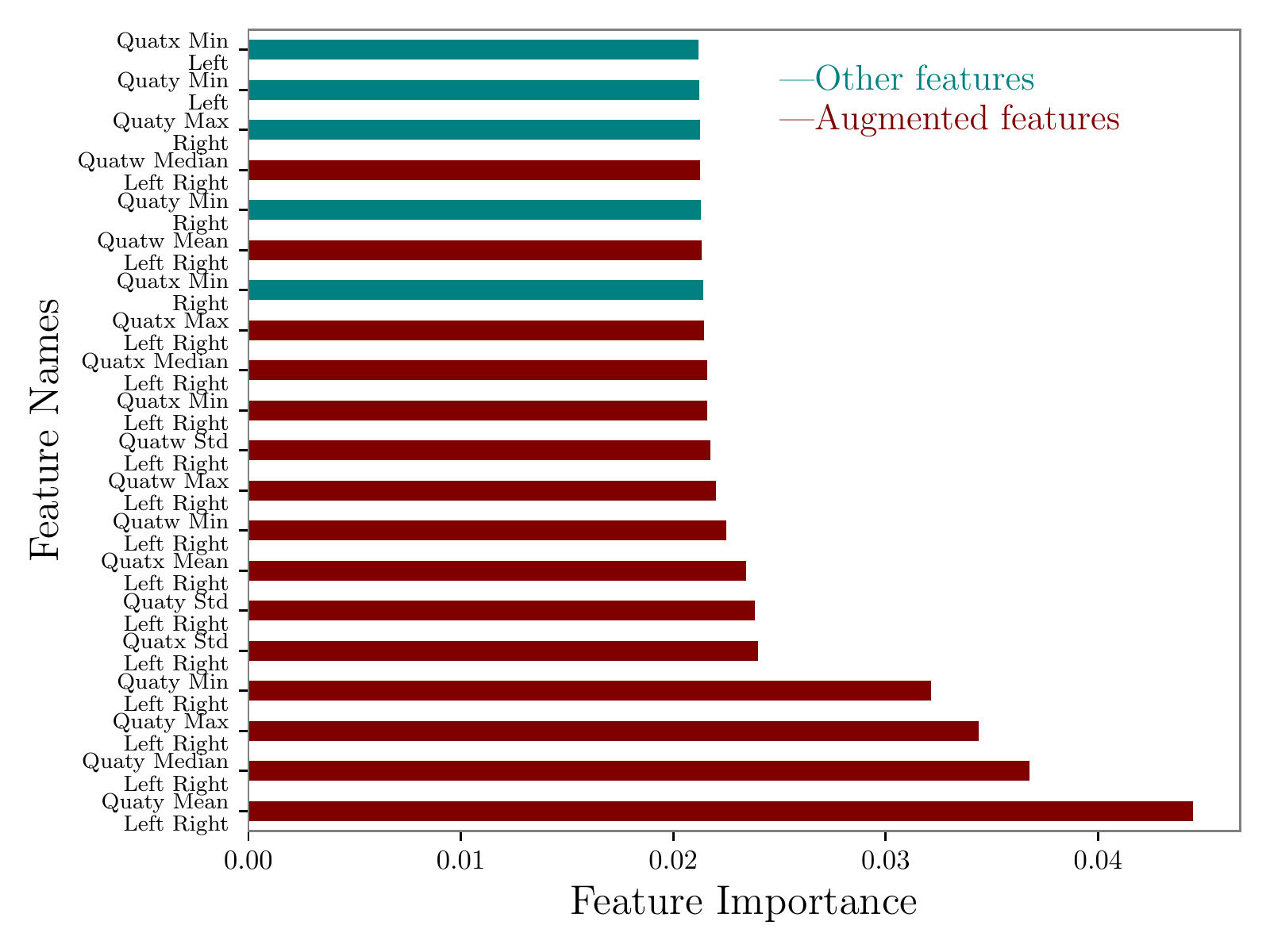}
        \caption{Top-20 features for EG}
        \label{fig:DeviceAdv_FeatureImp_Eye}
    \end{subfigure}\hfill
    \begin{subfigure}{0.24\textwidth}
        \centering
        \includegraphics[width=.95\linewidth]{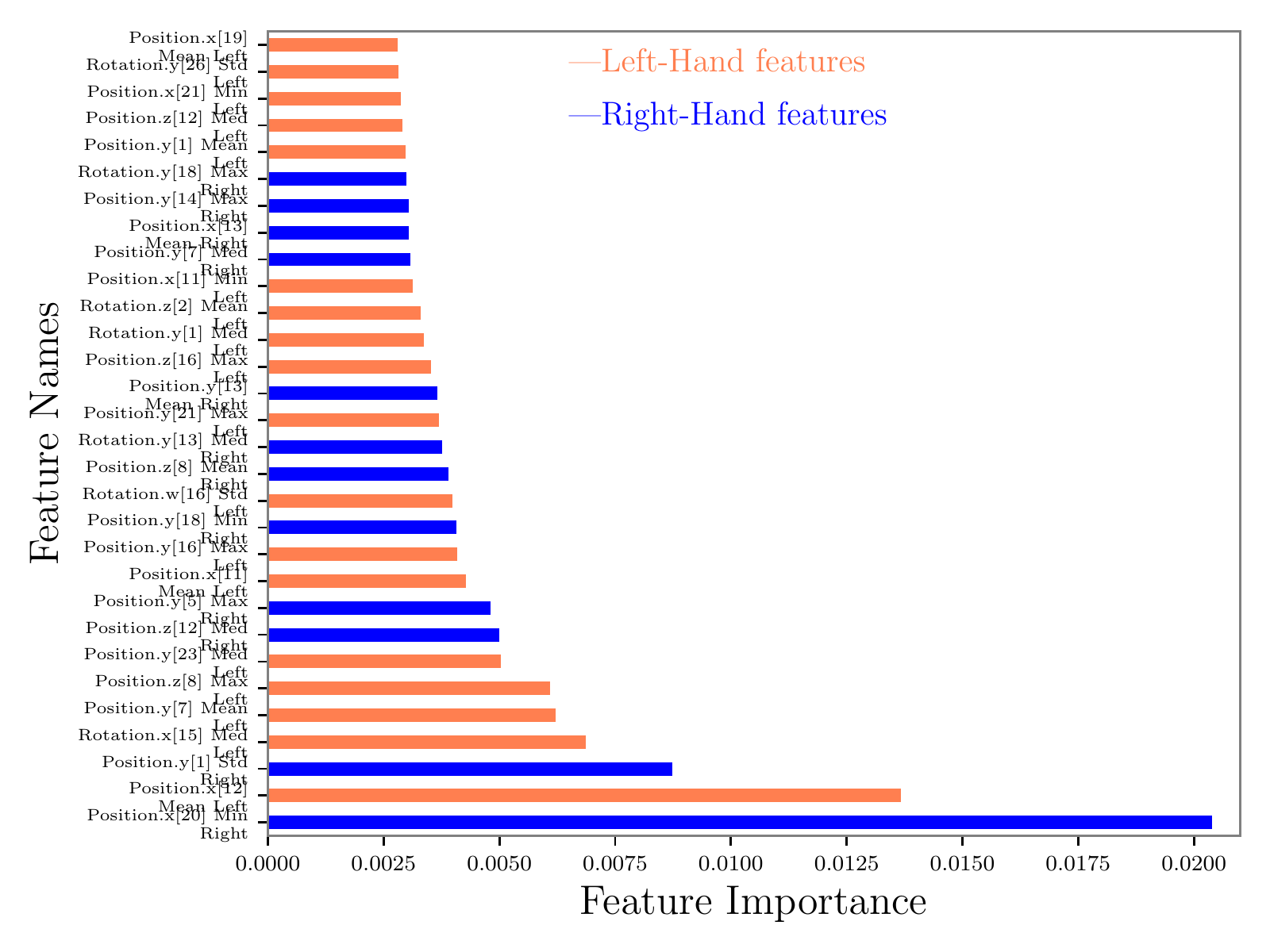}
        \caption{Top-30 features for HJ}
        \label{fig:DeviceAdv_FeatureImp_hand}
    \end{subfigure}\hfill
    \begin{subfigure}{0.245\textwidth}
        \centering
        \includegraphics[width=.95\linewidth]{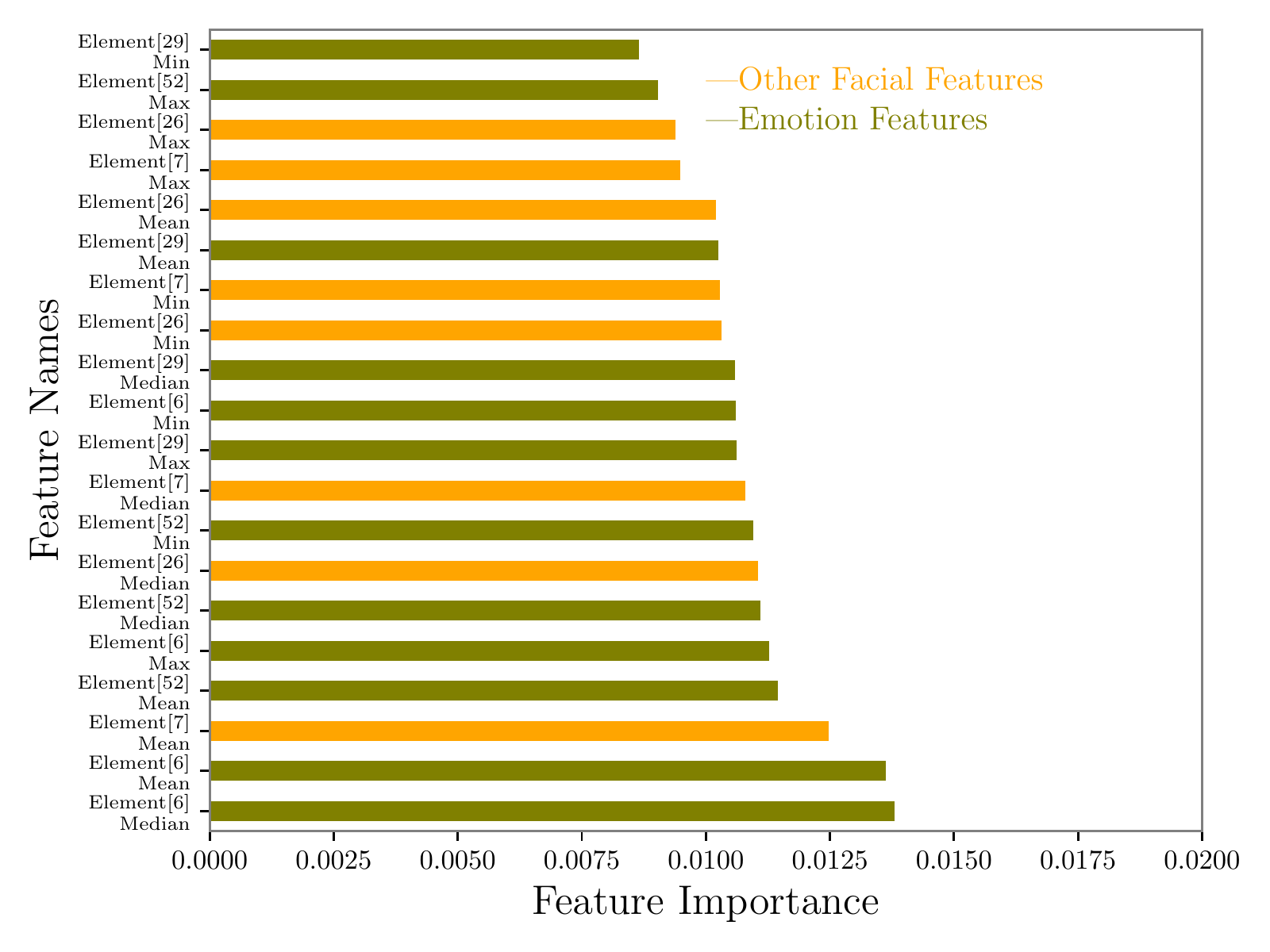}
        \caption{Top-20 features for FE}
        \label{fig:DeviceAdv_FeatureImp_face}
    \end{subfigure}
    \caption{Top features for user identification for \devadv{} \wrt{} each of the four sensor groups. }%
    \label{fig:DeviceAdv_FeatureImp_all}
\end{figure*}

\parheading{Data Summarization Output.}
Table~\ref{tab:abstraction-dim} presents the dimensions of the output of data summarization using FBA.
\subsection{Algorithm Selection}\label{app:algorithm-selection}
We initially explore various ML models such as Random Forest (RF)~\cite{RF}, Gradient Boosting (XGB)~\cite{XGB}, Support Vector Machine (SVM)~\cite{SVM}, and Long Short-Term Memory Networks (LSTM)~\cite{LSTM} across two apps : consist of one social app, namely Rec Room ($a_{15}$), and rhythm app, namely Beat Saber ($a_1$). We analyze the two apps (out of 20), which are among the most popular VR apps, as they contain common activities (\eg{} walking, waving, grabbing, \etc{}).
Table~\ref{tab:Alg_acc} shows that RF achieves the highest identification accuracy; We argue that LSTM is intended to perform sequence prediction, whereas \system{} focuses on identification (\ie{} a classification task); LSTM performs poorly ($\sim$$81\%$ accuracy for \bodydata{} in app $a_1$), thus, we do not consider LSTM further in our evaluation.

\begin{figure*}[t!]
    \centering
    \begin{subfigure}{0.245\textwidth}
        \centering
        \includegraphics[width=\linewidth]{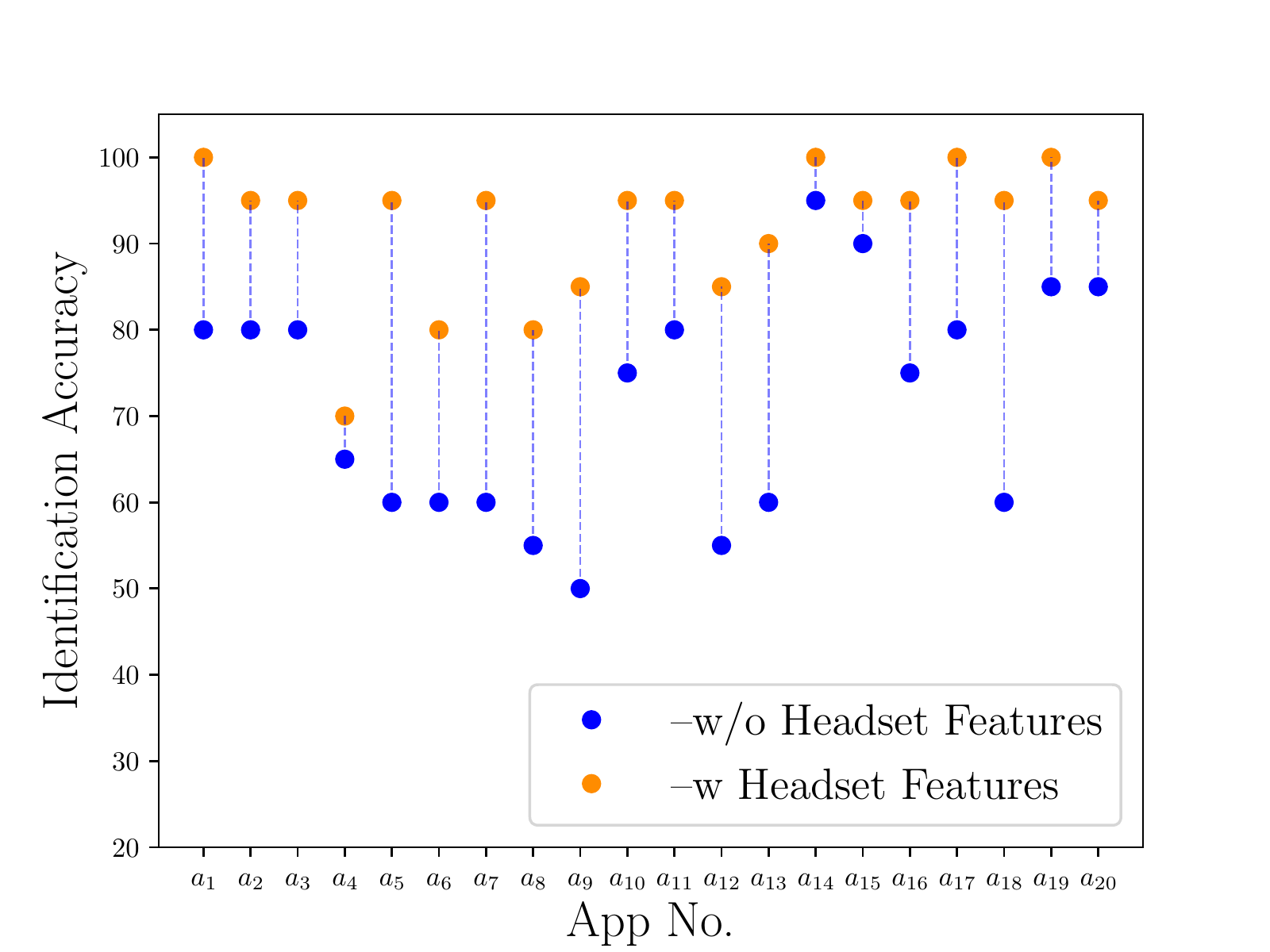}
\caption{Headset features of BM}
\label{fig:AppAdv_Acc_motion_cmp}
    \end{subfigure}
    \hfill
    \begin{subfigure}{0.245\textwidth}
        \centering
        \includegraphics[width=\linewidth]{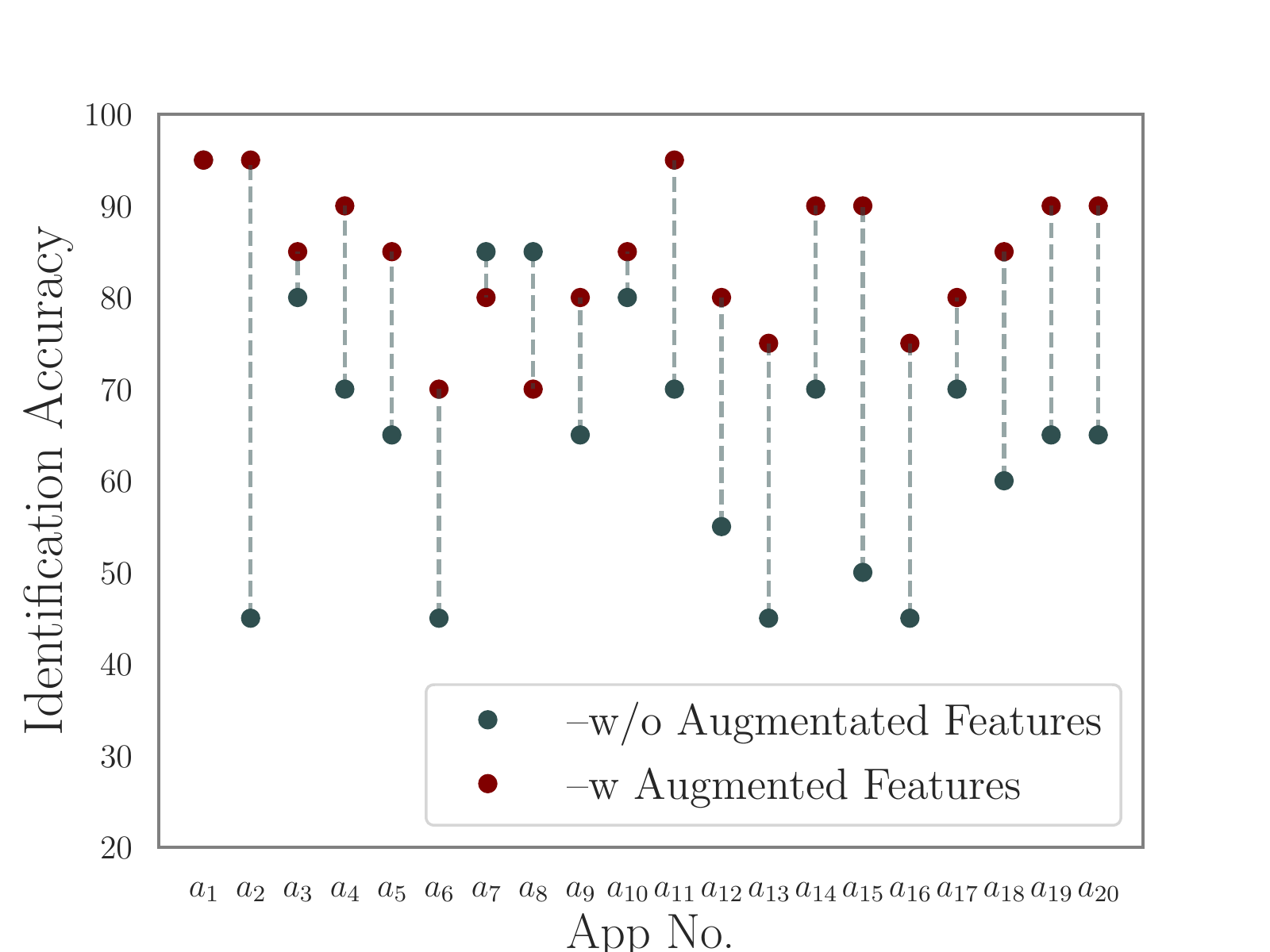}
\caption{Augmented features  of EG}
\label{fig:AppAdv_Acc_Eye_cmp}
    \end{subfigure}
    \hfill
    \begin{subfigure}{0.245\textwidth}
        \centering
        \includegraphics[width=\linewidth]{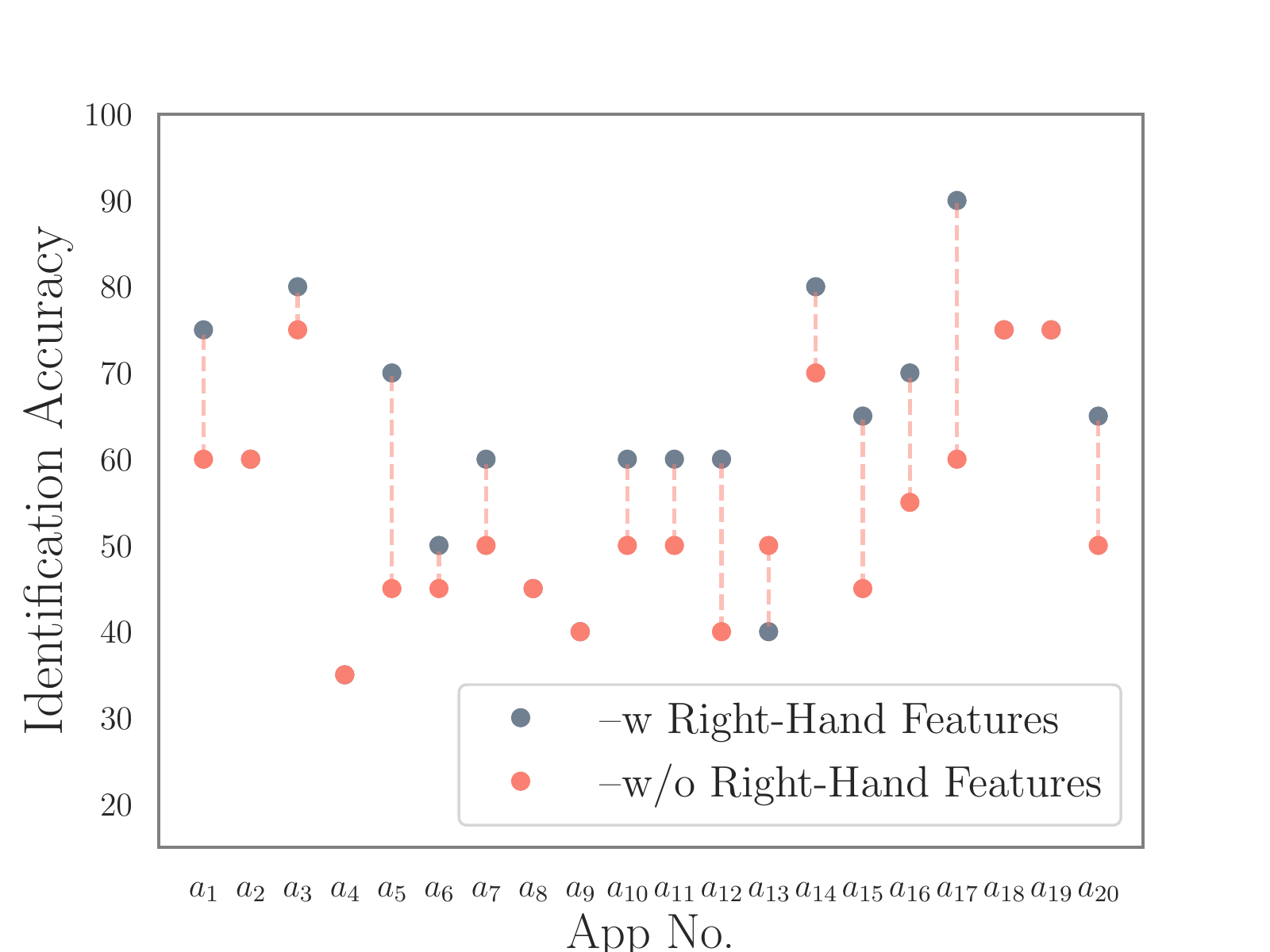}
\caption{Right-hand features of HJ}
\label{fig:AppAdv_Acc_hand_cmp}
    \end{subfigure}
    \hfill
    \begin{subfigure}{0.245\textwidth}
        \centering
        \includegraphics[width=\linewidth]{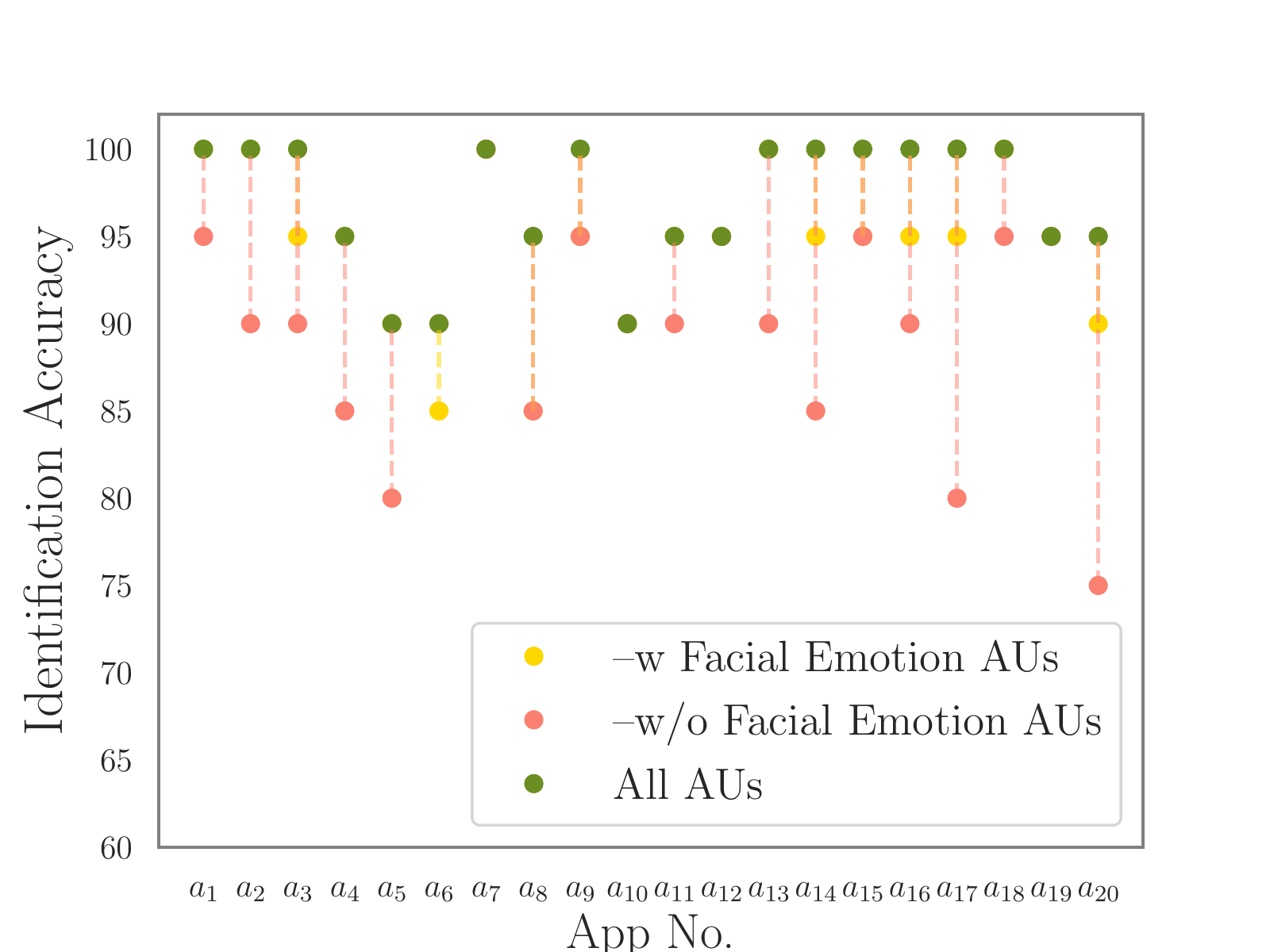}
\caption{Emotion AUs of FE}
\label{fig:AppAdv_Acc_face_cmp}
    \end{subfigure}
    \caption{Visualization of identification accuracy improvement for each of the four sensor groups \wrt{} top-features.}
    \label{fig:Feature_Importance_all}
\end{figure*}

\begin{table*}[ht!]
  \centering
  \scriptsize
  \caption{Top-3 features in user identification for app models for each of the four sensor groups.}
  \begin{tabular}{r | p{39mm} p{38mm} p{35mm} p{38mm}}
    \toprule
    \textbf{App No.} & \textbf{\Bodydata{}} & \textbf{\Eyedata{}} & \textbf{\Handdata{}} & \textbf{\Facedata{}} \\
    \midrule
    $a_1$ & Position.z Mean Left Controller, Position.z Min Headset', Position.y Median Right Controller & Quat.y Mean Left Right, Quat.y Median Left Right, Quat.y Max Left Right & Position.z[3] Max Right, Rotation.z[24] Mean Left, Position.z[1] Max Right & Element[23] Min, Element[5] Median, Element[6] Mean \\
    \hline
    $a_2$ & Position.x Max Headset, Position.y Max Headset, Position.x Mean Headset & Quat.y Mean Left Right, Quat.y Median Left Right, Quat.x Mean Right & Position.z[26] Med Left, Rotation.z[2] Med Right, Rotation.z[18] Min Left & Element[5] Min, Element[57] Median, Element[5] Median \\
    \hline
    $a_3$ & Position.x Mean Headset, Position.z Max Headset, Quat.y Median Headset & Quat.y Mean Left Right, Quat.y Mean Left, Quat.x Max Right & Quat.y Mean Left Right, Quat.y Max Left Right & Element[28] Min, Element[51] Min, Element[51] Median \\
    \hline
    $a_4$ & Position.z Min Headset, Position.y Max Headset, Position.z Max Headset & Quat.y Max Left Right, Quat.y Mean Left Right, Rotation.w Med Left & Position.y Mean Right, Rotation.z[25] Max Right, Position.x[26] Mean Right & Element[30] Min, Element[29] Mean, Element[29] Min \\
    \hline
    $a_5$ & Position.y Min Left Controller, Lin.0 Std Right Controller, Quat.z Mean Right Controller & Quat.y Min Left Right, Quat.y Mean Left Right, Quat.y Max Left Right & Position.x Mean Right, Rotation.z[3] Max Left, Rotation.z[11] Med Left & Element[6] Min, Element[57] Mean, Element[5] Mean \\
    \hline
    $a_6$ & Position.x Min Headset, Position.x Max Headset, Quatw Max Headset & Quat.y Mean Left Right, Quat.y Median Left Right, Quatw Mean Left & Rotation.z Min Left, Rotation.x Max Left, Position.y Max Right & Element[26] Min, Element[57] Mean, Element[5] Mean \\
    \hline
    $a_7$ & Quat.w Mean Headset, Position.x Mean Headset, Quat.x Min Right Controller & Quat.y Mean Left Right, Quat.y Median Left Right, Quat.y Max Left Right & Position.x Mean Left, Position.z Max Left, Rotation.y Min Left & Element[5] Median, Element[2] Min, Element[6] Median \\
    \hline
    $a_8$ & Position.y Max Headset, Position.y Mean Headset, Position.y Median Headset & Quat.y Mean Left Right, Quat.y Median Left Right, Quat.y Max Left Right & Rotation.z Max Right, Rotation.x Mean Right, Position.x Min Right & Element[30] Min, Element[29] Min, Element[6] Median \\
    \hline
    $a_9$ & Quat.y Min Right Controller, Position.x Mean Right Controller, Quat.x Max Right Controller & Quat.y Mean Left Right, Quat.y Median Left Right, Quat.y Max Left Right & Position.z Max Left, Position.y Mean Right, Position.z[2] Max Left & Element[30] Min, Element[6] Mean, Element[27] Max \\
    \hline
    $a_{10}$ & Position.x Median Headset, Position.x Max Headset, Position.x Mean Headset & Quat.y Mean Left Right, Quat.y Median Left Right, Quatw Mean Left & Position.y Mean Right, Rotation.y[12] Min Left, Rotation.z[6] Max Left & Element[29] Min, Element[25] Min, Element[2] Min \\
    \hline
    $a_{11}$ & Position.y Max Headset, Position.y Mean Headset, Position.x Min Headset & Quat.y Mean Left Right, Quat.y Median Left Right, Quat.y Max Left Right & Position.z Min Left, Position.x[11] Max Right, Position.x Mean Left & Element[51] Min, Element[51] Median, Position.x Mean Left \\
    \hline
    $a_{12}$ & Position.y Max Headset, Position.y Mean Headset, Position.y Min Headset & Quat.y Mean Left Right, Quat.y Median Left Right, Quat.y Max Left Right & Position.x Min Right, Position.x[24] Mean Right, Position.x[17] Min Right & Element[51] Median, Element[51] Min, Element[51] Mean \\
    \hline
    $a_{13}$ & Position.y Max Headset, Position.x Mean Headset, Position.y Mean Headset & Quat.y Mean Left Right, Quat.y Median Left Right, Quat.y Min Left Right & Position.x Mean Right, Position.z[15] Min Right, Position.x[5] Mean Right & Element[51] Min, Element[6] Min, Element[25] Max \\
    \hline
    $a_{14}$ & Quat.x Mean Right Controller, Position.z Min Left Controller, Quat.w Mean Left Controller & Quat.y Mean Left Right, Quat.x Max Left, Quatw Max Right & Position.x Mean Left, Position.y[24] Mean Left, Position.x[7] Med Right & Element[51] Median, Element[25] Median, Element[51] Min \\
    \hline
    $a_{15}$ & Position.y Max Headset, Position.x Max Headset, Position.x Mean Headset & Quat.y Mean Left Right, Quat.y Median Left Right, Quat.y Max Left Right & Position.x Mean Right, Position.x[14] Mean Right, Position.x[6] Med Left & Element[51] Min, Element[23] Min, Element[25] Median \\
    \hline
    $a_{16}$ & Position.y Max Headset, Position.x Min Headset, Position.x Mean Headset & Quat.y Mean Left Right, Quat.y Median Left Right, Quat.y Min Left Right & Position.z Min Left, Rotation.z[3] Med Left, Position.x[12] Mean Right & Element[25] Min, Element[5] Mean, Rotation.z[3] Med Left \\
    \hline
    $a_{17}$ & Position.x Min Headset, Position.x Mean Headset, Position.y Max Headset & Quat.y Mean Left Right, Quat.y Median Left Right, Quat.y Max Left Right & Position.y Med Left, Rotation.z[22] Min Right, Rotation.x[25] Mean Right & Element[50] Min, Element[41] Mean, Element[54] Mean \\
    \hline
    $a_{18}$ & Position.y Max Headset, Position.z Mean Headset, Position.y Median Headset & Quat.y Mean Left Right, Quat.y Median Left Right, Quat.x Mean Right & Position.z[13] Mean Right, Position.z[8] Max Left, Position.y[18] Med Left & Element[25] Median, Element[2] Min, Position.z[8] Max Left \\
    \hline
    $a_{19}$ & Quat.x Min Right Controller, Quatw Max Left, Position.y Median Right Controller & Quat.y Mean Left Right, Quat.x Min Right, Quatw Max Left & Position.z Max Left, Position.z[18] Med Left, Position.y[1] Med Left & Element[51] Min, Element[51] Mean, Element[25] Median \\
    \hline
    $a_{20}$ & Position.x  Min  Headset, Position.x  Median  Headset, Position.x  Mean  Headset & Quat.y Mean Left Right, Quat.y Median Left Right, Quat.y Max Left Right & position.x[11] Max Left, Position.x[12] Med Left, Position.x[4] Max Left & Element[30] Min, Element[30] Mean, Element[5] Mean \\
    \bottomrule
  \end{tabular}
  \label{tab:AppAdv_FeatureImp_All}
\end{table*}

\begin{table*}[t!]
   \scriptsize
   \centering 
   \caption{Identification accuracy (in \%) based on combinations of AUs that represent emotions \wrt{} app groups; %
   {\em Emotional States:} LA = low arousal, HA = high arousal, PV = positive valence, NV = negative valence.}
   \begin{tabular}{p{15mm}|l | r r | r  r| r r |r | r | r | r }
    \toprule
    \textbf{Emotion} & \textbf{Arousal/Valance} & \multicolumn{10}{c}{\textbf{Identification accuracy (\%) in App Groups}} \\
    \cline{3-12}
    & & \multicolumn{2}{c|}{\textbf{Social}} & \multicolumn{2}{c|}{\textbf{Flight Sim.}} & \multicolumn{2}{c|}{\textbf{Int. Nav.}} & \textbf{K.-walk.} & \textbf{Rhy.} & \textbf{Shooting} & \textbf{Archery} \\
    & & $a_{18}$ & $a_{15}$ & $a_{19}$ & $a_{20}$ & $a_{16}$ & $a_{10}$ & $a_{7}$ & $a_1$ & $a_{14}$ & $a_{5}$ \\
    \midrule
    Happiness & HA/PV & 100.0 & 100.0 & 85.0 & 70.0 & 80.0 & 75.0 & 95.0 & 95.0 & 80.0 & 70.0 \\
    Surprise & LA/PV & 100.0 & 95.0 & 85.0 & 80.0 & 80.0 & 85.0 & 100.0 & 100.0 & 90.0 & 85.0 \\
    Anger & HA/NV & 95.0 & 95.0 & 95.0 & 85.0 & 85.0 & 85.0 & 90.0 & 95.0 & 90.0 & 85.0 \\
    Disgust & HA/NV & 75.0 & 75.0 & 70.0 & 55.0 & 60.0 & 75.0 & 70.0 & 70.0 & 75.0 & 75.0 \\
    Fear & LA/NV & 90.0 & 95.0 & 90.0 & 90.0 & 90.0 & 95.0 & 100.0 & 100.0 & 95.0 & 90.0 \\
    Sadness & LA/NV & 85.0 & 90.0 & 100.0 & 90.0 & 80.0 & 80.0 & 90.0 & 90.0 & 95.0 & 85.0 \\
    All Emotion AUs & All & 95.0 & 100.0 & 95.0 & 90.0 & 95.0 & 90.0 & 100.0 & 100.0 & 100.0 & 85.0 \\
    All AUs & All & 95.0 & 100.0 & 100.0 & 95.0 & 100.0 & 90.0 & 100.0 & 100.0 & 100.0 & 90.0 \\
    \bottomrule
  \end{tabular}
  \label{tab:results_emotion}
\end{table*}

\section{More Evaluation Results}
\label{app:evaluation-tables-figures} %

In Section~\ref{sec:evaluation}, we presented evaluation results of \system{}'s app and device models for user identification. In this appendix, we present additional tables and figures related to evaluation.

\parheading{Sub-session Time Characterization.}
\label{app:subsession-time-characterization}
Figures~\ref{fig:AppAdv_AccBlockPerUser_motion},~\ref{fig:AppAdv_AccBlockPerUser_Eye},~\ref{fig:AppAdv_AccBlockPerUser_hand}, and~\ref{fig:AppAdv_AccBlockPerUser_face} (for \appadv{}); and Figures~\ref{fig:min-time-devadv} and~\ref{fig:min-time-devadv_hand} (for \devadv{}) show identification accuracy \wrt{} sub-session time.

\parheading{Top Features.}
\label{app:top-features}
We list the top features for user identification using app models trained with data from the four sensor groups in Table~\ref{tab:AppAdv_FeatureImp_All}.
Further, Figures~\ref{fig:DeviceAdv_FeatureImp_Motion},~\ref{fig:DeviceAdv_FeatureImp_Eye},~\ref{fig:DeviceAdv_FeatureImp_hand}, and~\ref{fig:DeviceAdv_FeatureImp_face} show the top features for user identification for \devadv{}.
In Figure \ref{fig:AppAdv_Acc_motion_cmp}, \ref{fig:AppAdv_Acc_Eye_cmp}, \ref{fig:AppAdv_Acc_hand_cmp} and \ref{fig:AppAdv_Acc_face_cmp} shows importance of headset features for BM, augmented features for EG, right-hand features of HJ and
Finally, AUs/elements of emotion for FE respectively.

\parheading{Identification Accuracy Based on Emotion Action Units.} \label{app:emotion-based-identification} In
Table \ref{tab:results_emotion} shows the identification accuracy based on combinations of AUs that represent emotions based on different app groups.

\parheading{Identification Accuracy for Open-World Settings.} \label{app:OW_identification} 
In Table \ref{tab:Open_world}, the identification accuracy for $5$ representative apps from five different app-groups is shown, given that the training and testing data are collected from different settings, difficulty levels, or songs (open world settings).

\parheading{Sensor Group Model Ensemble Results.} \label{app:model_ensemble} 
Table \ref{tab:ensemble} represents the identification accuracy for the attacker that ensemble multiple sensor group models and then calculates the final attack accuracy. The first three sub-columns of the Accuracy column represent individual sensor group accuracy (\eg{} either for BM, EG, or HJ). If individual sensor group identification accuracy is low, the attacker further ensemble those weak models of multiple sensor groups, as presented in the last two columns (BM\&EG and EG\&HJ). Any empty value on the table indicates that the individual sensor model provides high identification accuracy, the attacker further did not optimize it.

\begin{table}[t!]
    \centering 
    \scriptsize
    \caption{Evaluation Results for the Open-World Setting.}
    \begin{tabular}{r l| r r r r}
        \toprule
        \textbf{App No.} & \textbf{App Group} & \multicolumn{4}{c}{\textbf{Accuracy (\%)}}\\
        \cline{3-6}
        & & \textbf{BM} & \textbf{EG} & \textbf{HJ} & \textbf{FE} \\
        \midrule
       Social & $a_{15}$ &  80 & 60 & 60 & 100\\ %
       Int.Nav.& $a_{17}$ &  100 & 80 & 70 & 90  \\ %
       Knu.walk. &  $a_{7}$& 90 & 70 & 60& 90 \\ %
       Rhythm & $a_1$ &  80 & 60 & 60 & 80 \\
       Shoot.\& Arch.&  $a_{5}$ &  100& 70 &80 & 90\\ %
        \bottomrule
    \end{tabular}
    \label{tab:Open_world}
\end{table}

\begin{table}[t!]
    \centering 
    \scriptsize
    \caption{Evaluation Results for Model Ensemble (BM = \Bodydata{}, EG = \Eyedata{}, HJ = \Handdata{}, BM\&EG = Ensemble of \Bodydata{} and \Eyedata{} models, EG\&HJ = Ensemble of \Eyedata{} and \Handdata{} models). }
    \begin{tabular}{r | l| r | r| r| r|r}
        \toprule
        \textbf{App No.} & \textbf{App Group} & \multicolumn{4}{c}{\textbf{Accuracy (\%)}}\\
        \cline{3-7}
        & & \textbf{BM} &\textbf{EG} &\textbf{HJ} &\textbf{BM\&EG} & \textbf{EG\&HJ} \\
        \midrule
        Social & $a_{12}$ & 85 & 80 & 60 & 95& 80\\ 
        \hline
        Teleportation & $a_{4},a_{8}$ & 75,80 & 90,70 & 35,45 & 100, 90 & 90, 80\\ 
        \hline
        Flight Simulation & $a_{3},a_{20}$ & 95,95 & 85,75 & 80,75 & --,-- & 90,85 \\ 
        \hline
        Knu.Walking & $a_{7}$ &  95 & 80 & 65 & -- & 85\\ 
        \hline
        Int. Nav. & $a_2, a_{9}$ &  95,80 & 80,80 & 60,60 & --,80 & 90,80\\
        \hline
        Golfing & $a_6$ & 80 & 70 & 50 & 90 & 80\\
        \bottomrule
    \end{tabular}
    \label{tab:ensemble}
\end{table}

\end{document}